\documentclass[11pt]{article}
\RequirePackage{amsthm,amsmath,amsfonts,amssymb}
\RequirePackage[authoryear]{natbib}
\RequirePackage[colorlinks,citecolor=blue,urlcolor=blue,backref=page]{hyperref}
\RequirePackage{graphicx}
\usepackage{anyfontsize}
\usepackage{mathrsfs}
\usepackage[title]{appendix}%
\usepackage[dvipsnames,svgnames,x11names]{xcolor}%
\usepackage{textcomp}%
\usepackage{manyfoot}%
\usepackage{booktabs}%
\usepackage{algorithm}%
\usepackage{algorithmicx}%
\usepackage{algpseudocode}%
\usepackage{listings}%
\usepackage{bm}
\usepackage{longtable}
\usepackage{caption}
\usepackage{makecell}
\usepackage{authblk}
\usepackage{lmodern}

\makeatletter
\newenvironment{breakablealgorithm}
  {
   \begin{center}
   \refstepcounter{algorithm}
   \hrule height.8pt depth0pt \kern2pt
   \renewcommand{\caption}[2][\relax]{%
     {\raggedright\textbf{\ALG@name~\thealgorithm} ##2\par}%
     \ifx\relax##1\relax
       \addcontentsline{loa}{algorithm}{\protect\numberline{\thealgorithm}##2}%
     \else
       \addcontentsline{loa}{algorithm}{\protect\numberline{\thealgorithm}##1}%
     \fi
     \kern2pt\hrule\kern2pt
   }}
  {
   \kern2pt\hrule\relax
   \end{center}
  }
\makeatother

\newcommand{\bfb}{\mathbf{b}}
\newcommand{\bfc}{\mathbf{c}}

\newcommand{\bfk}{\mathbf{k}}

\newcommand{\bfm}{\mathbf{m}}

\newcommand{\bfp}{\mathbf{p}}

\newcommand{\bfw}{\mathbf{w}}
\newcommand{\bfx}{\mathbf{x}}
\newcommand{\bfy}{\mathbf{y}}

\newcommand{\bfB}{\mathbf{B}}
\newcommand{\bfC}{\mathbf{C}}
\newcommand{\bfD}{\mathbf{D}}

\newcommand{\bfF}{\mathbf{F}}

\newcommand{\bfI}{\mathbf{I}}

\newcommand{\bfK}{\mathbf{K}}
\newcommand{\bfL}{\mathbf{L}}

\newcommand{\bfS}{\mathbf{S}}

\newcommand{\bfV}{\mathbf{V}}
\newcommand{\bfW}{\mathbf{W}}
\newcommand{\bfX}{\mathbf{X}}

\newcommand{\bfbeta}{\bm{\beta}}

\newcommand{\bfepsilon}{\bm{\epsilon}}

\newcommand{\bfxi}{\bm{\xi}}
\newcommand{\bfpsi}{\bm{\psi}}
\newcommand{\bftheta}{\bm{\theta}}
\newcommand{\bfzeta}{\bm{\zeta}}

\newcommand{\bfmu}{\bm{\mu}}

\newcommand{\bfnu}{\bm{\nu}}

\newcommand{\bfPhi}{\bm{\Phi}}

\newcommand{\bfOmega}{\bm{\Omega}}

\newcommand{\iid}{\overset{\text{iid}}{\sim}}
\begin{document}

\title{Co-SIVI: A Correlated Semi-Implicit Variational Approach
for Spatial Models}

\author[1]{Sébastien Garneau}
\author[2]{Carlos T. P. Zanini}
\author[1]{Alexandra M. Schmidt}

\affil[1]{%
Department of Epidemiology, Biostatistics and Occupational Health,
McGill University, Montreal, Quebec, Canada\\
\href{mailto:sebastien.garneau@mail.mcgill.ca}
{\texttt{sebastien.garneau@mail.mcgill.ca}}\\
\href{mailto:alexandra.schmidt@mcgill.ca}
{\texttt{alexandra.schmidt@mcgill.ca}}
}

\affil[2]{%
Department of Statistical Methods,
Federal University of Rio de Janeiro, Rio de Janeiro, Brazil\\
\href{mailto:carloszanini@dme.ufrj.br}
{\texttt{carloszanini@dme.ufrj.br}}
}
\date{}
\maketitle

\begin{abstract}
We propose correlated semi-implicit variational inference (Co-SIVI) , a scalable approach for full posterior approximation in large spatial models with exponential-family likelihoods. Co-SIVI incorporates dependence directly into the conditional variational distribution of spatial random effects through an iterative weighted least squares algorithm that accommodates both Gaussian process and nearest-neighbor Gaussian process (NNGP) priors. For large samples, we further propose reparameterizing the variational family for the covariance parameters to better capture posterior dependence. Co-SIVI addresses an important limitation of semi-implicit variational inference (SIVI), for which dependence induced through the mixing distribution may be insufficient when spatial random effects are explicitly included in the variational family. In simulations with Gaussian, Poisson, Gamma, and Bernoulli outcomes, Co-SIVI closely reproduces Hamiltonian Monte Carlo (HMC) results at substantially lower computational cost, while SIVI performs similarly for marginalized Gaussian models. We apply SIVI and Co-SIVI, both with NNGP priors, to two large-scale datasets: temperature data modeled with a Gaussian likelihood and housing price data modeled with a Gamma likelihood. Overall, Co-SIVI provides a scalable and flexible alternative to HMC for \emph{full posterior} approximation in large non-Gaussian spatial models.
\end{abstract}

\noindent\textbf{Keywords:}
Exponential family; Gaussian process; geostatistics; variational distribution.

\section{Introduction}\label{sec:intro}

In geostatistics, observations are collected at a set of $n$ fixed locations $\mathcal{S} = \{s_1,\ldots,s_n\}$ within a region of interest $\mathcal{D} \subset \mathbb{R}^D$, where typically $D=2$ or $D=3$. Such data are commonly referred to as point-referenced data. At each location $s \in \mathcal{S}$, let $y(s)$ denote the observed outcome and let $\bfx(s)$ be the $p \times 1$ vector of covariates with corresponding vector of coefficients $\bfbeta \in \mathbb{R}^p$. Following \cite{Diggle_geostatistics}, a general model for point-referenced observations within the region $\mathcal{D}$ can be expressed as \begin{equation}
\label{eq:spatial_model}
    h(\mathbb{E}[y(s)\mid w(s)]) = \bfx(s)^\top \bfbeta + w(s), \quad s \in \mathcal{D},
\end{equation} where $h(\cdot)$ is a known link function and $w(s)$ is a spatial random effect that accommodates residual spatial dependence left after accounting for the fixed effects.

Conditionally on the spatial random effects $\bfw = (w(s_1), \ldots, w(s_n))^\top$, the observations $\bfy = (y(s_1), \ldots, y(s_n))^\top$ are typically assumed to be independent and to follow a distribution in the exponential family, with density 
\begin{equation}
    \label{eq:exp_fam}
    f(y(s_i)\mid \eta_{s_i}) = c(y(s_i),\alpha)\exp\left(\frac{y(s_i)\eta(s_i) - b(\eta(s_i))}{\alpha}\right),
\end{equation}
where $\eta(s_i)$ is the canonical parameter and $\alpha$ is a dispersion parameter. The mean is given by  $\mu(s_i) =b'(\eta(s_i))$, with $b'(\cdot)$ denoting the first derivative of $b(\cdot)$, and the variance by $\operatorname{Var}(y(s_i)) = \alpha \times b''(\eta(s_i))$, with $b''(\cdot)$ denoting the second derivative $b(\cdot)$. The mean is related to the covariates and the spatial random effects through \eqref{eq:spatial_model}.
 
Gaussian processes (GPs) have been widely used to model such spatial dependence. A stochastic process $\{w(s); \ s \in \mathcal{D}\}$ is said to be a GP with mean function $\mu:\mathcal{D} \rightarrow \mathbb{R}$ and covariance function $V:\mathcal{D}\times \mathcal{D} \rightarrow \mathbb{R}$, which we denote as $w (\cdot)\sim \operatorname{GP}(\mu, V)$,  if for any finite set of locations $s_1, \ldots, s_n \in \mathcal{D}$ and for any $n \in \mathbb{N},$ the random vector $\bfw = (w(s_1), \ldots, w(s_n))^\top$ follows a multivariate normal distribution $\bfw \sim N(\bfmu, \bfV_n)$ with mean vector $\bfmu = (\mu(s_1), \ldots, \mu(s_n))^\top$ and covariance matrix $\bfV_n$ given by $[\bfV_n]_{ij} = \operatorname{Cov}(w(s_i), w(s_j)) = V(s_i, s_j).$

In the spatial model described by \eqref{eq:spatial_model}, it is commonly assumed that $w(\cdot) \sim \operatorname{GP}(0, V)$, where the modeling choice for the covariance function $ V(s,s'); s,s' \in \mathcal{D}$ determines how spatial dependence varies as a function of the Euclidean distance between locations $s$ and $s'$ in $\mathcal{D}$. A common choice for $V$ is the Matérn covariance function \citep{Matern_cov} given by $V(s, s')= 
\sigma^2 \frac{2^{1-\nu}}{\Gamma(\nu)}\left(\sqrt{2\nu}d(s,s')\phi \right)^\nu K_\nu\left(\sqrt{2\nu}\phi d(s,s')\right)$, where $\sigma^2>0$ is the marginal variance of the spatial process, $\Gamma(\cdot)$ is the Gamma function, $K_\nu(\cdot)$ is the modified Bessel function of the second kind, $d(s, s')$ is the Euclidean distance between locations $s$ and $s'$, $\nu > 0$ controls the smoothness of the spatial process, and $\phi>0$ is the range parameter controlling the strength of the spatial correlation as distance between locations increases.  Throughout this paper, we consider two special cases of the Matérn covariance function: $\nu = 1/2$, corresponding to the exponential covariance function
, and $\nu = 5/2$, which we denote by Matérn $5/2$. These covariance functions induce different smoothness levels: the exponential process is not mean-squared differentiable, whereas the Matérn $5/2$ process is twice mean-squared differentiable.
The covariance parameters are denoted by $ \bfPhi = (\Phi_1, \ldots,\Phi_c)^\top$, where $c$ denotes the number of covariance parameters. For example, $\bfPhi = ( \sigma^2, \phi )^\top$ for the exponential covariance function.

However, the use of a GP comes with an important drawback: performing inference with a dense $n$-dimensional GP requires $\mathcal{O}(n^3)$ operations and $\mathcal{O}(n^2)$ memory. Hence, GP models become computationally intensive even for moderately large datasets. Common approaches to address this problem include stochastic partial differential equations \citep{Lindgren_SPDE, Rue_INLA}, low-rank approximations \citep{Banerjee_GPP}, and using conditional independence structures \citep{Vecchia_nngp}, among others. Another recent development is stochastic gradient Markov chain Monte Carlo (MCMC) methods, which use mini-batches instead of the full sample to speed up inference \citep{abba_stochastic_gradient}.

To alleviate this issue, \cite{Datta_NNGP} proposes an alternative prior for the spatial random effects known as the nearest-neighbor Gaussian process (NNGP). The main idea is to use Vecchia's approximation \citep{Vecchia_nngp} on the exact factorization on the prior of the spatial random effects $p(\bfw) = p(w(s_1))\prod_{i = 2}^n p(w(s_i) \mid w(s_1),\ldots,w(s_{i-1}))$. This approximation restricts the conditioning set of location $s_i$ to at most $M$ of its nearest neighbors: $p(\bfw) \approx \prod_{i = 1}^n p(w(s_i) \mid \bfw_{N(i)})$, where $N(i)$ denotes the set of at most $M$ nearest neighbors of location $s_i$ within $s_1, \ldots, s_{i-1}$. \cite{Datta_NNGP} shows that this formulation defines a valid Gaussian process with a sparse precision matrix. The main advantage of the NNGP prior is that computations only involve covariance matrices of dimension $M \times M$. Since $M$ is much smaller than the number of observed locations $n$, the NNGP prior greatly reduces computational and memory costs.

Variational inference (VI), a Bayesian technique that approximates the posterior distribution within a prespecified variational family of probability distributions, provides a computationally efficient alternative to MCMC for large-scale models with a GP component, such as those commonly used in spatial statistics. For a detailed introduction to VI, see \cite{Blei_reviewVI}.

VI methods for spatial data often focus on Gaussian outcomes, for which closed-form updates are available for most variational distributions of the model parameters. \cite{REN_VI} derives closed-form updates under a dense GP prior for all parameters other than the range parameter $\phi$, whose posterior distribution is approximated using sampling importance resampling \citep{Rubin_sampling}. Although \cite{REN_VI} is one of the first approaches to apply VI to spatial models, it struggles to scale to large datasets as it uses a dense GP. \cite{Wu_VNNGP} addresses this limitation by replacing the dense GP with an NNGP prior. However, they rely on a mean-field approximation, meaning they ignore posterior correlations among parameters. \cite{song_VINNGP} addresses this issue by using a correlated Gaussian distribution as the variational family, thereby allowing for posterior correlations. Recently, \cite{Nieman_Szabo_GP_reg} establishes theoretical guarantees for posterior contraction of covariance hyperparameters of Gaussian processes in the context of variational non-parametric Gaussian process regression. 

As these methods rely on closed-form updates, extending them to non-Gaussian outcomes is often challenging. Moreover, several existing approaches either require computationally expensive procedures for inference of the range parameter ($\phi$) or treat it as a {\em fixed} hyperparameter, providing limited or no uncertainty quantification.

Extensions to other likelihoods include \cite{lee_lee_infvb}, where a basis representation is used to approximate the dense GP, along with further approximations to obtain closed-form updates for non-Gaussian likelihoods. The method of \cite{Hensman-Classification} uses a low-rank approximation of the dense GP and relies on stochastic gradient optimization to update the variational distribution. However, since the range parameter is optimized as a hyperparameter instead of a model parameter, no variational distribution is obtained for it, and its uncertainty cannot be quantified. 

Therefore, there remains a need for scalable VI methods that accommodate non-Gaussian likelihoods and provide an accurate approximate joint posterior distribution for all model parameters, including covariance parameters such as the range parameter.

In this paper, we introduce correlated semi-implicit variational inference (Co-SIVI) for non-Gaussian spatial models, where strong posterior dependence among parameters may be present. Although standard semi-implicit variational inference (SIVI) \citep{SIVI} can, in principle, capture posterior correlations through its implicit mixing distribution, it can struggle to learn strong correlations in practice. Co-SIVI builds on SIVI by introducing model-informed dependence into the conditional variational distribution, facilitating the learning of strong posterior correlations. For large Gaussian spatial models, we propose SIVI-NNGP, which combines SIVI with an NNGP prior for the spatial random effects. SIVI-NNGP provides a scalable approach to accurately approximate the joint posterior distribution when the spatially correlated random effects are marginalized out from the model.

Recent work has also explored the use of SIVI with basis-function representations for large-scale spatial modeling \citep{lee_lee_sivi}. This provides a complementary strategy to the GP-based approach considered here. Basis-function methods reduce the dimension of the spatial process through a low-rank representation, whereas our approach preserves the spatial random effects at the observed locations and allows for inference on the range parameter $\phi$.

We evaluate Co-SIVI through simulation studies comparing it with wide\-ly available, general-purpose VI methods that estimate all model parameters for non-Gaussian spatial models and Gaussian spatial models under the conditional formulation. We similarly evaluate SIVI for Gaussian spatial models and one non-Gaussian spatial model. Specifically, we consider automatic differentiation variational inference (ADVI) \citep{ADVI} and Pathfinder \citep{Pathfinder}. Further, we compare these VI methods with the No-U-Turn sampler (NUTS) by \cite{hoffman2014}, an adaptive variant of Hamiltonian Monte Carlo (HMC) \citep{HMC}. To further improve computational efficiency, we consider the combination of Co-SIVI with an NNGP prior. The proposed Co-SIVI method requires a differentiable log joint density and, for the reparameterization developed here, a likelihood belonging to the exponential family. Therefore, it applies to a broad range of commonly used model specifications. We explore Gaussian, Poisson, Gamma, and Bernoulli outcomes to illustrate the proposed methodology. 

The remainder of this paper is organized as follows. In Section \ref{sec:VI} we review variational inference methods, focusing on the variational methods used in comparison to Co-SIVI. Section \ref{sec:CoSIVI} presents the proposed method and derives the specific computational steps needed for Poisson and Gamma spatial models. The corresponding derivations for Gaussian and Bernoulli spatial models are provided in Section A of the supplementary materials. The simulation studies are described in Section \ref{sec:Simulation}. We apply SIVI-NNGP and Co-SIVI to large-scale datasets in Section \ref{sec:large_scale}. We conclude this paper with a discussion in Section \ref{sec:discussion}.

\section{Review of Variational Inference Methods}\label{sec:VI}

Variational inference (VI) is an optimization-based approach for approximating a complex posterior distribution. Let $\bftheta \in \mathbb{R}^b$ denote the vector of model parameters. The Kullback-Leibler (KL) divergence between two densities $f_1(\bftheta)$ and $f_2(\bftheta)$ is defined as $\operatorname{KL}[f_1(\bftheta) \parallel f_2(\bftheta)] = \mathbb{E}_{f_1}(\log f_1(\bftheta) - \log f_2(\bftheta))= \int \left[\log f_1(\bftheta) - \log f_2(\bftheta) \right] f_1(\bftheta) d\bftheta$. The main idea of VI is to select a family $\mathcal{Q}$ of approximating distributions (called the variational family) and then find the member $q^*(\bftheta) \in \mathcal{Q}$ of the variational family that is closest to the posterior distribution $p(\bftheta\mid \bfy)$ in terms of the KL divergence: $q^*(\bftheta) = \operatorname{argmin}_{q\in \mathcal{Q}} \operatorname{KL}(q(\bftheta) \parallel p(\bftheta \mid \bfy))$. The mean-field approximation, one of the most commonly used restrictions when defining the variational family $\mathcal{Q}$, assumes that the variational distribution can be factorized into $B$ independent blocks of parameters: $q(\bftheta) = \prod_{i=1}^Bq(\bftheta_i), \quad \bftheta = (\bftheta_1^\top, \ldots, \bftheta_B^\top)^\top$.  

Direct evaluation of the KL divergence $\operatorname{KL}(q(\bftheta) \parallel p(\bftheta \vert \bfy))$ involves the marginal likelihood $p(\bfy)$, which is usually intractable. However, as it does not depend on $q(\cdot)$, minimizing the KL divergence over $q \in \mathcal{Q}$ is equivalent to maximizing the evidence lower bound (ELBO) defined as $\operatorname{ELBO} = -\mathbb{E}_q[\log q(\bftheta) ]+\mathbb{E}_q[\log p(\bftheta,\mathbf{y})] $, where $p(\bftheta, \bfy) = p(\bfy \mid \bftheta)p(\bftheta)$ corresponds to the joint density given by the likelihood multiplied by the prior. In fact, the KL divergence can be rearranged as
\begin{align*} 
    \operatorname{KL}(q(\bftheta) \parallel p(\bftheta  \mid \bfy) ) &= 
    \mathbb{E}_q[\log q(\bftheta) ]-\mathbb{E}_q[\log p(\bftheta\mid \bfy) ] \\
    &= \mathbb{E}_q[\log q(\bftheta) ]-\mathbb{E}_q[\log p(\bftheta,\bfy )-\log p(\bfy)] \\
    &= \mathbb{E}_q[\log q(\bftheta) ]-\mathbb{E}_q[\log p(\bftheta,\bfy) ] + \log p(\bfy)  \\
    &= - \operatorname{ELBO} + \log p(\bfy),
\end{align*} 
with $\log p(\bfy) $ being referred to as the log-evidence.

In Sections \ref{sec:advi}, \ref{sec:pathfinder}, and \ref{sec:sivi}, we review ADVI, Pathfinder, and SIVI, respectively. 

\subsection{Automatic Differentiation Variational Inference}
\label{sec:advi}

\cite{ADVI} introduces automatic differentiation variational inference (ADVI), a general framework that automates the derivation of variational inference algorithms. The key idea is to transform all model parameters into an unconstrained real-valued space. If $\bftheta = (\theta_1, \ldots, \theta_b)^\top$ represents the original parameters, an invertible and differentiable mapping $T: \bftheta \mapsto\bfxi \in \mathbb{R}^b$ is applied so that the joint distribution of the data and transformed parameters $p(\bfy,\bfxi)$ can be expressed as  $$p(\bfy,\bfxi):= p(\bfy,\bftheta)\bigg\rvert_{\bftheta = T^{-1}(\bfxi)} \times \vert J_{T^{-1}(\bfxi)}\vert,$$ where $\vert J_{T^{-1}(\bfxi)} \vert$ denotes the determinant of the Jacobian of the inverse transformation $T^{-1}$. This transformation defines the variational family, typically a multivariate Gaussian, on the unconstrained transformed parameter vector $\bfxi$ and reparameterizes it for gradient-based optimization.

Two common choices for the variational family in the context of ADVI are the mean-field Gaussian and the full-rank Gaussian families. The mean-field  Gaussian variational family assumes that the covariance matrix is diagonal: $q(\bfxi) = \prod_{i=1}^b N(\xi_i \vert \mu_i, \sigma_i^2)$ for $\bfxi = (\xi_1, \ldots, \xi_b)^\top$ with variational parameters $\bfnu = (\mu_1, \ldots, \mu_b, \sigma^2_1, \ldots, \sigma^2_b)^\top$. The full-rank Gaussian variational family assumes $q(\bfxi) = \operatorname{MVN}(\bfxi \mid \bfm, \bfS)$ with variational parameters $\bfnu = (\bfm, \bfL)$ where $\bfS = \bfL \bfL^{\top}$ and $\bfL$ denotes a lower triangular matrix. 

For the full-rank variational assumption, the Gaussian variational distribution can be reparameterized as $\bfxi = \bfm + \bfL\bfepsilon, \ \bfepsilon \sim \operatorname{MVN}(\bf0, \bfI)$ so that the ELBO can then be written as an expectation of a function of $\bfepsilon$ through $\bfxi$:
\begin{align*} 
    \operatorname{ELBO} &= \mathbb{E}_{\bfepsilon \sim N(\bf0, \bfI)}\{  \log p(\bfy, T^{-1}(\bfxi) ) +  \log \mid J_{T^{-1}(\bfxi)}\mid 
   - \log N(\bfxi ; \ \bfm, \bfL\bfL^{\top})\}.
\end{align*}
This expectation can be approximated via Monte Carlo, and the resulting ELBO estimate is optimized with respect to the variational parameters $\bfnu = (\bfm, \bfL)$ using gradient-based algorithms and automatic differentiation software. For example, \cite{kucukelbir} describes the implementation of ADVI in Stan \citep{Stan_Users_Guide}.

In general, the full-rank Gaussian variational family is a natural choice for models with highly correlated parameters, such as spatial models. However, if the dimension of $\bftheta$ is large, which is often the case for real spatial data applications, then this choice requires estimating and manipulating a large dense covariance matrix, making the algorithm computationally expensive.  

\subsection{Pathfinder}
\label{sec:pathfinder}

\cite{Pathfinder} introduces Pathfinder, a variational method for differentiable probability densities. Pathfinder begins from randomly generated initial points, typically located in the tails of the posterior distribution, and follows quasi-Newton optimization trajectories toward regions of high posterior density. Along each trajectory, Pathfinder constructs a sequence of local multivariate Gaussian approximations using gradient and curvature information. The approximation with the highest estimated ELBO is selected as the final variational distribution associated with that trajectory.

In multi-path Pathfinder, several optimization trajectories are run independently. Approximate posterior draws are then obtained by applying importance resampling to the pooled draws from the selected approximations across all trajectories (see \cite{Pathfinder} for the specifics of the implementation). Multi-path Pathfinder is highly parallelizable, as all trajectories are independent.  

Pathfinder and ADVI both use Gaussian approximations to the posterior distribution. They differ in how these approximations are obtained. ADVI directly optimizes the parameters of a Gaussian variational distribution by maximizing the ELBO, whereas Pathfinder constructs local Gaussian approximations along quasi-Newton optimization trajectories. Since Pathfinder depends on the quality of the local Gaussian approximations, it can struggle when the posterior distribution exhibits features not well captured by a Gaussian approximation.

\subsection{Semi-Implicit Variational Inference}
\label{sec:sivi}

\cite{SIVI} introduces semi-implicit variational inference (SIVI), a VI method that defines a flexible variational mixture distribution by combining explicit conditional densities with an implicit mixing distribution. The mixing distribution is often defined implicitly by transforming random noise through a neural network. The posterior distribution $p(\bftheta \mid \bfy)$ is approximated by a variational distribution of the form $ h_{\bfnu}(\bftheta) = \int q(\bftheta \mid\bfpsi)q_{\bfnu}(\bfpsi)d\bfpsi$. Here, $q(\bftheta \mid \bfpsi)$ is an explicit and reparameterizable conditional density, meaning that its analytical expression is known and samples from this distribution can be generated through a differentiable transformation of random noise, while $q_{\bfnu}(\bfpsi)$ is a mixing distribution, which may be implicitly defined. The augmented mixing variables $\bfpsi = (\psi_1, \ldots, \psi_{d_\psi})^\top \in \mathbb{R}^{d_\psi}$ are defined as $\bfpsi = g(\bfepsilon; \bfnu)$, where $g( \cdot \ ; \bfnu)$ typically denotes a feedforward neural network with variational parameters $\bfnu$ (its weights and biases), and $\bfepsilon \sim q(\bfepsilon)$ is a vector of random noise from a fixed distribution of dimension $n_\epsilon$. Therefore, the neural network transforms random noise $\bfepsilon$ into augmented mixing random variables $\bfpsi$, which implicitly defines the variational distribution of $\bfpsi$ as $\bfpsi \sim q_{\bfnu}(\bfpsi)$. The variational parameters $\bfnu$ would be optimized based on the ELBO,
\begin{align}
    \label{eq:elbo_sivi_1}
    \operatorname{ELBO} = \mathbb{E}_{\bftheta \sim h_{\bfnu}}[\log p(\bfy, \bftheta)-\log h_{\bfnu}(\bftheta) ].
\end{align}
However, the ELBO defined in \eqref{eq:elbo_sivi_1} is often intractable since the marginal density $h_{\bfnu}(\bftheta)$ does not usually have a closed form. To address this, \cite{SIVI} proposes a computable lower bound on the ELBO based on $K$ auxiliary variables $\widetilde{\bfpsi}_1,\ldots,\widetilde{\bfpsi}_K$ independently drawn from $q_{\bfnu}(\bfpsi)$, given by
\begin{align} \label{eq:SIVI_LowerBound}
    \underline{ \mathcal{L}}_{K} &=\mathbb{E}_{\bfpsi\sim q_{\bfnu}(\bfpsi)}\mathbb{E}_{\bftheta\sim q(\bftheta\mid\bfpsi)}\mathbb{E}_{\widetilde{\bfpsi}_1,\ldots,\widetilde{\bfpsi}_K \iid q_{\bfnu}(\bfpsi)}\Bigg\{\\
    &\log p(\bfy, \bftheta) 
    - \log\left( \frac{1}{K+1} \left( q(\bftheta \mid \bfpsi) + \sum^K_{k=1}q(\bftheta\mid\widetilde{\bfpsi}_k)\right) \right)\Bigg\} .\nonumber
\end{align}

The parameter vector $\bftheta$ is generated by first sampling $\bfpsi \sim q_{\bfnu}(\bfpsi)$ and then sampling $\bftheta \sim q(\bftheta \mid \bfpsi)$. Notice that the auxiliary variables are not used to generate $\bftheta$. Instead, they are used to evaluate the mixture density.

In practice, the expectations in \eqref{eq:SIVI_LowerBound} are approximated by Monte Carlo. For each Monte Carlo sample $j = 1, \ldots, J$, we first generate $\bfpsi_j = g(\bfepsilon_j; \bfnu)$, $\bfepsilon_j \iid q(\bfepsilon)$. Then, we sample $\bftheta_j \sim q(\bftheta \mid \bfpsi_j)$. We also generate $K$ auxiliary samples $\widetilde{\bfpsi}_k \sim g(\bfepsilon_k; \bfnu) ,\bfepsilon_k \sim q(\bfepsilon),  k=1,\ldots,K$. The resulting Monte Carlo approximation of the lower bound $\underline{ \mathcal{L}}_{K}$ from \eqref{eq:SIVI_LowerBound} is
\begin{align}
\label{eq:sivi_monte_carlo}
    \widehat{ \underline{\mathcal{L}}}_{K} 
    & = \frac{1}{J}\sum^J_{j=1} \left[\log p(\bfy, \bftheta_j) 
    -\log \left( \frac{1}{K+1}\left(q(\bftheta_j\mid\bfpsi_j)+\sum^K_{k=1} q(\bftheta_j\mid \widetilde{\bfpsi}_k) \right) \right) \right] \\
    &= \frac{1}{J}\sum^J_{j=1} \log\left\{\frac{p(\bfy,\bftheta_j)}{\frac{1}{K+1}\left[q(\bftheta_j\mid\bfpsi_j)+\sum^K_{k=1} q(\bftheta_j\mid \widetilde{\bfpsi}_k) \right]}\right\}. \nonumber
\end{align}
Recall that the components of $\bfepsilon,\bfepsilon_1, \ldots,\bfepsilon_K \iid q(\bfepsilon)$, and that the parameterizations $\bfpsi = g(\bfepsilon; \bfnu)$ and $\widetilde{\bfpsi}_k = g(\bfepsilon_k; \bfnu)$ are differentiable, at least almost everywhere, with respect to the variational parameters $\bfnu$. Therefore, the gradients for optimization of \eqref{eq:sivi_monte_carlo} with respect to $\bfnu$ can be calculated using an automatic differentiation software. Typically, $q(\bfepsilon)$ is a standard Gaussian distribution.

In practice, the conditional variational distribution $q(\bftheta \mid \bfpsi)$ is specified such that all elements of $\bftheta$ are conditionally independent given $\bfpsi$. For instance, if $\bftheta = \{\theta_1, \ldots, \theta_b\}$, then $q(\bftheta \mid \bfpsi) = \prod_{i=1}^b q(\theta_i \mid \bfpsi)$. It is important to note that this conditional independence does not imply that the overall variational distribution $h_{\bfnu}(\bftheta)$ is constrained to follow the mean-field assumption of independence since $h_{\bfnu}(\bftheta) = \int q(\bftheta \mid \bfpsi) q_{\bfnu}(\bfpsi) d\bfpsi$. In particular, the neural network defining $q_{\bfnu}(\bfpsi)$ almost surely generates dependent components of $\bfpsi$, which allows the marginal variational distribution $h_{\bfnu}(\bftheta)$ to capture posterior dependence among model parameters.

The specific choice of the component distributions $q(\theta_i \mid \bfpsi)$ can vary for each component $\theta_i$. The key requirement is that these distributions assign probability only to valid parameter values. For example, Gaussian distributions can be used for unconstrained parameters, while log-normal distributions can be used for strictly positive parameters. Appropriate transformations, such as exponential or softplus, should be applied to the raw neural network outputs when constraints must be satisfied by the different parameters of the conditional distributions $\bfpsi$.

Next, we illustrate an important limitation of SIVI through its implementation for Gaussian spatial models.

\subsubsection{Gaussian Spatial Model}\label{sec:Gaussian_model}

Gaussian spatial models are a special case of \eqref{eq:spatial_model}, where the link function $h(\cdot)$ is the identity function and the observed outcomes $\bfy$ are Gaussian. In this setting, the model can be expressed in two equivalent formulations. The first, referred to as the conditional model, keeps the random effects vector $\bfw$ explicitly: \begin{equation}
\mathbf{y}\mid \bfbeta, \bfw, \tau^2 \sim \operatorname{MVN}(\mathbf{X}\bfbeta + \mathbf{w}, \tau^2 \bfI_n),
\label{eq:Gaussian_spatial_model}
\end{equation}where $\bfX$ is an $n\times p$ matrix whose $i$-th row is the observed vector of covariates $\bfx(s_i)^\top$, $\bfI_n$ is the identity matrix of dimension $n$, and $\tau^2 > 0$ captures independent residual variance. The spatial random effects $\bfw$ are assigned a $\operatorname{GP}(0, V)$ prior, such that $\bfw = (w(s_1), \ldots, w(s_n)) \sim \operatorname{MVN}(\bf0, \bfV_n),$ with $[\bfV_n]_{ij} = V(s_i, s_j)$. The second formulation, referred to as the marginal model, integrates out the random effects vector $\bfw$, resulting in $\mathbf{y} \mid \bfbeta, \tau^2, \bfPhi \sim \operatorname{MVN}(\mathbf{X}\bfbeta, \bfV_n + \tau^2\bfI_n)$.

In both cases, we factorize the conditional variational distribution $q(\bftheta \mid \bfpsi)$ into a product of univariate conditional distributions for the components of $\bftheta$. Independent Gaussian distributions are used for the regression coefficients $\bfbeta$ and, under the conditional formulation, for the individual spatial random effects $w(s_i), i = 1,\ldots,n$. Independent log-normal distributions are used for the components of $\bfPhi$ and $\tau^2$ to ensure positivity.

To illustrate the conditional and marginal formulations, we selected the first simulated replicate with 500 locations from the simulation study described in Section \ref{sec:sim_gaussian}. Briefly, we generate observations from \eqref{eq:Gaussian_spatial_model} at 500 locations sampled over the square domain $[0,50]^2$. Then, we compare the posterior approximations obtained using SIVI under both formulations, conditional and marginal. The simulated data include one standardized covariate and a spatial random effect generated from an exponential covariance function with moderate spatial dependence. Specifically, the range parameter $\phi$ is fixed at $\phi=0.1$ when simulating the data, which implies that the correlation between two locations is approximately $0.05$ when their distance is approximately equal to $40\%$ of the maximum possible distance over the domain.

\begin{figure}[!ht]
    \centering 
    \includegraphics[width=0.8\textwidth,
    keepaspectratio]{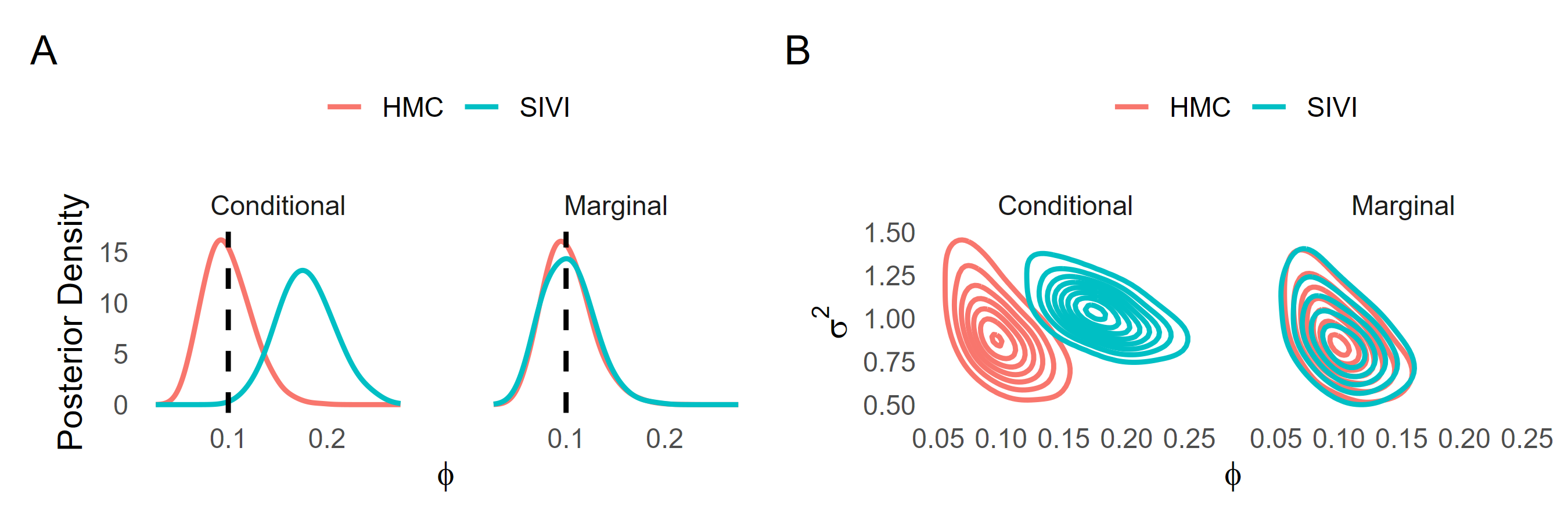}
\caption{\textbf{Posterior densities} of the range parameter $\phi$ (A) obtained with HMC and SIVI under the \textbf{conditional and marginal} formulations of a \textbf{Gaussian} spatial model with an \textbf{exponential} covariance function, for the first replicate with 500 locations. The dashed line indicates the true value, $\phi=0.1$. Corresponding \textbf{joint posterior} distributions of $\phi$ and the spatial process variance $\sigma^2$ (B).}
    \label{fig:gaussian_phi_comp}
\end{figure}

Panel A of Figure \ref{fig:gaussian_phi_comp} compares the posterior distributions of the range parameter $\phi$ obtained using SIVI under the conditional and marginal formulations with the corresponding posterior distributions obtained by HMC. While SIVI approximates the HMC posterior reasonably well under the marginal formulation, it struggles under the conditional formulation. Panel B of Figure \ref{fig:gaussian_phi_comp} compares the joint posterior distributions of the range parameter $\phi$ and the marginal variance of the spatial process $\sigma^2$ under both formulations. Once again, SIVI provides accurate approximations of the HMC posterior under the marginal formulation, but struggles under the conditional formulation. A comparison of the posterior distributions of the spatial random effects under both formulations is available in Section A of the supplementary materials.

It is clear from the panels of Figure \ref{fig:gaussian_phi_comp} that SIVI may have difficulties capturing the strong posterior dependence induced by explicitly retaining the spatial random effects. This limitation motivates the development of correlated SIVI (Co-SIVI), which is designed to better capture strong posterior correlation among parameters. This is especially important for non-Gaussian spatial models, such as Poisson, Gamma, and Bernoulli models, for which marginalizing over the spatial random effects is not analytically tractable. Details regarding the implementation of SIVI with an NNGP prior are available in Section B of the supplementary materials.

\section{Correlated Semi-Implicit Variational Inference}\label{sec:CoSIVI}

One of the main strengths of SIVI is that, even when the conditional distribution $q(\bftheta \mid \bfpsi)$ is specified using conditionally independent components, the resulting marginal variational distribution $h_{\bfnu}(\bftheta)$ can still capture posterior dependence among parameters. However, in this formulation, posterior dependence must be learned implicitly through the neural network defining the mixing distribution $q_{\bfnu}(\bfpsi)$. As a result, when the dependence structure is strong or difficult to learn, the performance of SIVI may deteriorate, as showcased in Section \ref{sec:Gaussian_model}. To address this limitation, we propose correlated semi-implicit variational inference (Co-SIVI), a framework that incorporates explicit dependence among parameters so the neural network is no longer the sole mechanism through which posterior correlations are learned. 

We assume that the conditional distribution of $y(s_i)$ belongs to the exponential family with density described by \eqref{eq:exp_fam}. The core idea of Co-SIVI is to use a joint conditional variational distribution $q(\bfw \mid \bfpsi)$ for the spatial random effects, centered at an iterative weighted least squares (IWLS) approximation, rather than modeling the components of $\bfw$ as conditionally independent. The conditional variational distribution then factorizes as: $q(\bftheta \mid \bfpsi) = q(\bfw \mid \bfpsi)q(\alpha \mid \bfpsi)\prod_{i=1}^pq(\beta_i \mid \bfpsi)\prod_{j=1}^cq(\Phi_j \mid \bfpsi)$. 

The univariate component distributions for the remaining parameters can be specified using the same principles as in SIVI. That is, each conditional distribution $q(\theta_i \mid \bfpsi)$ is chosen to respect the support of the corresponding model parameter. For the random effect block, a natural choice is a multivariate normal distribution $q(\bfw \mid \bfpsi) = \operatorname{MVN}(\bfmu_w , \bfOmega)$ with $\bfpsi_w = (\bfmu_w, \bfOmega)$ being the output units of the neural network $g(\bfepsilon; \bfnu)$ corresponding to the distribution of the random effects $q(\bfw \mid \bfpsi)$.

Although allowing the neural network to learn both the mean vector $\bfmu_w$ and an unstructured covariance matrix $\bfOmega$ may seem attractive at first, this approach scales poorly with the number of random effects. The number of entries in $\bfOmega$, or in its Cholesky factorization, grows quadratically with the dimension of $\bfw$. Therefore, additional structure must be imposed to obtain a scalable method.

Instead of directly specifying a structured covariance matrix for $q(\bfw \mid \bfpsi)$, we introduce dependence through a transformation. Let $\bfzeta = (\zeta(s_1), \ldots$, $\zeta(s_n))^\top$ denote a vector of auxiliary latent variables with conditionally independent components under the SIVI approach. For example, we may take $q(\bfzeta \mid \bfpsi) = \operatorname{MVN}(\bfmu_\zeta,\bfD_\zeta)$ with mean vector $\bfmu_{\zeta} =  (\mu_{\zeta}(s_1), \ldots, \mu_{\zeta}(s_n))^\top$ and diagonal covariance matrix $\bfD_\zeta= \mbox{diag}(\sigma^2_{\zeta(s_1)}, \ldots, \sigma^2_{\zeta(s_n)})$, where the vector $\bfpsi_{\zeta} = (\bfmu_\zeta^\top, \sigma^2_{\zeta(s_1)}, \ldots, \sigma^2_{\zeta(s_n)})$ is the output from the neural network $g(\bfepsilon; \bfnu)$ corresponding to the conditional variational distribution of $\bfzeta$.

Now we transform the variational distribution $q(\bfzeta\mid \bfpsi)$ into the variational distribution of random effects $q(\bfw\mid \bfpsi)$ by defining $\bfw = \bfm + \bfL \bfzeta$, where $\bfL$ is a lower diagonal matrix such that $\bfC = \bfL \bfL^{\top}$ represents a covariance matrix and $\bfm$ is a location vector. Notice that this formulation includes the SIVI representation as a particular case when $\bfm = \bf0$ and $\bfL = \bfI$. More generally, this transformation is capable of inducing dependence among the components of $\bfw$ as long as $\bfL$ is not a diagonal matrix, even though the components of $\bfzeta$ are conditionally independent given $\bfpsi$ due to the diagonal structure of $\bfD_{\zeta}$. 

It is essential to select good values for $\bfm$ and $\bfL$ for Co-SIVI to produce a conditional variational distribution $q(\bfw \mid \bfpsi)$ that, together with $q_{\bfnu}(\bfpsi)$, results in the marginal $q(\bfw)$ being an accurate approximation for the posterior distribution of the random effects $p(\bfw \mid \bfy)$. To obtain these quantities, we use IWLS proposals for a Metropolis-Hastings algorithm for generalized linear mixed models \citep{Gamerman_proposal} to construct the location vector $\bfm$ and the covariance matrix $\bfC = \bfL\bfL^\top$. Given $\bfm$ and $\bfC = \bfL\bfL^\top$, the transformation $\bfw = \bfm + \bfL \bfzeta, \ q(\bfzeta \mid \bfpsi) = \operatorname{MVN}(\bfmu_\zeta,\bfD_\zeta)$ proposed by Co-SIVI can be seen as an offset from the multivariate Gaussian approximation given by the IWLS proposal to $p(\bfw \mid \bfy)$. For instance, if the neural network $\bfpsi = g(\bfepsilon ; \bfnu)$ learns that $\bfmu_{\zeta} = \bf0$ and $\sigma^2_{\zeta(s_1)} = \ldots = \sigma^2_{\zeta(s_n)} = 1$ independently of its inputs $\bfepsilon$, then $q(\bfw \mid \bfpsi)$ reduces to the IWLS proposal. Analogously, if it learns $\bfmu_{\zeta} \approx \bf0$ and $\sigma^2_{\zeta(s_1)} \approx \ldots \approx \sigma^2_{\zeta(s_n)} \approx 1$, it means that slightly modifying the IWLS proposal improves the variational approximation $q(\bfw\mid \bfpsi)$ to the posterior $p(\bfw \mid \bfy)$. Finally, of course, eventually large deviations from the IWLS proposal can also be learned by the neural network under the Co-SIVI formulation.

The core idea of the IWLS proposal is to approximate the likelihood contribution using a Gaussian model based on pseudo-outcomes. First, assume $y(s_i)$ is in the exponential family, as described in \eqref{eq:exp_fam}. In the context of spatially correlated random effects, the linear predictor is related to the mean $\bfmu = (\mu(s_1), \ldots, \mu(s_n))^\top;\mu(s_i) = E(y(s_i) \mid \eta(s_i), \alpha)$ via \eqref{eq:spatial_model}.

At a given iteration, the IWLS proposal defines the pseudo-outcomes $\widetilde{\bfy} = (\widetilde{y}(s_1),\ldots,\widetilde{y}(s_n))^\top$ as $\widetilde{y}(s_i) = \bfx(s_i)^\top\bfbeta + w(s_i) + \left(y(s_i) - \mu(s_i)\right)h'(\mu(s_i))$ where $\mu(s_i) = h^{-1}(\bfx(s_i)^{\top}\bfbeta + w(s_i))$ is evaluated at the sampled value of $\bfbeta$. The distribution of the pseudo-outcomes is then approximated by $\widetilde{\bfy} \sim \operatorname{MVN}(\bfX\bfbeta +\bfw, \bfW^{-1})$, where $\bfW^{-1}$ is diagonal with entries $[\bfW]_{ii}^{-1} = \operatorname{Var}(y(s_i))(h'(\mu(s_i)))^2, \ i = 1, \ldots, n$. Our construction of $\bfW^{-1}$ differs slightly from the original IWLS construction in \cite{Gamerman_proposal}, where the diagonal entries are instead defined as $[\bfW^{-1}]_{ii} = b''(\eta(s_i))h'(\mu(s_i))^2$. Note that both constructions are identical when the dispersion parameter $\alpha$ is equal to 1 since $\operatorname{Var}(y_i) = \alpha b''(\eta(s_i))$. In preliminary analyses, incorporating the dispersion parameter $\alpha$ through $\operatorname{Var}(y(s_i))$ improved performance for models with $\alpha \neq 1$. Combining this approximated likelihood with the Gaussian prior of $\bfw \sim\operatorname{MVN} (\boldsymbol{0}, \bfV_n)$ results in a Gaussian approximation for the proposal distribution of $\bfw$. Following the derivation in \cite{Gamerman_proposal}, the corresponding covariance matrix and mean vector at iteration $t$ are given recursively by
\begin{align} \label{eq:gamerman_general}
    \bfC^{(t)} = \left(\bfV_n^{-1} + \bfW^{(t-1)} \right)^{-1} \, \text{ and } 
    \bfm^{(t)} = \bfC^{(t)}\bfW^{(t-1)} \left(\widetilde{\bfy}^{(t-1)} - \bfX\bfbeta^{(t-1)} \right). 
\end{align}
In Co-SIVI, the construction in \eqref{eq:gamerman_general} cannot be used directly as in the original Metropolis-Hastings algorithm, since the variational optimization does not generate a chain with previous random-effect values $\bfw^{(t-1)}$. Instead, for each Monte Carlo draw of the model parameters, we construct a deterministic approximation to the conditional distribution of $\bfw$. We initialize a temporary vector of random effects at $\bfw^{(0)} = \bf0$ and, using the current sampled values of the other parameters, run a fixed number of IWLS-style updates from \eqref{eq:gamerman_general} by iterating $t = 1, \ldots, T$. Alternatively, the updates could be continued until a prespecified convergence criterion is met. Algorithm \ref{alg:gamerman} summarizes these updates in pseudo-code. The final iteration provides the location vector $\bfm$ and covariance matrix $\bfC$, with Cholesky decomposition $\bfL$. Then, these quantities are used to obtain a sample for $\bfw$ with the transformation $\bfw = \bfm + \bfL\bfzeta$. With this construction, the conditional variational distribution factorizes as $q(\bftheta \mid \bfpsi) = q(\bfw\mid\bfpsi, \alpha, \bfbeta, \bfPhi)q(\alpha\mid\bfpsi) q(\bfbeta \mid \bfpsi)q(\bfPhi\mid\bfpsi)$ with $q(\bfw \vert \bfpsi, \alpha, \bfbeta, \bfPhi) = \operatorname{MVN}(\bfm + \bfL\bfmu_{\bfzeta}, \bfL \bfD_{\zeta}\bfL^\top)$.

Although not strictly necessary, we adopt a factorized structure for  $q(\bfbeta \vert \bfpsi) = \prod_{i=1}^p q(\beta_i \vert \bfpsi)$ and for $q(\bfPhi \vert \bfpsi) = \prod_{j=1}^cq(\Phi_j\vert\bfpsi)$ in our simulation studies without compromising the variational approximations to the posterior distributions (see Section \ref{sec:Simulation}). Other parameterizations are also possible. For example, under the exponential covariance function, the variational distribution may instead be defined on $\kappa = \sigma^2\phi$ and $\sigma^2$, where $\kappa$ is the identifiable combination of the covariance parameters \citep{Zhang_identifiability}. Given samples of $\log\kappa$ and $\log\sigma^2$, samples of $\phi$ are obtained from $\log\phi = \log\kappa - \log\sigma^2$. This reparameterization can induce dependence between $\sigma^2$ and $\phi$ within the variational family and improve uncertainty quantification for both parameters at larger sample sizes (see Section \ref{sec:Housing}), where the number of parameters learned by the neural network becomes large. This reparameterization applies only to the variational family; the model remains parameterized in terms of $\sigma^2$ and $\phi$. The implementation of Co-SIVI with an NNGP prior for the spatial random effects is available in Section B of the supplementary materials.
 
\begin{breakablealgorithm}
\caption{IWLS-Style Updates}\label{alg:gamerman}
\begin{algorithmic}
\State{\bf {Inputs}}: outcome data $\bfy$ of length $n$, observed covariates $\bfX$, link function $h(\cdot)$, number of IWLS-style updates $T \in \mathbb{N}$, prior covariance matrix $\bfV_n$, regression coefficients $\bfbeta$.
\State{Initialize} $\bfw^{(0)}=\bf{0}$
\For{$t = 1, \ldots, T$}
    \State{Calculate $\bfmu^{(t-1)} = h^{-1}(\bfX\bfbeta + \bfw^{(t-1)})$}
    \State{Compute the pseudo-outcomes:}
    \State $\widetilde{\bfy}^{(t-1)} = \bfX\bfbeta + \bfw^{(t-1)} + (\bfy - \bfmu^{(t-1)})h'(\bfmu^{(t-1)})$
    \State{Compute the entries of $\bfW^{-1}$:}
    \State $[\bfW]_{ii}^{-1} = \operatorname{Var}(y(s_i))\left(h'(\mu(s_i)^{(t-1)})\right)^2,\ i = 1,\ldots,n$
    \State{Compute the updated moments:}
    \State $\bfC^{(t)} = \left(\bfV_{n}^{-1} + \bfW^{(t-1)} \right)^{-1}$ and $\bfm^{(t)} = \bfC^{(t)}\bfW^{(t-1)} \left(\widetilde{\bfy}^{(t-1)} - \bfX\bfbeta \right)$  
    \State $\bfw^{(t)} = \bfm^{(t)}$
\EndFor   
\State {Compute the Cholesky decomposition of $\bfC^{(T)}$ as} $\bfC^{(T)} = \bfL^{(T)}\bfL^{(T)\top}$
\State{\bf{Output:}} Final approximation parameters $\bfm^{(T)}$ and $\bfL^{(T)}$
\end{algorithmic}
\end{breakablealgorithm}

Algorithm \ref{alg:Co-SIVI} provides a schematic procedure for implementing Co-SIVI with the Adam optimizer \citep{kingma2017_adam}. Sections \ref{sec:Poisson_model} and \ref{sec:Gamma_model} detail the derivation of Co-SIVI for Poisson and Gamma spatial models, respectively. Section C of the supplementary materials provides the derivation for Gaussian and Bernoulli outcomes, with an application to one artificial dataset each. 

\begin{breakablealgorithm}
\caption{Correlated Semi-Implicit Variational Inference (Co-SIVI)}\label{alg:Co-SIVI}
\begin{algorithmic}
\State{\bf {Inputs}}: outcome data $\bfy$ of length $n$, observed covariates $\bfX$, link function $h(\cdot)$,  number of Monte Carlo samples $J \in \mathbb{N}$, number of auxiliary samples $K \in \mathbb{N}$, number of IWLS-style updates $T \in \mathbb{N}$, initial variational parameters values $\bfnu_0$, Adam learning rate $\eta \in (0,1)$, neural network architecture $g( \ \cdot \ ; \ \bfnu)$
\State{Initialize the neural network parameters $\bfnu = \bfnu_0$}
\For{each training iteration}
    \For{$j = 1, \ldots, J$}
        \State{Simulate a random noise vector:} $\bfepsilon_j \iid q(\bfepsilon)$
        \State{Compute $\bfpsi_j$ from the neural network:} $\bfpsi_j = g(\bfepsilon_j; \bfnu)$
        \State{Sample the non-random effect parameters and latent variables $\bfbeta_{j} \sim q(\bfbeta \mid \bfpsi_j), \ \bfPhi_{j} \sim q(\bfPhi \mid \bfpsi_j), \ \alpha_j \sim q(\alpha \mid \bfpsi_j), \ \bfzeta_{j} \sim q(\bfzeta \mid \bfpsi_j)$.}
        \State{Using the sampled vector $\bfPhi_j$
        , construct the prior covariance matrix of $\bfw$: $\bfV_{n,j}$.} 
        \State {Using the sampled values, construct the approximation parameters $\bfm_j$ and $\bfL_j$ using Algorithm \ref{alg:gamerman}} with inputs $(\bfy, \bfX, h(\cdot), T, \bfV_{n,j}, \bfbeta_j)$.
        \State {Compute the random effects:} $\bfw_j = \bfm_j + \bfL_{j}\bfzeta_j$
        \EndFor
    \State{Generate $K$ auxiliary variables}:
    \State $\widetilde{\bfpsi}_k = g(\widetilde{\bfepsilon}_k; \bfnu), \widetilde{\bfepsilon}_k \iid q(\bfepsilon), \ k = 1, \ldots, K$    
    \State {Approximate the ELBO lower bound by:}
    \State $\widehat{ \underline{\mathcal{L}}}_{K}= \frac{1}{J}\sum^J_{j=1} \log\left\{\frac{p(\bfy,\bftheta_j)}{\frac{1}{K+1}\left[q(\bftheta_j\mid\bfpsi_j)+\sum^K_{k=1} q(\bftheta_j\mid \widetilde{\bfpsi}_k) \right]}\right\}$ where the random effects contribution on $q(\bftheta_j\vert \widetilde{\bfpsi_k})$ (and analogously for $q(\bftheta_j\vert \bfpsi_j)$) is evaluated as $q(\bfw_j \vert \widetilde{\bfpsi}_k,\alpha, \bfbeta,\bfPhi) = q_{\bfzeta}(\bfzeta_j \vert \widetilde{\bfpsi}_k)\left|\det(\bfL_{j}) \right|^{-1}.$
    \State Compute the gradient $\nabla_{\bfnu}\widehat{ \underline{\mathcal{L}}}_{K}$ via automatic differentiation.
    \State {Update the neural network parameters using Adam.}
\EndFor
\State Using the final neural network parameters $\bfnu$, repeat the sampling steps to obtain $N$ approximate posterior samples $\bftheta_1, \ldots \bftheta_N$ 
\State{\bf{Output:}} Final neural network parameters $\bfnu$ and samples  $\bftheta_1, \ldots, \bftheta_N$
\end{algorithmic}
\end{breakablealgorithm}

\subsection{Poisson Spatial Model} \label{sec:Poisson_model}

First, we consider a Poisson likelihood. Let $y(s_i) \mid \lambda(s_i) \sim \operatorname{Poisson}(\lambda(s_i))$, $s_i \in \mathcal{D}$, with canonical logarithmic link function $h(u) = \log(u)$, such that $\log\lambda(s_i) = \bfx(s_i)^\top\bfbeta + w(s_i), \ w(\cdot) \sim \operatorname{GP}(0, V)$. The Poisson likelihood can be written in the exponential family form of \eqref{eq:exp_fam} with $\alpha = 1, \ \eta(s_i) = \log\lambda(s_i), \ b(\eta(s_i)) = \exp(\eta(s_i))$ and $c(y(s_i),\alpha) = \frac{1}{y(s_i)!}$.

For this model, the parameter vector is $\bftheta = \left(\bfbeta^\top, \bfw^\top, \bfPhi^\top \right)^\top$. 
To obtain the conditional variational distribution for the random effects $\bfw$, we construct the pseudo-outcomes. Since $\mu(s_i) = \lambda(s_i)$ and $h'(\mu(s_i)) = \frac{1}{\mu(s_i)}$, the pseudo-outcomes at the current update can be written as $\widetilde{y}^{(t-1)}(s_i) = \bfx(s_i)^\top\bfbeta + w^{(t-1)}(s_i) + \frac{y(s_i) - \mu^{(t-1)}(s_i)}{\mu^{(t-1)}(s_i)}$, where $\mu^{(t-1)}(s_i)$ is given by \newline $\exp\left( \bfx(s_i)^\top \bfbeta + w(s_i) \right)$. The diagonal matrix $\bfW^{-1}$ is defined as $\left([\bfW]_{ii}\right)^{-1}=\operatorname{Var}(y(s_i))  h'(\mu(s_i) ) ^2 = \frac{\mu(s_i)}{\mu(s_i)^2} = \frac{1}{\mu(s_i)}$ for $i = 1, \ldots, n$. Hence, $\bfW^{(t-1)} = \mbox{diag}( \bfmu^{(t-1)})$ with $\bfmu^{(t-1)} = \exp(\bfX\bfbeta + \bfw^{(t-1)})$, where the exponential function is applied element-wise.

Plugging these quantities into \eqref{eq:gamerman_general} gives the update mechanism for the location vector $\bfm$ and the covariance matrix $\bfC$. Specifically, we find $\bfC^{(t)} = \left(\bfV_n^{-1} + \operatorname{diag}(\bfmu^{(t-1)})\right)^{-1}$. The corresponding location vector is $\bfm^{(t)} =$ \newline $\bfC^{(t)}\operatorname{diag}(\bfmu^{(t-1)})(\widetilde{\bfy}^{(t-1)}-\bfX\bfbeta) = \bfC^{(t)}\operatorname{diag}(\bfmu^{(t-1)})\left(\bfw^{(t-1)}+\frac{\bfy-\bfmu^{(t-1)}}{\bfmu^{(t-1)}}\right) = \bfC^{(t)}\left(\bfmu^{(t-1)} \odot \bfw^{(t-1)} + \bfy-\bfmu^{(t-1)}\right)$,  where the division is element-wise and $\odot$ denotes an element-wise multiplication. 

After specifying the prior distributions and conditional variational distributions for the remaining model parameters, these updates can be incorporated into Algorithm \ref{alg:Co-SIVI} to obtain a complete Co-SIVI implementation for the Poisson spatial model.

\subsection{Gamma Spatial Model}\label{sec:Gamma_model}

Second, we consider a Gamma likelihood.  Let $y(s_i) \mid \delta,\mu(s_i) \sim \operatorname{Ga}\left(\delta, \frac{\delta}{\mu(s_i)}\right)$ where the Gamma distribution is parametrized using shape and rate parameters resulting in $\mathbb{E}(y(s_i)) = \frac{\delta}{\delta/\mu(s_i)}=\mu(s_i)$ and $\operatorname{Var}(y(s_i)) = \frac{\delta}{\delta^2/\mu(s_i)^2} = \frac{\mu(s_i)^2}{\delta}$. The mean is linked to the linear predictors using a logarithmic link function. This parametrization of the Gamma likelihood can be written in the exponential family form of \eqref{eq:exp_fam} with $\alpha = \frac{1}{\delta}, \eta(s_i) = \frac{-1}{\mu(s_i)}, b(\eta(s_i)) = -\log(-\eta(s_i))$ and $c(y(s_i), \alpha) = \frac{\alpha^{\alpha}}{\Gamma(\alpha)}y(s_i)^{\alpha-1}$.

For this model, the vector of parameters is $\bftheta = \left(\bfbeta^\top, \bfw^\top, \bfPhi^\top, \delta \right)^\top$. The pseudo-outcomes can be written as $ \widetilde{y}^{(t-1)}(s_i) = \bfx(s_i)^\top\bfbeta + w^{(t-1)}(s_i) + \frac{y(s_i) - \mu^{(t-1)}(s_i)}{\mu^{(t-1)}(s_i)}$, where $\mu^{(t-1)}(s_i) = \exp\left(\bfx^\top(s_i)\bfbeta + w(s_i)\right)$. The diagonal matrix $\bfW^{-1}$ satisfies $\left([\bfW]_{ii}\right)^{-1}=\operatorname{Var}(y(s_i)) h'(\mu(s_i))^2 = \frac{\mu(s_i)^2}{\delta} \frac{1}{\mu(s_i)^2} = \frac{1}{\delta}$. Hence, $\bfW = \delta \bfI_n$ and stays constant during the updates.   

Plugging these quantities into \eqref{eq:gamerman_general} gives the update mechanism for the location vector $\bfm$ and the covariance matrix $\bfC$. Specifically, we find $\bfC = \left(\bfV_n^{-1} + \delta\bfI_n \right)^{-1}$, which also stays constant during the updates. The corresponding location vector is $\bfm^{(t)} = \delta \bfC(\widetilde{\bfy}^{(t-1)}-\bfX\bfbeta) = \delta \bfC\left(\bfw^{(t-1)}+\frac{\bfy-\bfmu^{(t-1)}}{\bfmu^{(t-1)}}\right)$ $= \delta \bfC \left( \bfw^{(t-1)}+\frac{\bfy}{\bfmu^{(t-1)}} - \mathbf{1}\right)$, where $\mathbf{1}$ denotes a vector of 1s. 

\section{Simulation Studies} \label{sec:Simulation}

We evaluate the performance of Co-SIVI for non-Gaussian spatial models, focusing on Poisson and Gamma outcomes, and SIVI for Gaussian outcomes under the marginal formulation. To further highlight SIVI's limitations under conditional formulations, we also apply it to Poisson outcomes with an exponential covariance function. For Gaussian spatial models under the marginal formulation, the random effects are integrated out and therefore do not require the correlated conditional variational distribution used by Co-SIVI. The proposed methods are compared with HMC, ADVI, and Pathfinder, as justified in Section \ref{sec:intro}. Pathfinder results are reported only for the marginal Gaussian formulation, as preliminary experiments showed that most of its paths failed for models in which the spatial random effects are sampled explicitly, even after tuning Pathfinder's hyperparameters. These failures led to unstable posterior approximation and poor predictive performance, and the resulting fits did not satisfy Pathfinder's own diagnostic checks. 

We first describe the general simulation settings shared by all models, including sample sizes, covariance functions, software and hardware, and the comparison criteria. Then, in Sections \ref{sec:sim_gaussian} and \ref{sec:sim_non_gaussian} we present the model-specific details, including the data-generating mechanisms and results for the Gaussian and non-Gaussian spatial models, respectively.

We consider three sample sizes, $n_{total} \in \{70, 170, 520\}$. For each sample size, we split the data into a training set and a validation set of size 20, resulting in training sample sizes of 50, 150, and 500, respectively. For each setting, 50 independent replicates are generated. In each replicate, spatial locations are randomly sampled over a square spatial domain $\mathcal{D}=[0,50]^2$. 

We consider two covariance functions for the spatial random effects: the exponential and the Matérn $5/2$ covariance functions. Results for the Matérn $5/2$ covariance function are presented in Section D of the supplementary materials. We also consider both dense GP and NNGP priors for the random effects. Methods using an NNGP prior are identified by appending a "-NNGP" to the method name. For instance, Co-SIVI-NNGP denotes Co-SIVI implemented with an NNGP prior instead of a dense GP prior. For all NNGP implementations, we set the maximum neighborhood size to $M=15$ and order locations by the first coordinate, then by the second.

We assign independent Gaussian priors to the intercept and fixed effect coefficient, and independent log-normal priors to the strictly positive parameters, such as $\phi$ and $\sigma^2$. Except for the range parameter $\phi$, the log-normal priors are chosen to be weakly informative, allowing for a wide range of prior uncertainty around their true values. For $\phi$, we assign a more informative prior, with its variance chosen such that the prior probability of $\phi$ exceeding twice its prior median is less than approximately $1\%$. In general, it is recommended to use an informative prior for the range parameter $\phi$ in both the Matérn $5/2$ and exponential covariance functions to address well-known identifiability issues in geostatistics \citep{Zhang_identifiability, Tang_identifiability}. The specific prior distributions and model specifications are provided in the corresponding model-specific sections.  

SIVI and Co-SIVI are implemented in Python using TensorFlow \citep{tensorflow}, whereas HMC, ADVI, and Pathfinder are implemented in R via CmdStanR \citep{cmdstanR}. For details on implementing GP in Stan, see the Stan User's Guide \citep{Stan_Users_Guide}. All computations are performed on Rorqual, a high-performance computation cluster operated by {\em Calcul Québec} and part of the Digital Research Alliance of Canada.  

The HMC algorithm is run using four parallel chains, each with 2,000 iterations, including 1,000 warm-up iterations. For ADVI, we use the full-rank Gaussian variational family, as preliminary testing showed poor performance for the mean-field Gaussian family. For Pathfinder, we use 20 independent paths, each with up to 1,000 steps along the optimization trajectory. The L-BFGS history size used for the inverse-Hessian approximation is set to 5, and 50 Monte Carlo samples are used to evaluate the ELBO.

SIVI and Co-SIVI are implemented with a neural network that has three hidden layers of 256 neurons each and ReLU activation functions, and are optimized with the Adam optimizer \citep{kingma2017_adam}. We use a 50-dimensional standard Gaussian noise vector as input. The neural networks are trained for 5,000 iterations, with an Adam learning rate of $0.001$ for the first 2,000 iterations, $0.0005$ for the next 2,000 iterations, and $0.0001$ for the final 1,000 iterations. We use 20 Monte Carlo samples and 100 auxiliary variables, corresponding to $J =20$ and $K = 100$ in \eqref{eq:sivi_monte_carlo}. To accelerate computations, we enable TensorFlow graph mode with the XLA compiler using \verb|@tf.function(jit_compile=True)|. For Co-SIVI, we use $T=3$ IWLS-style updates. For the conditional variational distributions, we use independent Gaussian distributions for the regression coefficients and independent log-normal distributions for the positive model parameters. For the spatial random effects, we use the conditional variational distributions derived in Section \ref{sec:CoSIVI}.  

HMC uses four CPU cores, with one core per chain. ADVI is restricted to a single-core CPU implementation, since it does not currently support parallel execution across multiple cores. Although Pathfinder is highly parallelizable, we restrict it to six CPU cores to represent a realistic computational setting. SIVI and Co-SIVI use one-eighth of an NVIDIA H100 GPU. The CPU-only methods are run on an AMD EPYC 9654 (Zen 4) processor.

The comparison between methods is threefold. First, we assess predictive performance on the validation set using three metrics: the continuous ranked probability score (CRPS), the interval score (IS), and the negative log-predictive density (NLPD) (see \cite{Gneiting_proper_scoring} for more details). Together, these metrics assess predictive accuracy, uncertainty quantification, and calibration. Second, we evaluate parameter estimation by comparing posterior means across replicates and by examining the posterior distributions of selected parameters for a randomly selected replicate. Third, we compare the runtime of the different inference procedures.   
       
\subsection{Marginal Gaussian Spatial Model} \label{sec:sim_gaussian}

Observations are generated from the conditional Gaussian spatial model described in Section \ref{sec:Gaussian_model} with one covariate $x(s)$. To simulate the data, we set the parameter values $\beta_0 = 0$, $\beta_1 = 0.5$, $\sigma^2 = 1$, $\tau^2 = 0.25$. The range parameter $\phi$ is set to $0.1$ for the exponential covariance function and $0.084$ for the Matérn $5/2$ covariance function, such that the correlation between two locations is approximately equal to $0.05$ when they are separated by the same distance under both covariance functions. Under these covariance specifications, the correlation between two locations separated by a distance of approximately 30 ( $\approx40\%$ of the maximum possible distance) is about $0.05$. The covariate $x(s)$ is generated as  $x^*(s) \sim \operatorname{Poisson}(3)$ and then standardized $x(s) = \frac{x^*(s) - \bar{x^*}(s)}{\operatorname{SD}(x^*(s))}$.

We assign independent $N(0,1)$ priors to the intercept and fixed effect coefficient. We assign independent log-normal priors to the covariance parameters. Specifically, we use $\sigma^2 \sim LN(0,0.5)$, $\tau^2 \sim LN(\log(0.25), 0.5)$, and $\phi \sim LN(\log(0.1),0.09)$, where $LN(m,v)$ denotes a log-normal distribution with log-scale mean $m$ and log-scale variance $v$. These distributions yields approximate 95\% prior intervals of $0.25-4.00$, $0.06 - 1.00$ and $0.06 - 0.18$ for $\sigma^2, \tau^2$, and $\phi$, respectively.

\begin{figure}[!hbt]
    \centering
    \begin{minipage}[t]{0.49\textwidth}
        \centering
        \includegraphics[
            width=\linewidth,
            keepaspectratio
        ]{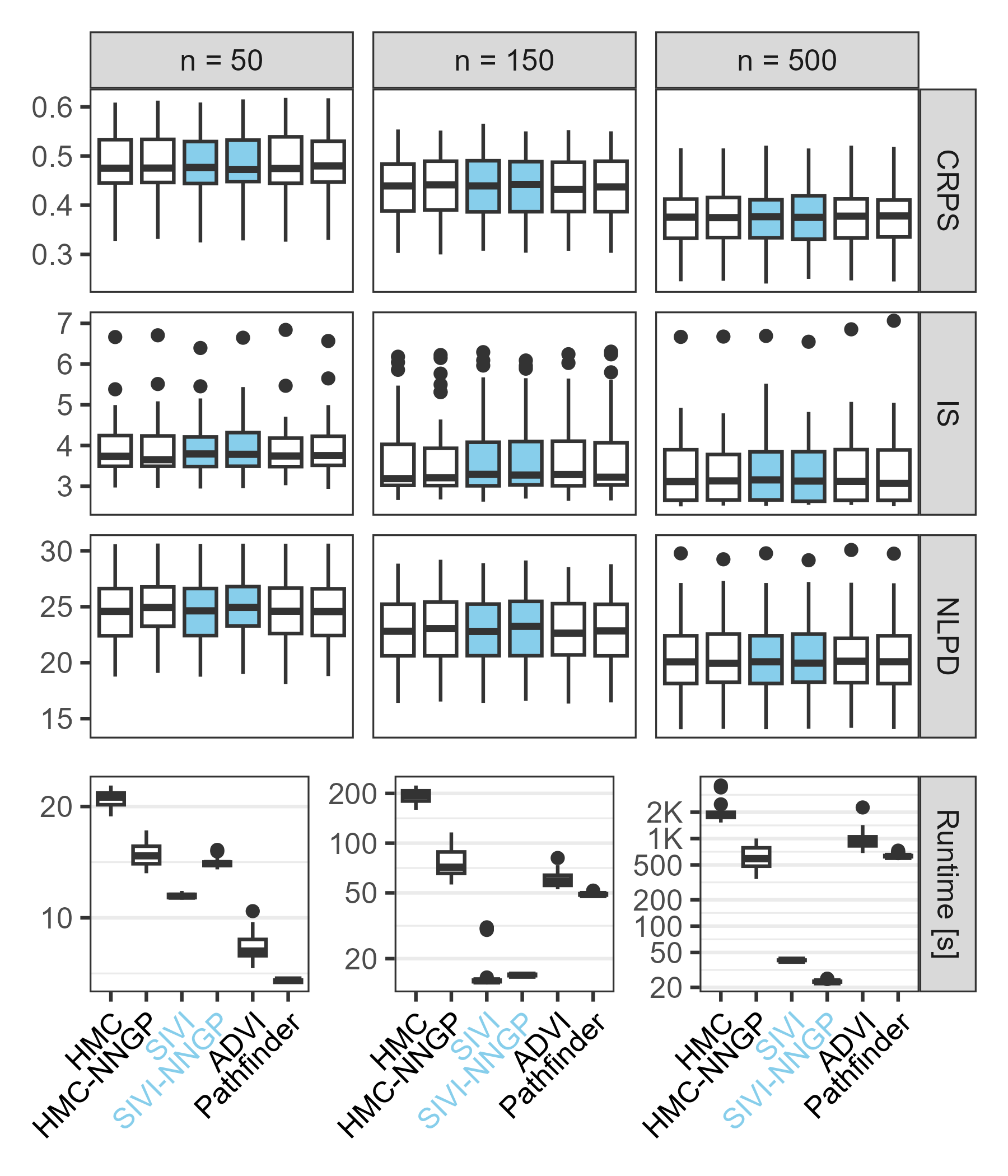}
    \end{minipage}
    \begin{minipage}[t]{0.49\textwidth}
        \centering
        \includegraphics[
            width=\linewidth,
            keepaspectratio
       ]{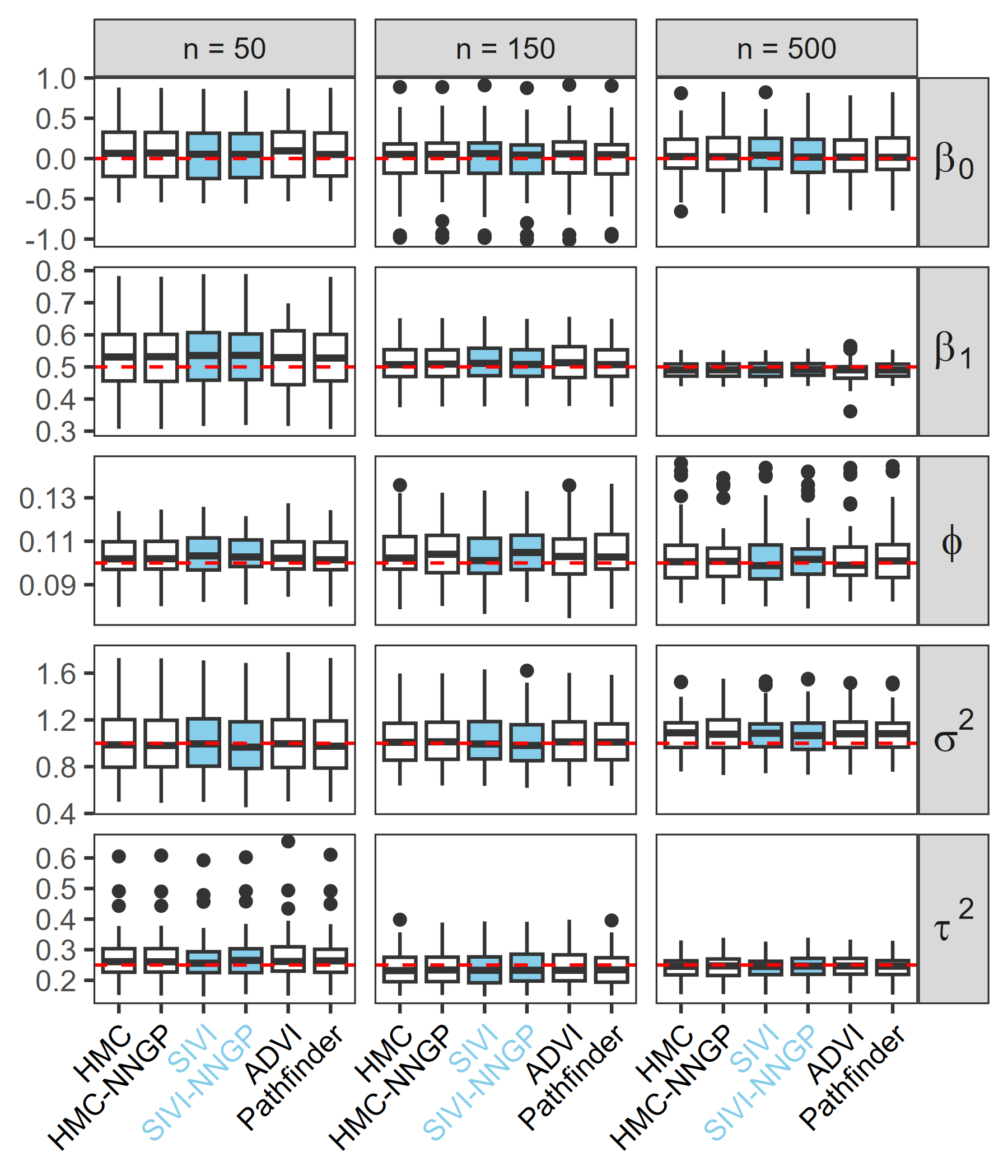}
    \end{minipage}
    \caption{Results across 50 replicates for a \textbf{marginal Gaussian} spatial model with an \textbf{exponential covariance} function. The left panel presents \textbf{predictive scores} and \textbf{runtime} (log scale), while the right panel presents distributions of the \textbf{posterior means} of model parameters. Predictive scores are based on 20 held-out locations. SIVI methods are highlighted in light blue, and dashed lines indicate the true parameter values. CRPS: continuous ranked probability score; IS: interval score; NLPD: negative log-predictive density. Methods with the \texttt{-NNGP} suffix use a nearest-neighbor Gaussian process prior.}
    \label{fig:Results_gaussian_exp}
\end{figure}

The left panel of Figure ~\ref{fig:Results_gaussian_exp} presents the predictive scores and runtime of the methods considered. The last row shows that SIVI becomes the fastest method once the training sample size reaches 150, with the computational advantage increasing at the largest training sample size of 500. For $n_{train} = 500$, SIVI takes, on average, 41 seconds with a dense GP prior and 23 seconds with an NNGP prior. On the other hand, HMC takes, on average, 1967 seconds ($\approx$ 30 minutes) with a dense GP prior and 642 seconds ($\approx 10$ minutes) with an NNGP prior. All methods yield similar prediction scores across sample sizes. The distributions of the posterior means across the 50 replicates are shown in the right panel of Figure ~\ref{fig:Results_gaussian_exp}. The estimates are similar across methods for all sample sizes.

Section D of the supplementary materials compares approximations to the posterior density of the range parameter $\phi$ from one randomly selected replicate, and shows that SIVI yields a posterior distribution similar to HMC.

Overall, the considered methods perform similarly across the predictive and pa\-ra\-me\-ter-estimation criteria. The main difference is the computational time, with SIVI, under either a dense GP  or an NNGP prior distribution, being the fastest method for the two largest sample sizes considered.

\subsection{Non-Gaussian Spatial Models} \label{sec:sim_non_gaussian}

Observations for the Poisson and Gamma spatial models are generated according to the models described in Sections \ref{sec:Poisson_model} and \ref{sec:Gamma_model}, respectively, with one covariate. For both models, we set $\beta_1 = 0.25$ and $\sigma^2 = 0.1$. The range parameter $\phi$ is set to $0.1$ for the exponential covariance function and $0.084$ for the Matérn $5/2$ covariance function. The covariate is independently generated as $x \sim \operatorname{N}(0,1)$. For the Poisson spatial model, we set the intercept to $\beta_0 = 1.5$, whereas for the Gamma spatial model, we set it to $\beta_0 = 1$. The dispersion parameter in the Gamma spatial model is set to $\delta = 2$.

For both models, we assign independent $N(0,1)$ priors to the intercept and fixed effect coefficient. We also assign independent log-normal priors to the covariance parameters. Specifically, we use $\sigma^2 \sim LN(\log(0.1),0.5)$, and $\phi \sim LN(\log(0.1),0.09)$. These prior distributions yield approximate 95\% prior intervals of $0.03-0.40$, and $0.06 - 0.18$ for $\sigma^2$ and $\phi$, respectively. For the dispersion parameter of the Gamma spatial model, we use $\delta \sim LN(\log(2),1)$, which yields approximate 95\% prior intervals of $0.28-14.20$.

\begin{figure}[!ht]
    \centering
    \begin{minipage}[t]{0.47\textwidth}
        \centering
        \includegraphics[
            width=\linewidth,
            keepaspectratio
        ]{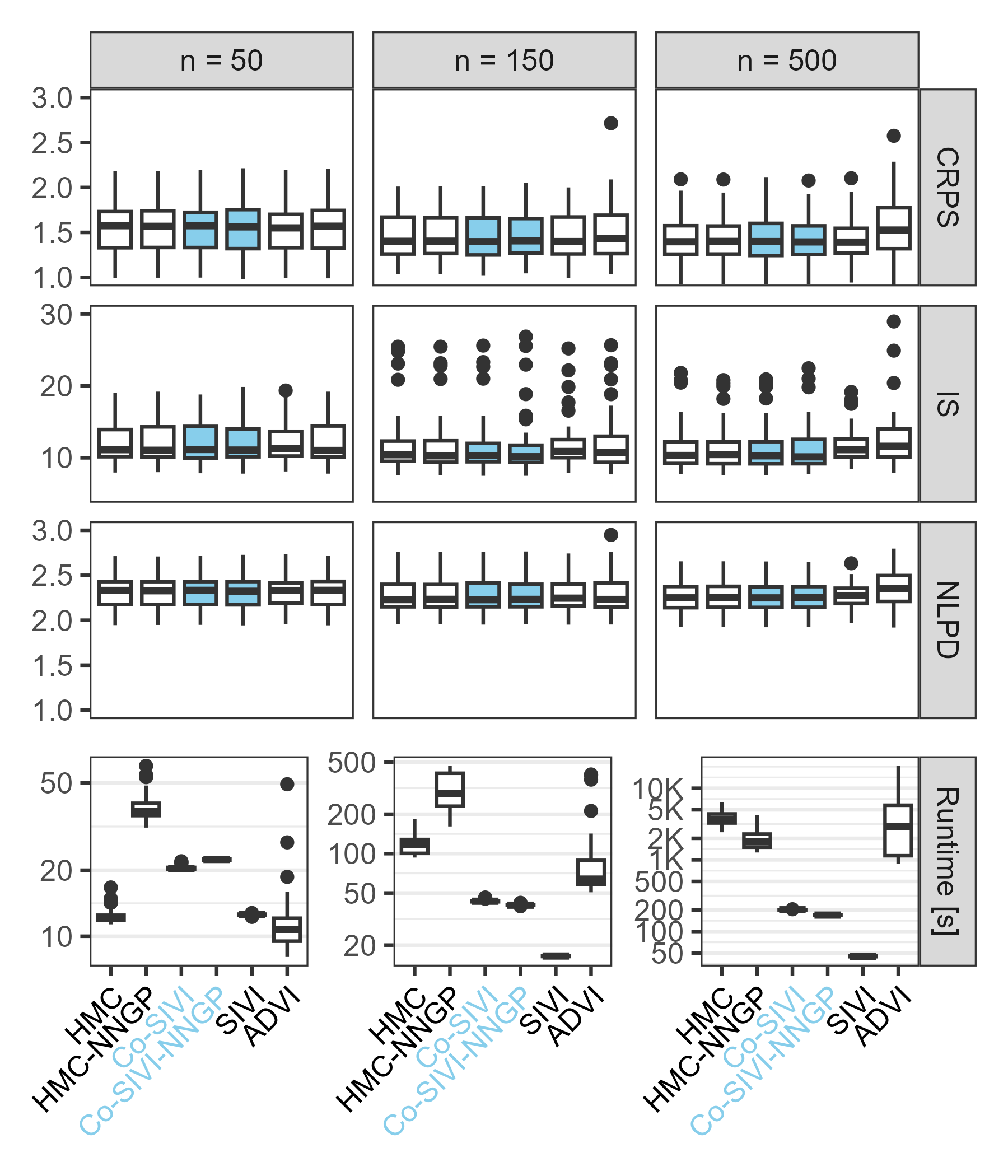}
    \end{minipage}
    \begin{minipage}[t]{0.47\textwidth}
        \centering
        \includegraphics[
            width=\linewidth,
            keepaspectratio
        ]{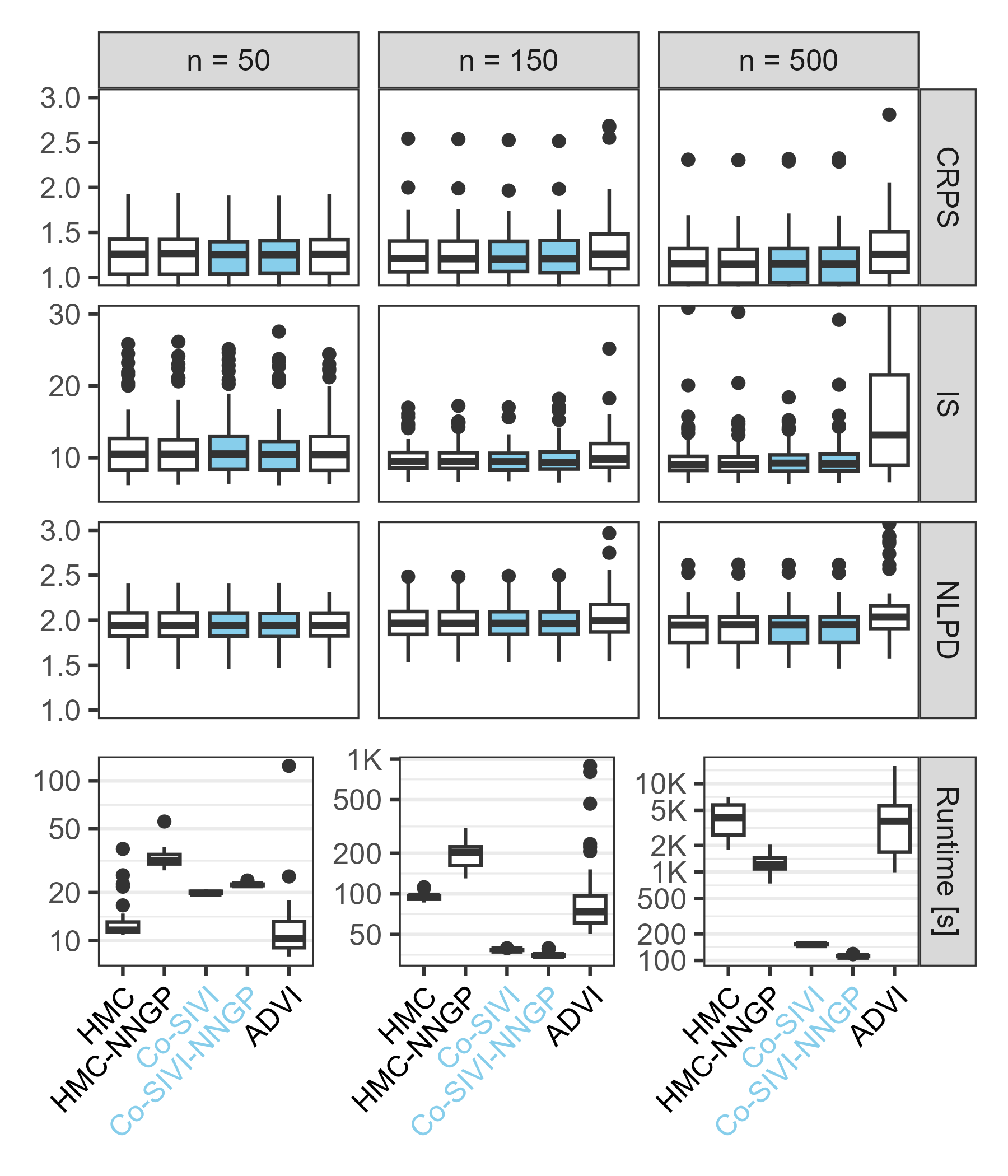}
    \end{minipage}

    \begin{minipage}[t]{0.47\textwidth}
        \centering
        \includegraphics[
            width=\linewidth,
            keepaspectratio
        ]{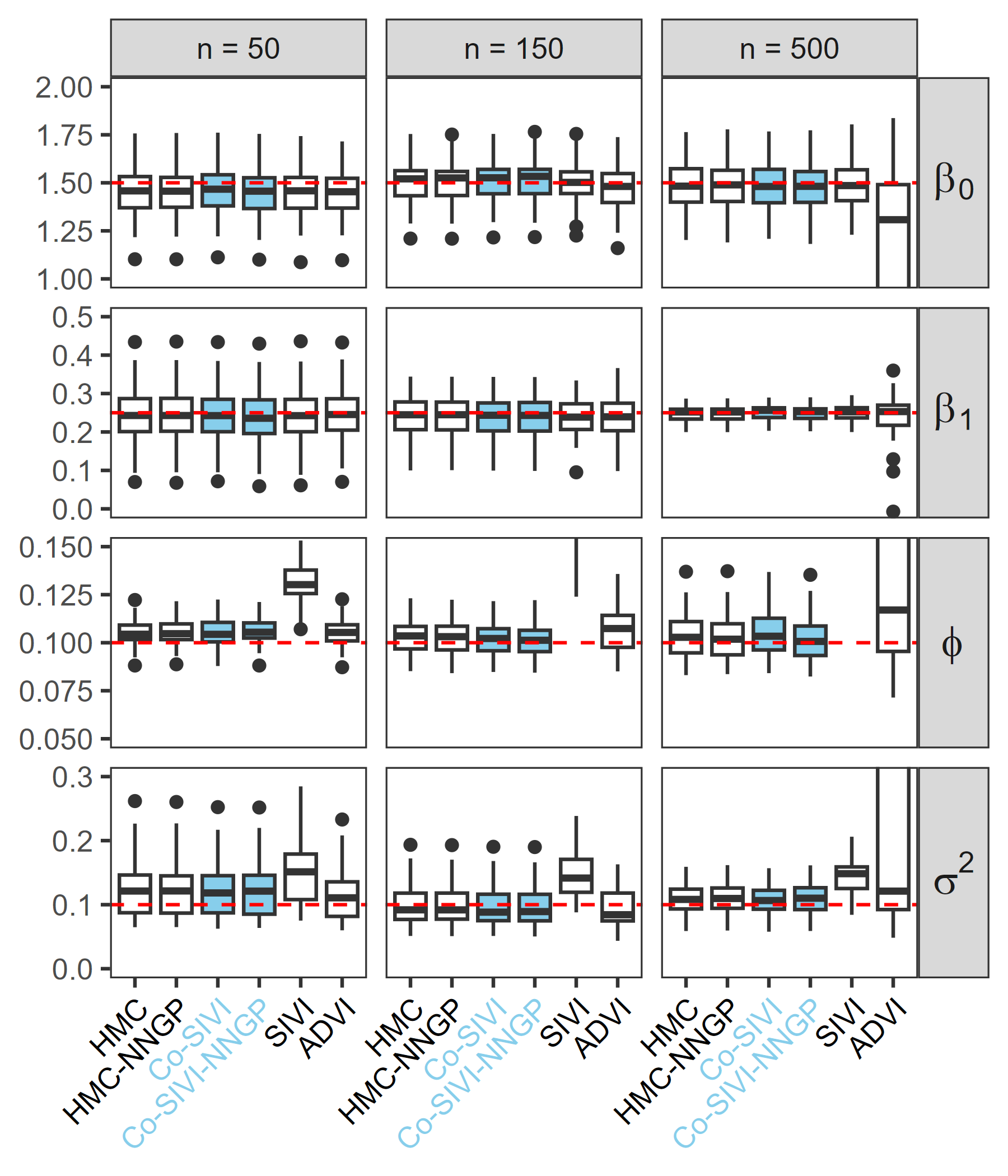}
    \end{minipage}
    \begin{minipage}[t]{0.47\textwidth}
        \centering
        \includegraphics[
            width=\linewidth,
            keepaspectratio
        ]{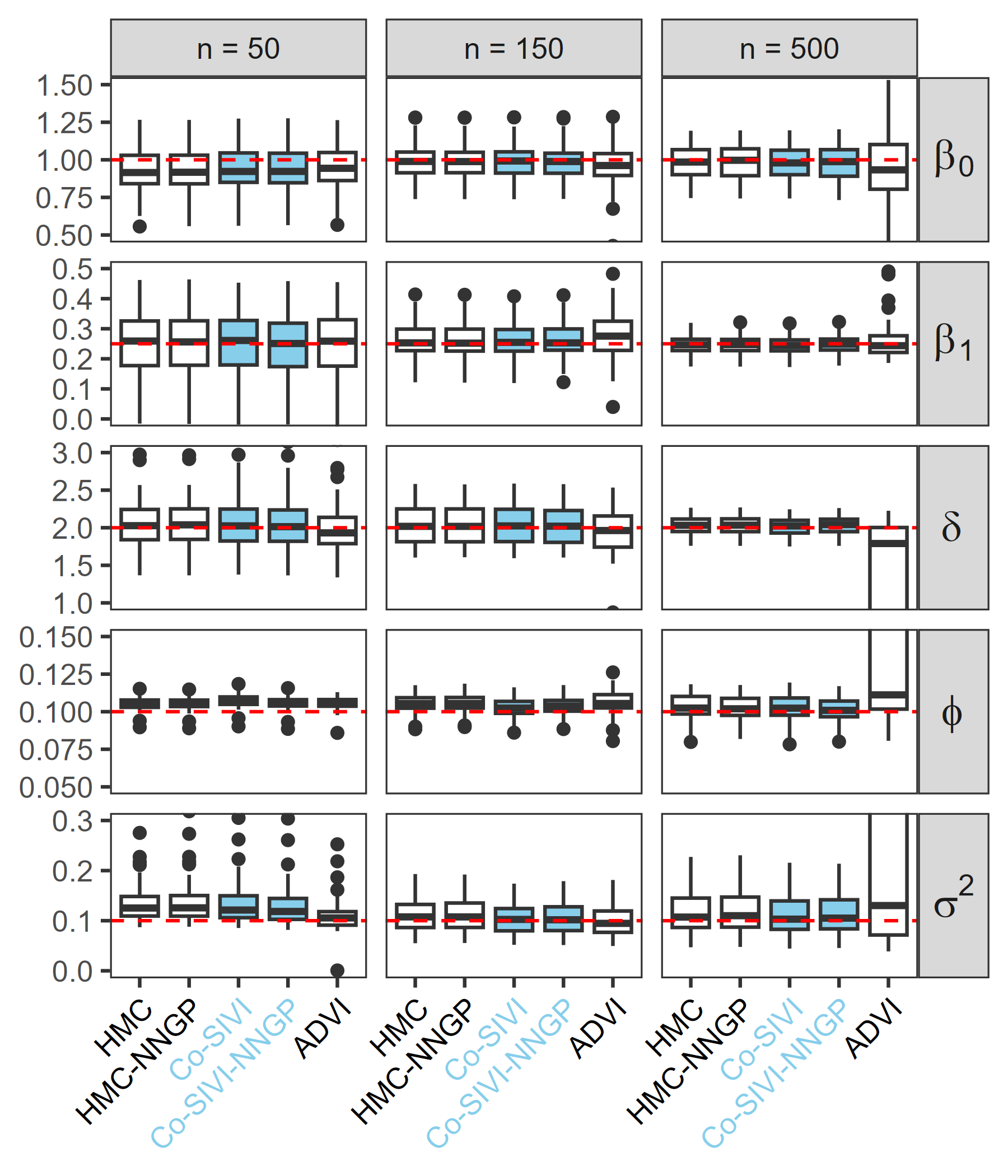}
    \end{minipage}

    \caption{Results across 50 replicates for the \textbf{Poisson} (left) and \textbf{Gamma} (right) spatial models with an \textbf{exponential covariance} function. The top row shows \textbf{predictive scores} and \textbf{runtime} (log scale), and the bottom row shows distributions of \textbf{posterior means}. Predictive scores are based on 20 held-out locations. Proposed methods are highlighted in light blue, and dashed lines indicate true parameter values. CRPS: continuous ranked probability score; IS: interval score; NLPD: negative log-predictive density. Methods with the \texttt{-NNGP} suffix use a nearest-neighbor Gaussian process prior.}
    \label{fig:Results_non_gaussian_exp}
\end{figure}

The predictive scores and the runtime of the considered methods are presented on the top row of Figure ~\ref{fig:Results_non_gaussian_exp}, with the results from the Poisson spatial model on the left column, and results from the Gamma spatial model on the right. The runtime results show that Co-SIVI becomes the fastest method after SIVI once the training sample size reaches 150, with the computational advantage increasing at the largest training sample size of 500. For the Poisson model and $n_{train} = 500$, Co-SIVI takes, on average, 201 seconds with a dense GP prior and 169 seconds with an NNGP prior. On the other hand, HMC takes, on average, 3921 seconds ($\approx$ 65 minutes) with a dense GP prior and 1983 seconds ($\approx 30$ minutes) with an NNGP prior. For the Gamma spatial model and $n_{train}= 500$, Co-SIVI takes, on average,  152 seconds with a dense GP prior and 112 seconds with an NNGP prior. On the other hand, HMC takes, on average, 4178 seconds ($\approx$ 70 minutes) with a dense GP prior and 1303 seconds ($\approx $ 20 minutes) with an NNGP prior. For both models, Co-SIVI achieves predictive scores similar to HMC, whereas ADVI performs slightly worse on average for the largest training sample size considered, with some replicates yielding much larger predictive scores that are truncated in the figure. For the Poisson spatial model, SIVI achieves similar CRPS and NLPD scores, but slightly larger IS for the largest sample size.

\clearpage

The distributions of the posterior means across the 50 replicates are shown on the bottom row of Figure \ref{fig:Results_non_gaussian_exp}, with the results from the Poisson spatial model on the left and results from the Gamma spatial model on the right. For both models, the estimates are similar for Co-SIVI and HMC. However, for the largest sample sizes, ADVI struggles to recover posterior means comparable to those from HMC for the range parameter $\phi$ and the marginal variance $\sigma^2$, which likely helps explain the larger prediction scores observed for ADVI. For the Poisson spatial model, SIVI struggles with both the range parameter and the marginal variance across all sample sizes, with the differences increasing as sample size grows. A comparison of the posterior densities of the range parameter $\phi$ and the first 30 spatial random effects, illustrating the close agreement between Co-SIVI and HMC, is presented in Section D of the supplementary materials for two randomly selected replicates, one from each spatial model.

Overall, because it is much less flexible than the neural network-based variational family used in Co-SIVI, ADVI struggles to match HMC results at the largest sample sizes and shows greater variability in its estimates across replicates. Furthermore, for the Poisson spatial model, SIVI also struggles to achieve similar results across all sample sizes considered. Co-SIVI performs similarly to HMC on predictive and parameter-estimation criteria, while being much faster for larger sample sizes, even though it runs more iterations than HMC.

\section{Large-Scale Examples}\label{sec:large_scale}

In this section, we illustrate the performance of the proposed methods SIVI-NNGP and Co-SIVI-NNGP on large-scale datasets. First, we apply SIVI-NNGP to a land surface temperature dataset of 150,000 locations from \cite{Heaton_case_study} using a Gaussian spatial model under the marginal formulation. The predictive performance of SIVI-NNGP is compared with the results in \cite{Heaton_case_study}. Then, we apply both SIVI-NNGP and Co-SIVI-NNGP to the housing price dataset of approximately 12,000 locations in California from \cite{Geron_housing_df} using a Gamma spatial model. The predictive performance and parameter inference are compared with those obtained by an HMC-NNGP implementation.

\subsection{Land Surface Temperature}\label{sec:temperature}

\cite{Heaton_case_study} conducts a comprehensive comparison of approximate spatial prediction methods for large-scale Gaussian spatial models. The methods are evaluated on two large datasets with $150,000$ locations: one real and one simulated. The real dataset consists of daytime land surface temperature measured by the Terra instrument onboard the MODIS satellite on August 4, 2016, covering longitudes of $-95.91$ to $-91.28$ and latitudes of $34.30$ to $37.07$.

The simulated dataset is generated from a Gaussian process with a constant mean, an exponential covariance function, and independent measurement errors such that it mimics observations from the real dataset. In both datasets, locations are split into a training set of $105,569$ locations and a validation set of $44,431$ locations. 

We apply SIVI-NNGP to the simulated dataset, since the data-ge\-ne\-ra\-ting model is known and used by all methods. The neural network architecture is identical to that described in Section \ref{sec:Simulation}. It is trained for $1,000$ iterations, with an Adam learning rate of $0.0001$. We use 10 Monte Carlo samples and 20 auxiliary variables. Lastly, we use a maximum neighborhood size of $M=15$. The implementation uses three-eighth of an NVIDIA H100 GPU. 

Table \ref{tab:land_surface_results} compares the predictive scores obtained by SIVI-NNGP with the 13 methods reported in \cite{Heaton_case_study} at the validation locations. The SIVI-NNGP implementation yields a CRPS of 0.46, an IS of 3.83, and a root mean-squared error (RMSE) of 0.88, placing SIVI-NNGP among the best-performing methods. Furthermore, only one method (Gapfill from \cite{Gerber_Gapfill}) runs faster than SIVI-NNGP, with a runtime under 119 seconds. However, Gapfill's speed comes at the cost of worse predictive scores (CRPS of 0.64, RMSE of 1.00, and IS of 18.01). Among all methods considered in \cite{Heaton_case_study}, only one, reported as NNGP, achieved similar predictive scores (CRPS of 0.46, RMSE of 0.88, and IS of 3.76) in a similar runtime (approximately two minutes) to SIVI-NNGP. However, their NNGP implementation \textit{fixes some parameters} in the covariance structure, whereas the SIVI-NNGP implementation provides a full joint posterior distribution for all model parameters. 
\begin{table*}[!th]
    \centering
    \caption{Continuous ranked probability score (CRPS), root mean-squared error (RMSE), interval score (IS), and runtime for semi-implicit variational inference with a nearest-neighbor Gaussian process prior (SIVI-NNGP) and the methods reported in \cite{Heaton_case_study}. Predictive scores are evaluated on the 44,431 validation locations.}
    \begin{tabular}{lcc}
        \hline
        Score & SIVI-NNGP & Other Methods (best - worst) \\
        \hline
        CRPS & 0.46 & (0.43 - 0.74) \\
        RMSE & 0.88 & (0.83 - 1.31) \\
        IS & 3.83 & (3.64 - 18.01) \\
        Runtime [min] & 1.98 & (0.63 - 2888.89) \\
        \hline
    \end{tabular}
    \label{tab:land_surface_results}
\end{table*}

\begin{figure}[!bth]
\centering{
\includegraphics[width=0.7\textwidth,
keepaspectratio]{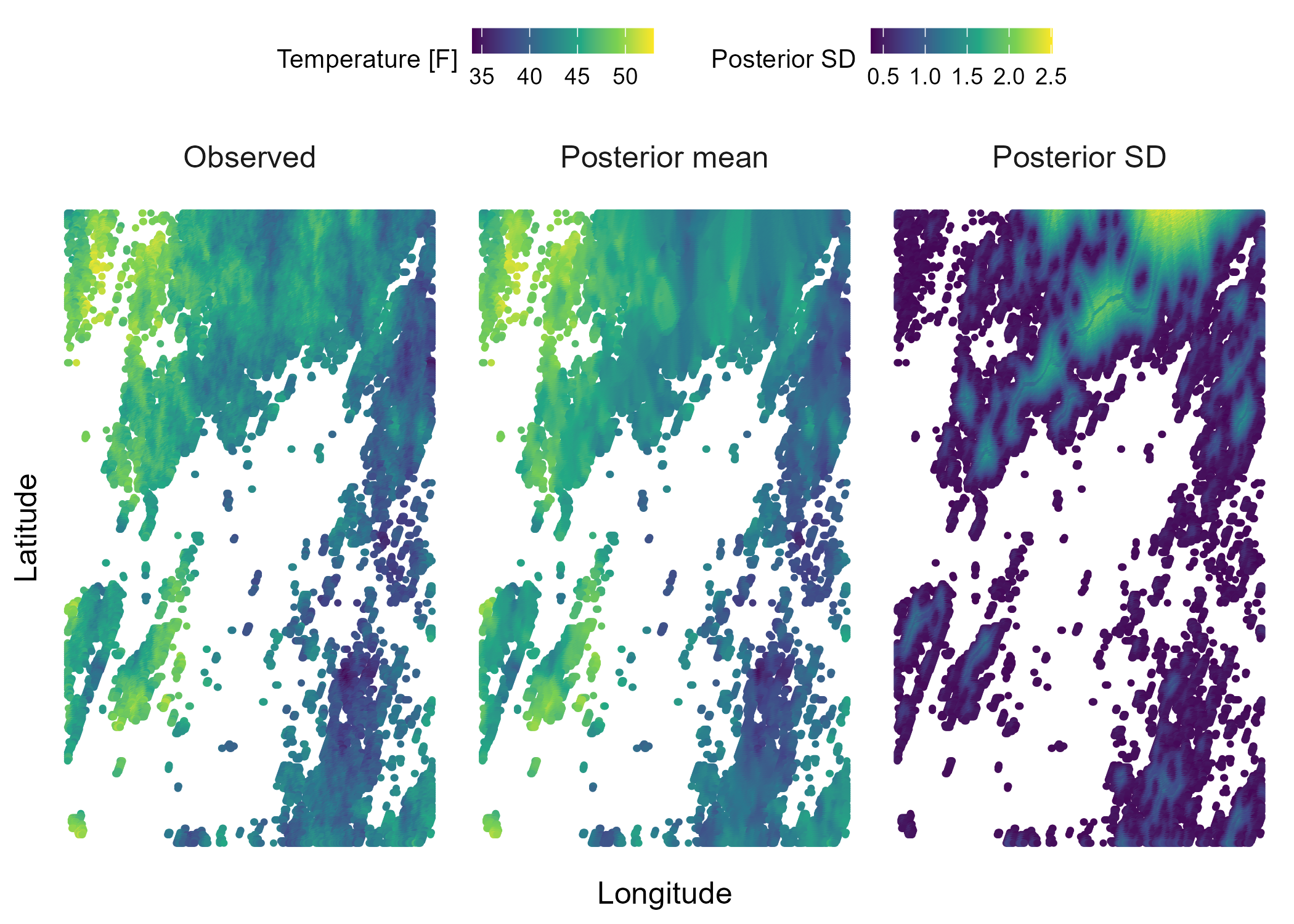}
}
\caption{Simulated land surface temperature from \cite{Heaton_case_study}, posterior predictive means obtained with SIVI-NNGP, and corresponding posterior predictive standard deviations (SDs). The spatial domain covers longitudes $-95.91$ to $-91.28$ and latitudes $34.30$ to $37.07$\label{fig:Temp_map}}
\end{figure}

Figure \ref{fig:Temp_map} shows the posterior means and predictive standard deviations at each location in the validation set, along with the observed values. The posterior means obtained by SIVI-NNGP are very similar to the observed values.

\subsection{California Housing Price}\label{sec:Housing}

The dataset from \cite{Geron_housing_df} contains information from the 1990 California Census. Each observation corresponds to a housing block and includes its longitude and latitude, along with several block-level characteristics: the median age of houses, the total number of rooms, bedrooms, people, and households, the median household income in tens of thousands of US dollars, and the median house value in US dollars. The dataset also includes a categorical variable describing the block's proximity to the ocean.  Our objective is to model the median house value in tens of thousands of US dollars using a Gamma spatial model, as described in Section \ref{sec:Gamma_model}, with spatial dependence modeled by an exponential covariance function.

The dataset contains $20,640$ observations, many of which share the same pair of coordinates. Because the spatial models considered in this paper require a single observation at each location, we keep only the last observation for each unique coordinate pair. After removing duplicate locations and observations with missing values, the sample size drops to $12,507$. To make the Gamma distribution more appropriate for the analysis, we exclude the observations censored at $ 500,000 \$ $, representing approximately 5 \% of observations.  This results in a final sample of $11,874$ unique locations. We then split this sample into a training set of $10,000$ locations and a validation set of $1,874$ locations. 

Further preprocessing steps include projecting longitude and latitude coordinates to UTM coordinates in kilometers, recoding the categories related to proximity to the ocean into a binary variable indicating whether a location is close to the ocean, and selecting the covariates to include in the model. Based on boxplots and scatter plots for each covariate, we include median household income, total number of bedrooms, and binary proximity to the ocean as covariates. Other covariates, including total number of rooms and total number of households, are excluded due to their strong correlation with total number of bedrooms.

We use four methods to fit this spatial model: HMC-NNGP, Co-SIVI-NNGP, Co-SIVI, and SIVI-NNGP, each with a maximum of $M=10$ neighbors for the NNGP-based methods. To illustrate the importance of accounting for spatial dependence, we also fit a Bayesian Gamma regression model without spatial random effects. Co-SIVI-NNGP and SIVI-NNGP each use three-eights of an NVIDIA H100 GPU, whereas Co-SIVI uses a full NVIDIA H100 GPU due to the increased memory required to fit a dense GP. All SIVI-based methods are trained for $2,000$ iterations, with an Adam learning rate of $0.001$ for the first 1,000 iterations, $0.0005$ for the next 500 iterations, and $0.0001$ for the final 500 iterations. The neural network architectures are the same as those described in Section \ref{sec:Simulation}. We use 10 Monte Carlo samples, 50 auxiliary variables, and 10 IWLS-style updates for the Co-SIVI methods. We also initialize the neural network biases for the means of the conditional variational distributions near the prior means for all model parameters, so the initial iterations produce parameters on an appropriate scale. Instead of directly sampling $\sigma^2$ and $\phi$, we parameterize the variational distribution in terms of $\kappa$ and $\sigma^2$ (Section \ref{sec:CoSIVI}), as preliminary results showed poor uncertainty quantification for $\sigma^2$ and $\phi$ under the direct parameterization. 

The HMC-NNGP implementation uses 4 CPU cores on an AMD EPYC 9654 (Zen 4) processor, with one core allocated to each of the four chains. Each chain is run for $2,000$ iterations, including $1,000$ warm-up iterations. The target acceptance probability is set to $0.7$, and the maximum tree depth is set to 12. These settings produce no divergent transitions, effective sample sizes of at least 700 for all parameters, and R-hat values smaller than 1.01. Taken together, these diagnostics provide no evidence of convergence issues for the HMC-NNGP implementation.

To ensure HMC-NNGP can run in a reasonable amount of time, some model simplifications are required. Specifically, the median household income covariate is included as a linear term on the log scale, even though it exhibits a nonlinear relationship with the log-transformed outcome. Furthermore, the prior for the dispersion parameter $\delta$ is centered at $8$, which is close to the point estimate obtained using a frequentist Gamma regression without random effects. These model simplifications and prior specifications are applied identically across all methods to ensure a fair comparison. Without these simplifications and hyperparameter tuning, the HMC-NNGP implementation would take several days to approximately a week to complete. 

The methods are compared using the CRPS, IS, and NLPD evaluated on the validation set. We also compare the posterior distributions of the model parameters and the predicted spatial random effects.

\begin{figure}[!ht]
\centering{
\includegraphics[width=\textwidth,
keepaspectratio]{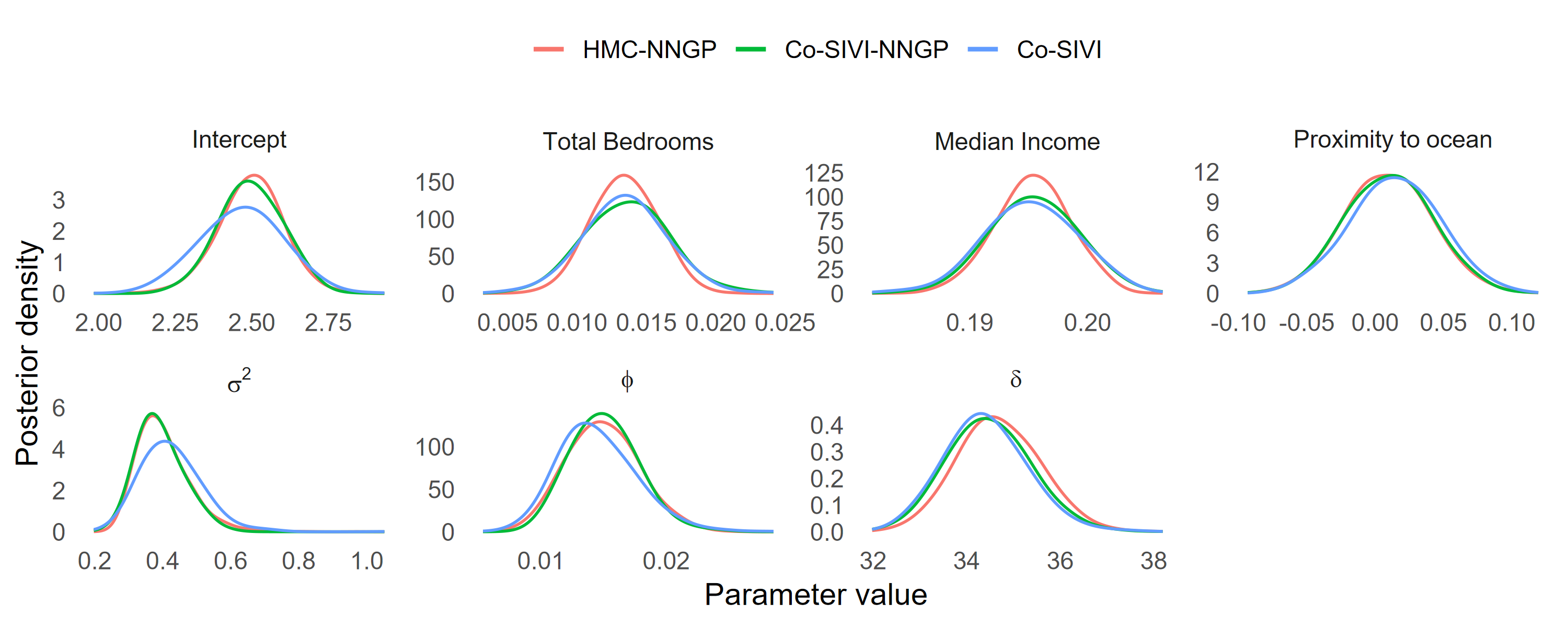}
}
\caption{\textbf{Posterior densities} for all model parameters included in the Gamma model (besides random effects) obtained using HMC-NNGP, Co-SIVI-NNGP, and Co-SIVI with a dense GP prior.\label{fig:housing_dens}}
\end{figure}

Figure \ref{fig:housing_dens} presents the posterior densities of the model parameters obtained by the considered methods, {\em excluding the Gamma regression and SIVI-NNGP}, whose posterior densities {\em differ substantially} from those obtained by HMC-NNGP. A comparison of the posterior densities obtained by all methods is available in Section E of the supplementary materials. Overall, Co-SIVI-NNGP produces posterior densities similar to those of HMC-NNGP for all model parameters, except for the regression coefficients associated with the total number of bedrooms and the household median income, for which Co-SIVI-NNGP produces slightly wider densities. Co-SIVI produces posterior densities similar to those obtained by Co-SIVI-NNGP, except for the intercept and $\sigma^2$. These differences may be partly attributable to the use of a dense GP prior for Co-SIVI and an NNGP prior for Co-SIVI-NNGP. Additional results, including scatter plots comparing the posterior means and SDs obtained with each method against those obtained with HMC-NNGP, and a comparison of the predicted values, are available in Section E of the supplementary materials.

Table \ref{tab:housing_results} presents the predictive scores obtained by HMC-NNGP, Co-SIVI-NNGP, Co-SIVI, SIVI-NNGP, and the Gamma regression without spatial random effects at the validation locations in the California housing dataset. The Gamma regression model without spatial random effects yields much worse predictive scores than the spatial models, indicating that accounting for spatial dependence improves predictive performance. SIVI-NNGP is the fastest method, taking less than one minute, and achieves similar CRPS and NLPD scores, but has slightly higher IS.  Co-SIVI-NNGP, on the other hand, achieves predictive scores similar to those obtained by HMC-NNGP while being \emph{17 times} faster than it. Specifically, Co-SIVI-NNGP takes 134 minutes (2.2 hours) compared to 2316 minutes (38 hours) for HMC-NNGP.

\begin{table*}[!h]
    \centering
    \caption{Continuous ranked probability score (CRPS), interval score (IS), negative log-predictive density (NLPD), and runtime for HMC-NNGP, Co-SIVI-NNGP, Co-SIVI with a dense GP prior, SIVI-NNGP, and a Gamma regression without spatial random effects. Predictive scores are evaluated on the 1,874 locations in the validation set.}
    \begin{tabular}{lccccc}
        \hline
        Score & \makecell{HMC-\\NNGP}  & \makecell{Co-SIVI-\\NNGP} & Co-SIVI & \makecell{SIVI-\\NNGP}  & \makecell{Gamma\\Regression}\\
        \hline
        CRPS &2.13 &2.13  & 2.13 &2.11 &3.32  \\
        IS & 21.30 &21.47  & 21.31 & 21.64 &33.23 \\
        NLPD & 2.61 & 2.61 & 2.61 &2.63 &3.11 \\
        Runtime [min] & 2316.50 & 134.08 & 213.07 &0.89 &0.23 \\
        \hline
    \end{tabular}
    \label{tab:housing_results}
\end{table*}

\section{Conclusion} \label{sec:discussion}

In this paper, we introduce Co-SIVI as a scalable variational approach for Bayesian spatial models in which the variational approximation explicitly represents spatial random effects. By incorporating dependence directly into the conditional variational distribution of the spatial random effects, Co-SIVI improves posterior recovery over SIVI for non-Gaussian spatial models. For Gaussian spatial models, where the random effects can be marginalized analytically, SIVI provides an efficient alternative to HMC. Together, these approaches offer complementary strategies for scalable variational inference in Bayesian geostatistical models, and both can be combined with NNGP priors to further improve scalability. In our experiments, relatively few IWLS-style updates sufficed, with no improvement beyond $T=10$ for larger sample sizes; alternatively, a convergence criterion based on $\bfm^{(t)}$ can terminate the updates.

Empirical results and large-scale applications show that these methods provide inference beyond point estimates and predictions across models with both exponential and Matérn 5/2 covariance functions. Unlike other methods \citep{Hensman-Classification, song_VINNGP}, Co-SIVI yields approximate posterior distributions for \emph{all model parameters}, including covariance parameters, that closely match those obtained by HMC for non-Gaussian spatial models, while substantially reducing computational cost; SIVI provides similar benefits under the Gaussian marginal formulation.

We did not evaluate memory usage in our comparisons, and both Co-SIVI and SIVI can still require large amounts of memory at large sample sizes. Although NNGP priors help mitigate this issue, further implemen\-tation-level improvements, including stochastic-gradient updates based on minibatches of data, could improve both memory efficiency and computational speed. Future work could also investigate more advanced neural network architectures, anisotropic covariance structures, and extensions to spatiotemporal models.

\subsubsection*{Data Availability Statement}\label{data-availability-statement}

The temperature dataset used in Section \ref{sec:temperature}, from \cite{Heaton_case_study}, is freely available on \href{https://github.com/finnlindgren/heatoncomparison}{Lindgren's GitHub}. The California housing dataset used in Section \ref{sec:Housing}, from \cite{Geron_housing_df}, is freely available on \href{https://www.kaggle.com/datasets/camnugent/california-housing-prices}{Kaggle}.

\subsubsection*{Acknowledgments}

This research was enabled in part by support provided by Calcul Québec (\url{www.calculquebec.ca}) and the Digital Alliance of Canada \newline (\url{www.alliancecan.ca/en}).

\subsubsection*{funding}
Schmidt is grateful for financial support from the Natural Sciences and Engineering Research Council (NSERC) of Canada (Discovery Grant RGPIN-2024-04312).

\clearpage
\appendix
\setcounter{figure}{0}
\renewcommand{\appendixname}{}
\renewcommand{\thefigure}{\thesection\arabic{figure}}
\section{Random Effect Uncertainty Under Conditional and Marginal Gaussian Formulations} \label{sec:re_comp_sivi_gaussian}

Under the conditional formulation, posterior samples of the spatial random effects $\bfw$ are obtained directly. On the other hand, the marginal formulation integrates out $\bfw$ and, therefore, does not directly provide posterior samples of the random effects. However, even in the marginal formulation, the posterior samples of $\bfw$ can be reconstructed from the posterior samples of the other model parameters using the full-conditional posterior distribution. Using the conditional model specification $\mathbf{y}\mid \bfbeta, \bfw, \tau^2 \sim \operatorname{MVN}(\mathbf{X}\bfbeta + \mathbf{w}, \tau^2 \bfI_n)$ and properties of the multivariate Gaussian distribution, the conditional distribution of $\bfw$ given $\bfy$ and $\bftheta =(\bfbeta^\top,\sigma^2,\phi,\tau^2)^\top$ is given by $\bfw \mid \bfy,\bftheta \sim \operatorname{MVN}(\bfm_{\bfw}, \bfC_{\bfw})$, where $\bfm_{\bfw} = \bfV_n(\bfV_n + \tau^2\bfI_n)^{-1}(\bfy-\bfX\bfbeta)$ and $\bfC_{\bfw} = \bfV_n - \bfV_n(\bfV_n + \tau^2\bfI_n)^{-1}\bfV_n$. Therefore, for each posterior sample of $\bftheta$, a corresponding posterior sample of $\bfw$ can be generated from this conditional distribution.

\begin{figure}[!ht]
\centering{
\includegraphics[width=0.8\textwidth,
keepaspectratio]{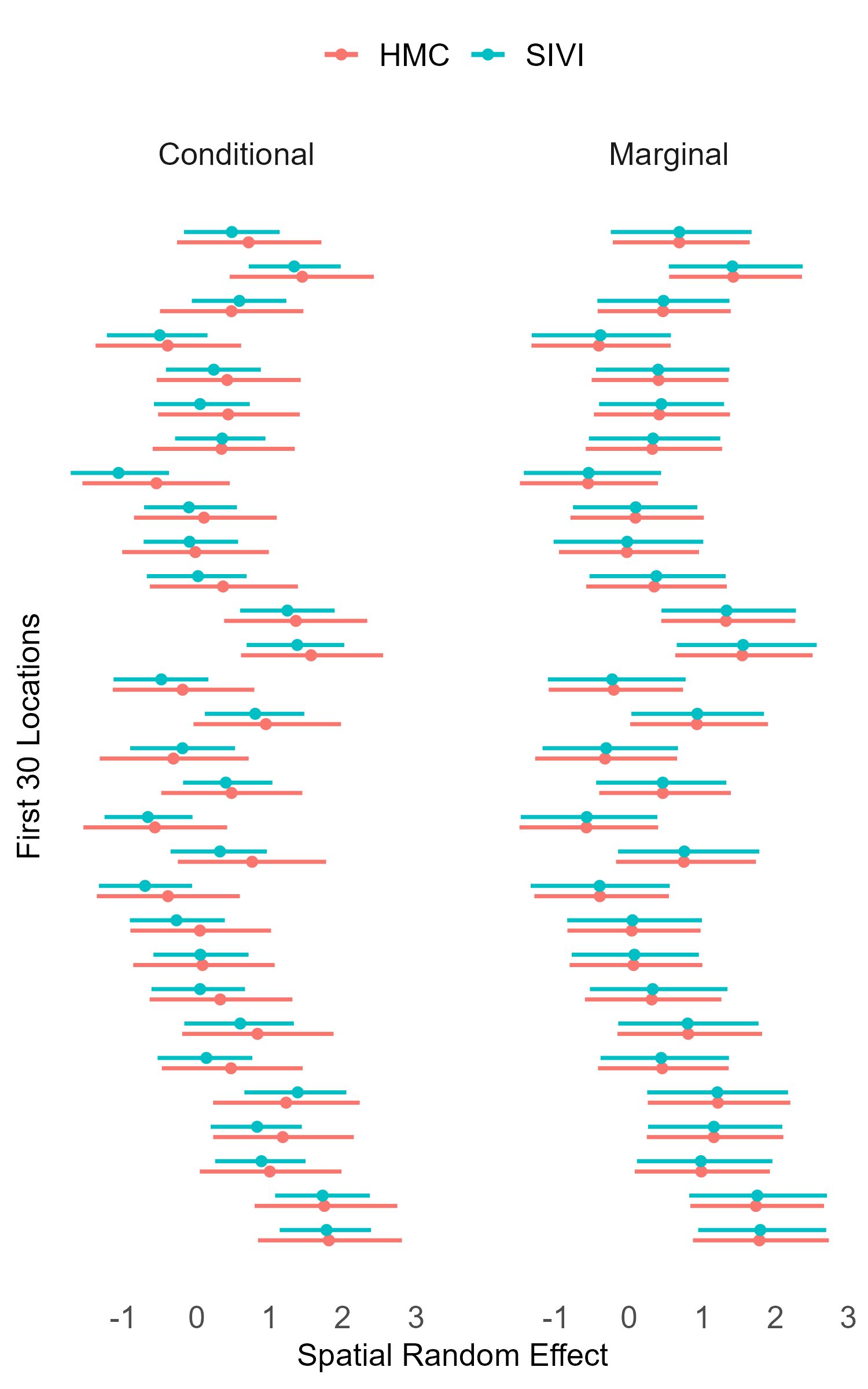}
}
\caption{Posterior means and 95\% credible intervals of the first 30 random effects obtained from Hamiltonian Monte Carlo (HMC) and semi-implicit variational inference (SIVI) under the \textbf{conditional and marginal} formulations of a \textbf{Gaussian} spatial model with \textbf{exponential} covariance function for the first replicate with 500 locations \label{fig:Comp_gaussian_re}  }
\end{figure}

Figure \ref{fig:Comp_gaussian_re} compares the posterior means and 95\% credible intervals of the first 30 random effects obtained for SIVI and HMC under both the conditional and marginal formulation for the first replicate with 500 locations. SIVI approximates the HMC posterior means and credible intervals under the marginal formulation well, but does not recover point estimates and intervals similar to HMC under the conditional formulation.  

\clearpage

\section{Integrating Nearest-Neighbor Gaussian Process Into SIVI}
\label{sec:SIVI_NNGP}

We describe how to incorporate NNGP into SIVI for the conditional and marginal formulations of a Gaussian spatial model. We first order the locations, typically by one coordinate, and specify a maximum number of neighbors $M$. Starting from such ordering $s_1, \ldots, s_n$ of all observed locations within the region of interest $\mathcal{S}$, for any location $s_i \in \mathcal{S}$, let $N(i) = \{s_{i1}, \ldots, s_{i,d_i}\} \subset \{s_1, \ldots, s_{i-1}\},$ $d_i = \min\{M, i-1\}$ be the set of neighbors associated with $s_i$. The set $N(i)$ contains at most $M$ locations among ${s_1, \ldots,s_{i-1}}$, corresponding to those with the smallest Euclidean distances to $s_i$. Following \cite{Datta_NNGP}, let $\bfc_{i,N(i)}=(\mbox{Cov}(w(s_i), w(s_{i1}) ), \ldots,\\ \mbox{Cov}(w(s_i), w(s_{id_i}) ) )^\top = (V(s_i, s_{i1}), \ldots, V(s_i, s_{id_i}))^\top$ denote the $d_i \times 1$ vector of covariances between $w(s_i)$ and its neighbors in $N(i)$, and let $\bfV_{N(i)}$ denote the $d_i \times d_i$ covariance matrix corresponding to the locations in $N(i)$, i.e., the $(\ell,j)-$th entry of $\bfV_{N(i)}$ is $[\bfV_{N(i)}]_{\ell j} = V(s_{i\ell}, s_{ij})$. Then, the GP prior for $\bfw = (w(s_1), \ldots, w(s_n))^ \top$ is approximated by the NNGP prior $$p(\bfw) \approx \prod_{i=1}^n N\left(w(s_i) \vert \bfb_i^\top \bfw_{N(i)},f_i\right),$$ where $\bfb_i =  \bfV_{N(i)}^{-1} \bfc_{i,N(i)}$ and $f_i = \operatorname{Var}(w(s_i)) - \bfc_{i,N(i)}^{\top} \bfV_{N(i)}^{-1} \bfc_{i,N(i)}$. 

Under the conditional formulation, the dense GP prior for the spatial random effect is replaced by the NNGP prior, whereas, under the marginal formulation, the NNGP approximation is applied directly to the likelihood $\mathbf{y} \mid \bfbeta, \tau^2, \bfPhi \sim \operatorname{MVN}(\mathbf{X}\bfbeta, \bfV_n + \tau^2\bfI_n)$. We refer to the latter as the response NNGP formulation.

To obtain the response NNGP formulation, the approximation is applied directly to the likelihood. Let $\bfK = \bfV_n + \tau^2 \bfI_n$ denote the covariance matrix of the outcomes under the marginal formulation. Analogously to the conditional formulation, $\bfK_{N(i)}$ denotes the $d_i \times d_i$ covariance matrix among the neighbors and $\bfk_{i, N(i)}$ denotes the $d_i \times 1$ vector of covariances between $y(s_i)$ and its neighbors in $N(i)$.  Then, the likelihood is approximated by $p(\bfy \mid \bfbeta, \bfPhi, \tau^2) \approx \prod_{i=1}^n N(y(s_i) \mid \bfx(s_i)^\top\bfbeta + \widetilde{\bfb}_i^\top(\bfy_{N(i)}-\bfX_{N(i)}\bfbeta), \widetilde{f}_i)$, where $\widetilde{\bfb}_i = \bfK_{N(i)}^{-1} \bfk_{i,N(i)}$ and $\widetilde{f}_i = \operatorname{Var}(y(s_i)) - \bfk_{i,N(i)}^{\top} \bfK_{N(i)}^{-1} \bfk_{i,N(i)}$. 

The two NNGP formulations modify different components of the joint density $p(\bfy,\bftheta_j)$ appearing in the Monte Carlo approximation of the lower bound $\widehat{ \underline{\mathcal{L}}}_{K}$. Under the conditional formulation, the NNGP approximation replaces the dense GP prior contribution $p(\bfw)$, whereas, under the marginal formulation, the NNGP approximation replaces the likelihood $p(\bfy \mid \bfbeta, \bfPhi, \tau^2)$.

\subsection[
  Integrating Nearest-Neighbor Gaussian Processes into Co-SIVI
]{%
  Integrating Nearest-Neighbor Gaussian Processes into\protect\\
  Co-SIVI
}
Implementation of Co-SIVI with an NNGP prior proceeds similarly to the conditional formulation described in Section \ref{sec:SIVI_NNGP}. An additional modification is required because the dense covariance matrix $\bfV_n$ also appears in the IWLS-style updates. Let $\bfB$ denote a $n \times n$ strictly lower-triangular matrix whose $i$th row contains the coefficients $\bfb_i^\top$ in the columns corresponding to $N(i)$ and zeros elsewhere, and let $\bfF = \operatorname{diag}(f_1,\ldots,f_n)$, where the $\bfb_i^\top$ and $f_i$ are defined in Section \ref{sec:SIVI_NNGP}. Then, the covariance matrix can be approximated by $\widetilde{\bfV}_n = (\bfI_n - \bfB)^{-1}\bfF(\bfI_n-\bfB)^{-\top}$, or equivalently the sparse precision matrix $\widetilde{\bfV}^{-1}_n = (\bfI_n - \bfB)\bfF^{-1}(\bfI_n-\bfB)^{\top}.$

\clearpage

\setcounter{figure}{0}
\section[Co-SIVI for Gaussian and Bernoulli Spatial Models
]{%
Co-SIVI for Gaussian and Bernoulli Spatial \protect\\ Models
}
\subsection{Gaussian Spatial Model}
We illustrate the application of Co-SIVI for a Gaussian likelihood. Let $y(s_i)\mid \bfbeta, w(s_i), \tau^2 \sim \operatorname{N}(\bfx(s_i)^\top\bfbeta + w(s_i), \tau^2)$. Gaussian outcomes are a special case for Co-SIVI since the likelihood is already Gaussian. Therefore, no pseudo-outcomes approximation is required. In this case, $ \widetilde{y}^{(t-1)}(s_i) = y(s_i)$ and the diagonal matrix $\bfW^{-1}$ satisfies $\left([\bfW]_{ii}\right)^{-1} = \tau^2$. Thus, the location vector $\bfm$ and the covariance matrix $\bfC$ used by Co-SIVI do not require iterative IWLS-style updates and are given by $\bfC = \left[\bfV_n^{-1} + \operatorname{diag}\left(\frac{1}{\tau^2}\right) \right]^{-1}$ and $\bfm = \frac{\bfC}{\tau^2}(\bfy - \bfX\bfbeta)$. These correspond to the covariance matrix and mean vector of the full conditional distribution of the spatial random effects. 

We briefly clarify the differences between Co-SIVI and traditional SIVI in the context of the conditional formulation for Gaussian responses. Co-SIVI uses $q(\bfzeta \vert \bfpsi)=\mbox{MVN}(\bfmu_{\zeta}, \bfD_{\zeta})$ with a diagonal covariance matrix $\bfD_{\zeta}$ and the random effects are $\bfw = \bfm + \bfL \bfzeta$ with $\bfL\bfL^\top = \bfC$ being the Cholesky decomposition $\bfC$ and the neural network $g_{\bfnu}(\bfepsilon)$ outputting $(\bfmu_{\zeta}, \bfD_{\zeta}).$ On the other hand, SIVI assumes $q(\bfw\mid \bfpsi) = \mbox{MVN}(\bfm^*, \bfS^*)$, with a diagonal covariance matrix $\bfS^*$, where the neural network outputs $\bfm^*$ and $\bfS^*$ directly.

\begin{figure}[!ht]
\centering{
\includegraphics[width=0.8\textwidth,
keepaspectratio]{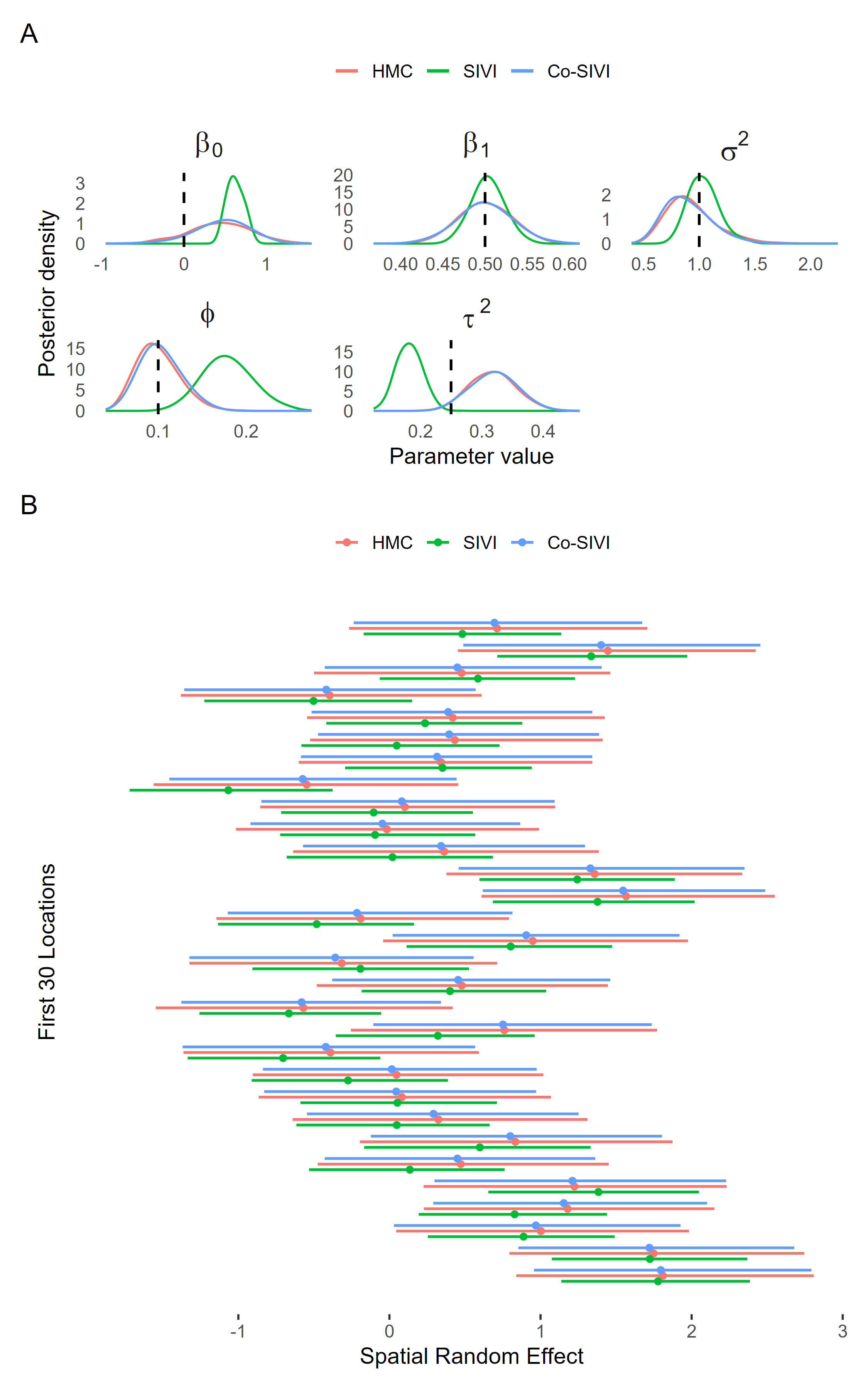}
}
\caption{\textbf{Posterior distributions} of the model parameters (A) and $95\%$ credible intervals for the first 30 spatial random effects (B) for a \textbf{Gaussian} spatial model under the \textbf{conditional formulation} with an \textbf{exponential covariance} function. Vertical dashed lines indicate the true parameter values.\label{fig:gaussian_cond}}
\end{figure}

We apply HMC, SIVI, and Co-SIVI on the same dataset considered in Section \ref{sec:re_comp_sivi_gaussian} using the conditional formulation of a Gaussian spatial model. We assign the same priors as in the marginal Gaussian simulation in the main manuscript. The resulting posterior distributions of all model parameters and of the first 30 locations are presented in Figure \ref{fig:gaussian_cond}. Overall, Co-SIVI produces posterior distributions for all model parameters, including the spatial random effects, that are similar to those obtained with HMC, whereas SIVI fails to recover comparable posterior distributions. Co-SIVI completes the analysis in 134 seconds, compared with 40 seconds for SIVI and 7580 seconds ($\approx126$ minutes) for HMC. 

\clearpage
\subsection{Bernoulli Spatial Model}
\label{sec:sim_bernoulli}

We illustrate the application of Co-SIVI for a Bernoulli likelihood. Let $y(s_i) \mid p(s_i) \sim \operatorname{Bern}\left(p(s_i)\right)$, where $\bfp=(p(s_1),\ldots,p(s_n))^\top$ denotes the probability of successes. This results in $\mathbb{E}(y(s_i)) = p(s_i)$ and $\operatorname{Var}(y(s_i)) = p(s_i)(1-p(s_i))$. The probability of success is linked to the linear predictors using a logit link function, such that $\operatorname{logit}(p(s_i)) = \bfx^\top(s_i)\bfbeta + w(s_i), w(\cdot) \sim \operatorname{GP}(0, V)$. This Bernoulli likelihood can be written in the exponential family with $\alpha = 1, \eta(s_i) = \log(\frac{p(s_i)}{1-p(s_i)}), b(\eta(s_i)) = \log\left[1+\exp(\eta(s_i))\right]$ and \newline $c(y(s_i), \alpha) = 1$.

For this model, the vector of parameters is $\bftheta = \left(\bfbeta^\top, \bfw^\top, \bfPhi^\top \right)^\top$. The pseudo-outcomes can be written as $ \widetilde{y}^{(t-1)}(s_i) = \bfx(s_i)^\top\bfbeta + w^{(t-1)}(s_i) + \frac{y(s_i) - p^{(t-1)}(s_i)}{p^{(t-1)}(s_i)(1-p^{(t-1)}(s_i))}$, where $p^{(t-1)}(s_i) = \frac{1}{1+\exp\left(-\bfx^\top(s_i)\bfbeta - w^{(t-1)}(s_i)\right)}$. The diagonal covariance matrix $\bfW^{-1}$ satisfies $\left([\bfW]_{ii}\right)^{-1}=\operatorname{Var}(y(s_i)) \left( h'(p^{t-1}(s_i))  \right)^2 $ $= \left[ p^{(t-1)}(s_i)(1-p^{(t-1)}(s_i))\right]^{-1}$. Hence, $\bfW^{(t-1)} = \operatorname{diag}(\bfp^{(t-1)} \odot (\mathbf{1}-\bfp^{(t-1)}))$.

Plugging these quantities into Equation 7 of the main manuscript gives the update mechanism for the location vector $\bfm$ and the covariance matrix $\bfC$. Specifically, we find $\bfC^{(t)} = \left[\bfV_n^{-1} + \operatorname{diag}\left(\bfp^{(t-1)}(1-\bfp^{(t-1)})\right) \right]^{-1}$. The corresponding location vector is $\bfm^{(t)} =  \bfC^{(t)}\operatorname{diag}\left(\bfp^{(t-1)}(1-\bfp^{(t-1)})\right)$ $(\widetilde{\bfy}^{(t-1)}-\bfX\bfbeta) = \bfC^{(t)}\left(\bfp^{(t-1)}\odot(1-\bfp^{(t-1)})\odot\bfw^{(t-1)} + \bfy - \bfp^{(t-1)}   \right)$.

Observations are generated from 500 locations sampled over the square region $[0,50]^2$ from the Bernoulli spatial model described above with one covariate $x(s) \sim N(0,1)$ and the exponential covariance function. Specifically, we use the parameter values $\beta_0 = 0.2$, $\beta_1 = -1$, $\sigma^2 = 0.1$, and $\phi = 0.1$. Under this model, the observed locations with a success are approximately $50\%$. 

We assign independent $N(0,1)$ priors to the intercept and fixed effect coefficient. We assign independent log-normal priors to the covariance parameters. Specifically, we use $\sigma^2 \sim LN(\log(0.1),1)$, and $\phi \sim LN(\log(0.1),0.09)$. These distributions yields approximate 95\% prior intervals of $0.01-0.71$, and $0.06 - 0.18$ for $\sigma^2$ and $\phi$, respectively.

\begin{figure}[!ht]
\centering{
\includegraphics[width=0.8\columnwidth]{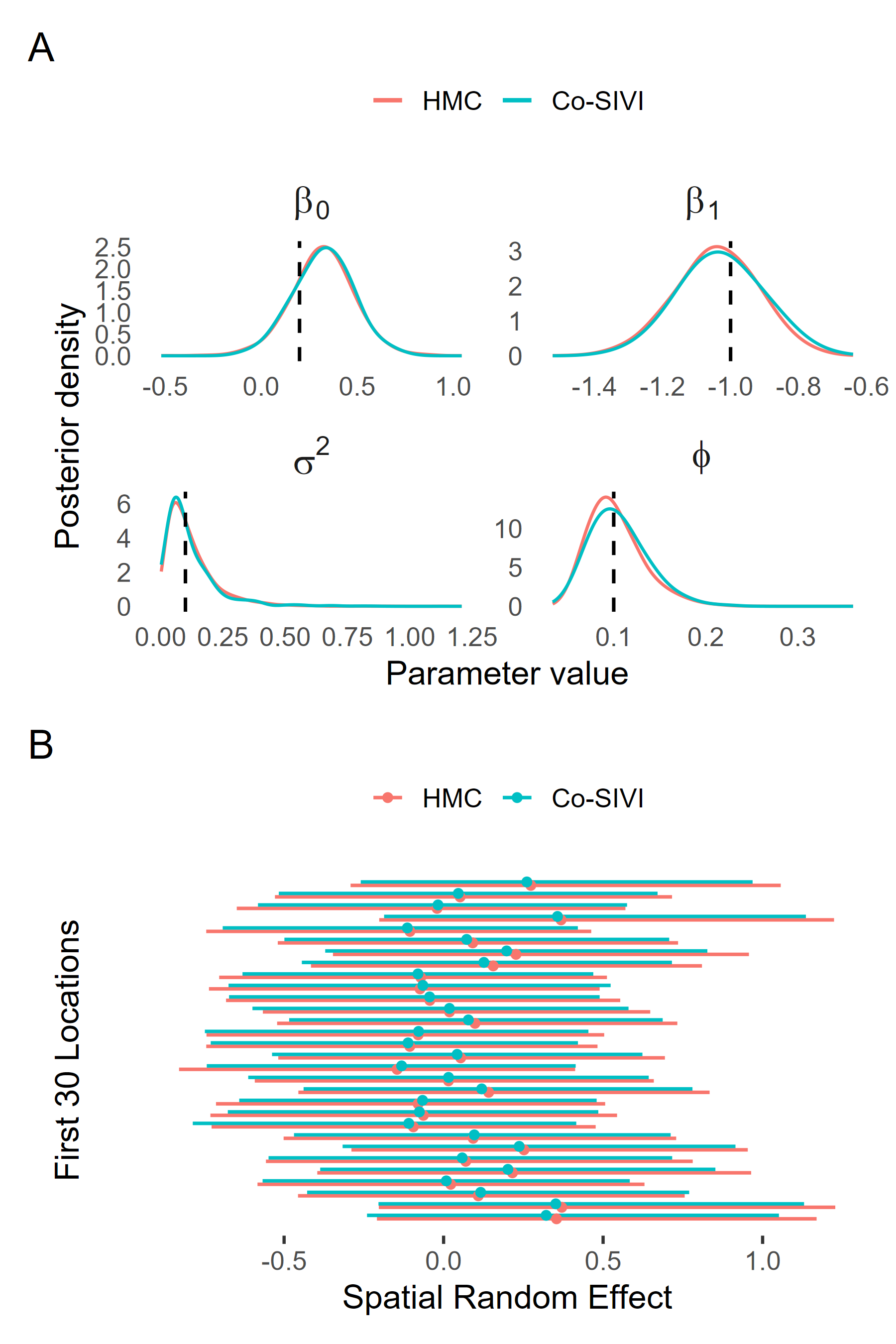}
}
\caption{\textbf{Posterior distributions} (A) of the model parameters and $95\%$ credible intervals for the first 30 spatial random effects (B) for a \textbf{Bernoulli} spatial model with an \textbf{exponential covariance} function. Vertical dashed lines indicate the true parameter values.\label{fig:bernoulli}}
\end{figure}

We compare Co-SIVI and HMC with dense GP. The posterior distributions for both Co-SIVI and HMC are presented in Figure \ref{fig:bernoulli}. Overall, Co-SIVI yields posterior distributions similar to those from HMC for the model parameters and random effects. Co-SIVI takes 199 seconds to complete while HMC takes 1790 seconds.

\clearpage
\setcounter{figure}{0}
\section{Additional Simulation Results}

\subsection{Additional Posterior Results for Simulated Data Under Exponential Covariance Function} \label{sec:post_dens_exp}
\subsubsection{Marginal Gaussian Spatial Model}

\begin{figure}[!ht]
\centering{
\includegraphics[width=0.8\textwidth,
keepaspectratio]{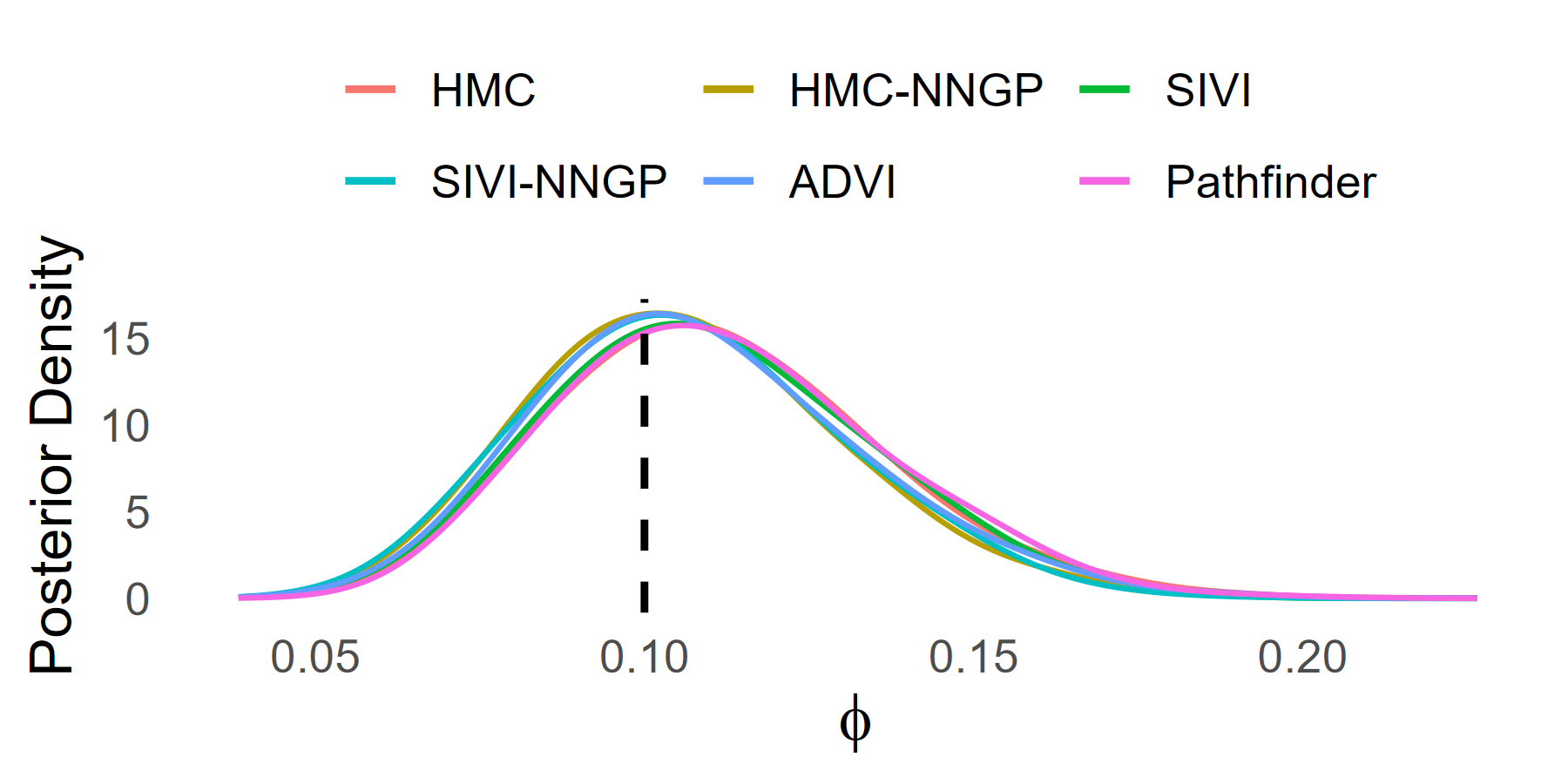}
}
\caption{\textbf{Posterior density} of the range parameter $\phi$ for the sixth replicate of the \textbf{marginal Gaussian} spatial model with an \textbf{exponential covariance} function. Methods with the \texttt{-NNGP} suffix use a nearest-neighbor Gaussian process prior.\label{fig:Dens_phi_gaussian_exp}}
\end{figure}

Figure \ref{fig:Dens_phi_gaussian_exp} shows the posterior density of the range parameter $\phi$ for the sixth replicate (selected at random from replicates 1 to 50) across all considered models. All models yield similar posterior distributions, with a small difference for the methods using an NNGP prior. 

\clearpage

\subsubsection{Non-Gaussian Spatial Models}

\begin{figure}[!ht]
\centering{
\includegraphics[width=0.5\columnwidth]{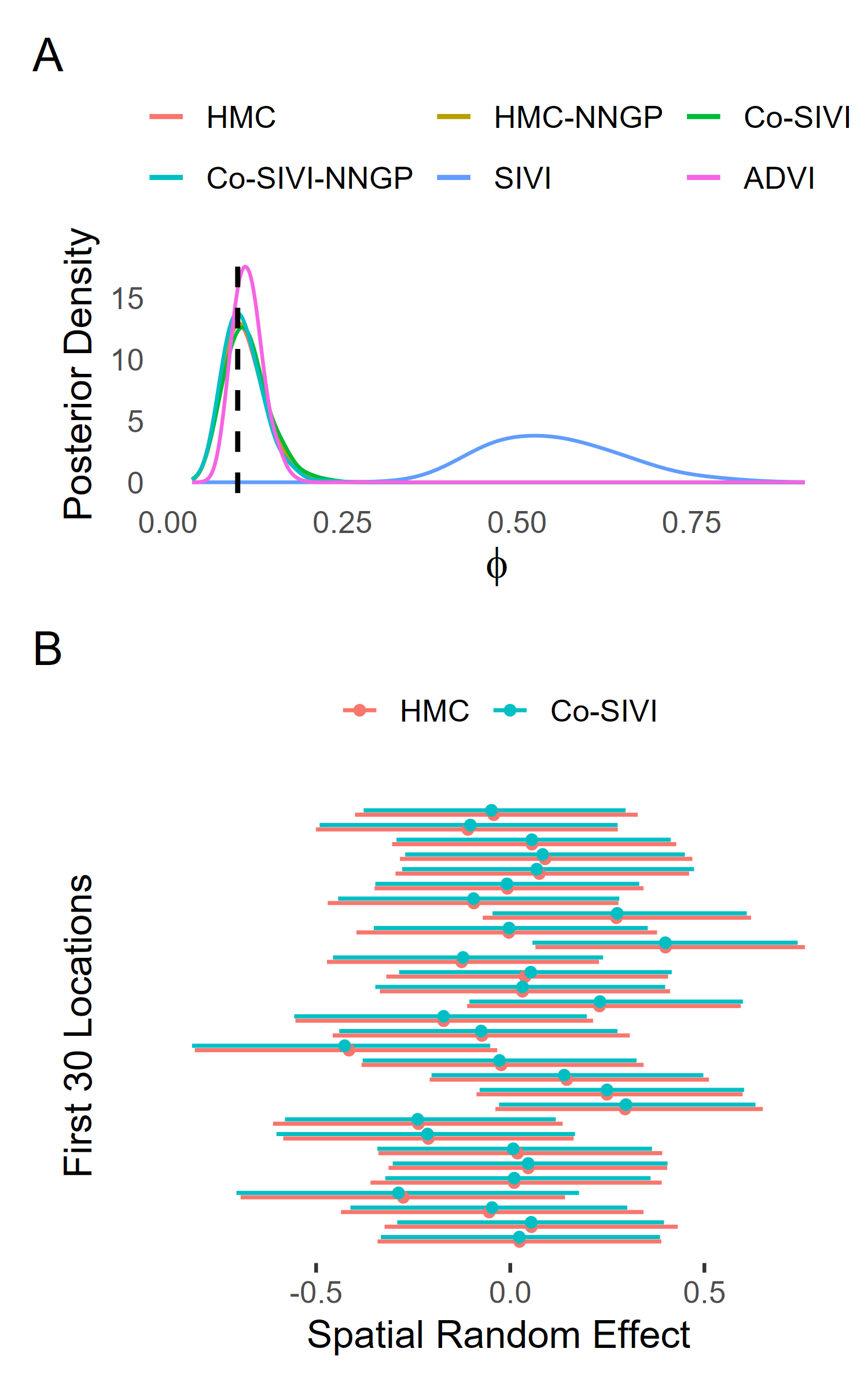}
}
\caption{\textbf{Posterior density} of the range parameter $\phi$ (A) and $95\%$ credible intervals for the first 30 spatial random effects (B) for replicate 31 of the \textbf{Poisson} spatial model with an \textbf{exponential covariance} function. The \texttt{-NNGP} suffix indicates use of a nearest-neighbor Gaussian process prior.\label{fig:Dens_poisson_exp}}
\end{figure}

Panel A of Figure \ref{fig:Dens_poisson_exp} shows the posterior density of the range parameter $\phi$ for replicate 31 (selected at random from replicates 1 to 50) of the simulated datasets with Poisson responses across all considered inference methods.  Co-SIVI and HMC yield similar posterior distributions, whereas ADVI and SIVI do not closely approximate the posterior distribution obtained by HMC. Panel B of Figure \ref{fig:Dens_poisson_exp} shows the $95\%$ credible intervals of the first 30 random effects for Co-SIVI and HMC. Co-SIVI yields credible intervals very similar to those from HMC. 

\begin{figure}[!ht]
\centering{
\includegraphics[width=0.5\columnwidth]{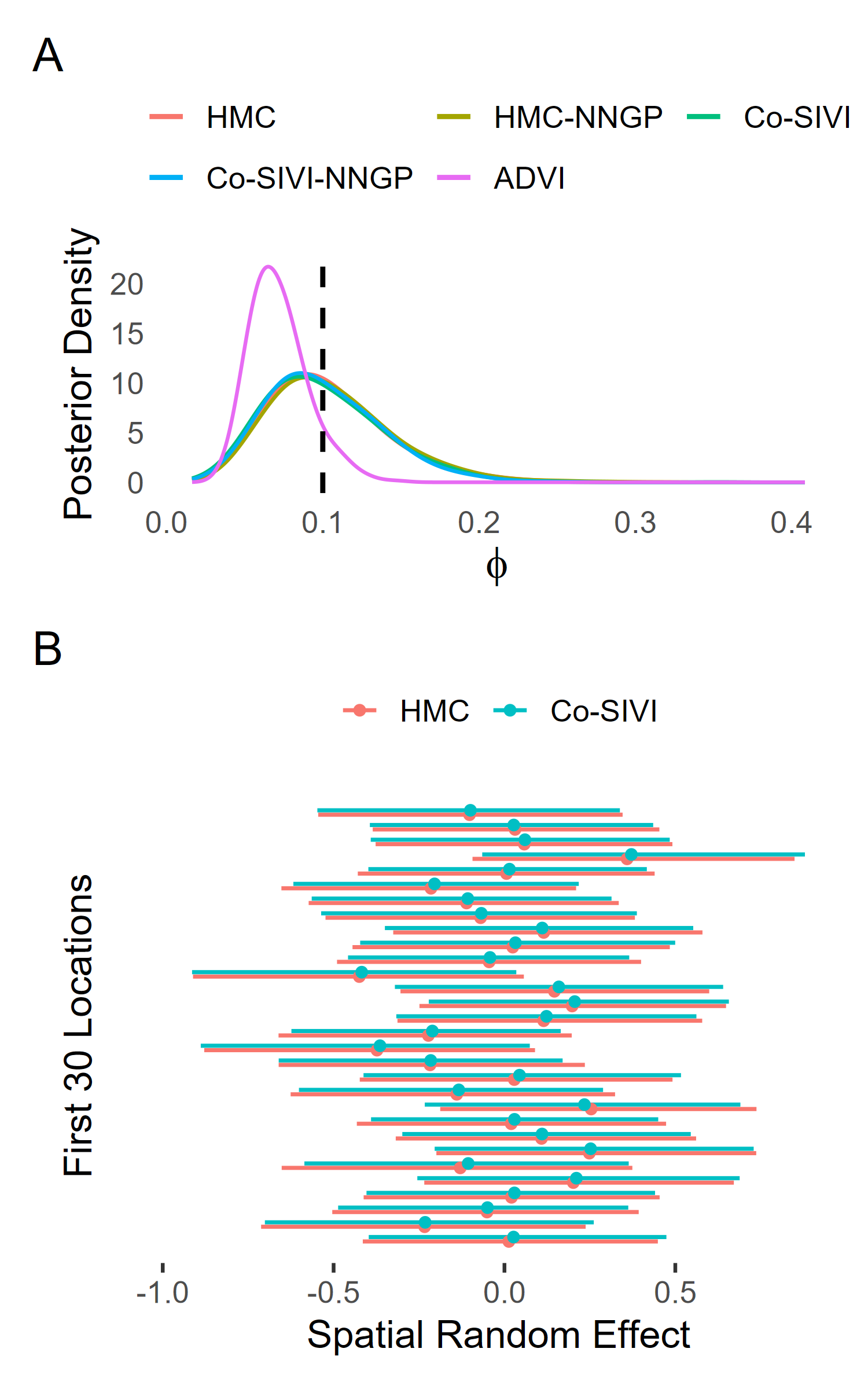}
}
\caption{\textbf{Posterior density} of the range parameter $\phi$ (A) and $95\%$ credible intervals for the first 30 spatial random effects (B) for replicate 31 of the \textbf{Gamma} spatial model with an \textbf{exponential covariance} function. The \texttt{-NNGP} suffix indicates use of a nearest-neighbor Gaussian process prior.\label{fig:Dens_gamma_exp}}
\end{figure}

Panel A of Figure \ref{fig:Dens_gamma_exp} shows the posterior density of the range parameter $\phi$ for replicate 40 (randomly selected from replicates 1 to 50) of the simulated datasets with Gamma responses across all inference methods considered. Co-SIVI and HMC yield similar posterior distributions, whereas ADVI deviates from the other methods and provides a tighter posterior distribution. Panel B of Figure \ref{fig:Dens_poisson_exp} shows the $95\%$ credible intervals of the first 30 random effects for Co-SIVI and HMC. Co-SIVI yields credible intervals very similar to those from HMC. 

\subsection{Simulation Results for Matérn 5/2 Covariance Function}
\label{sec:sim_matern}

\subsubsection{Marginal Gaussian Spatial Model}

We use the same prior distributions as in the main manuscript for the exponential covariance function, except for $\phi$, whose prior is $\phi \sim LN(\log(0.084)$, $0.09)$. This gives a 95\% prior interval of $0.05-0.15$. 

\begin{figure}[!ht]
   \centering
   \begin{minipage}[t]{0.49\textwidth}
       \centering
        \includegraphics[
            width=\linewidth,
            keepaspectratio
        ]{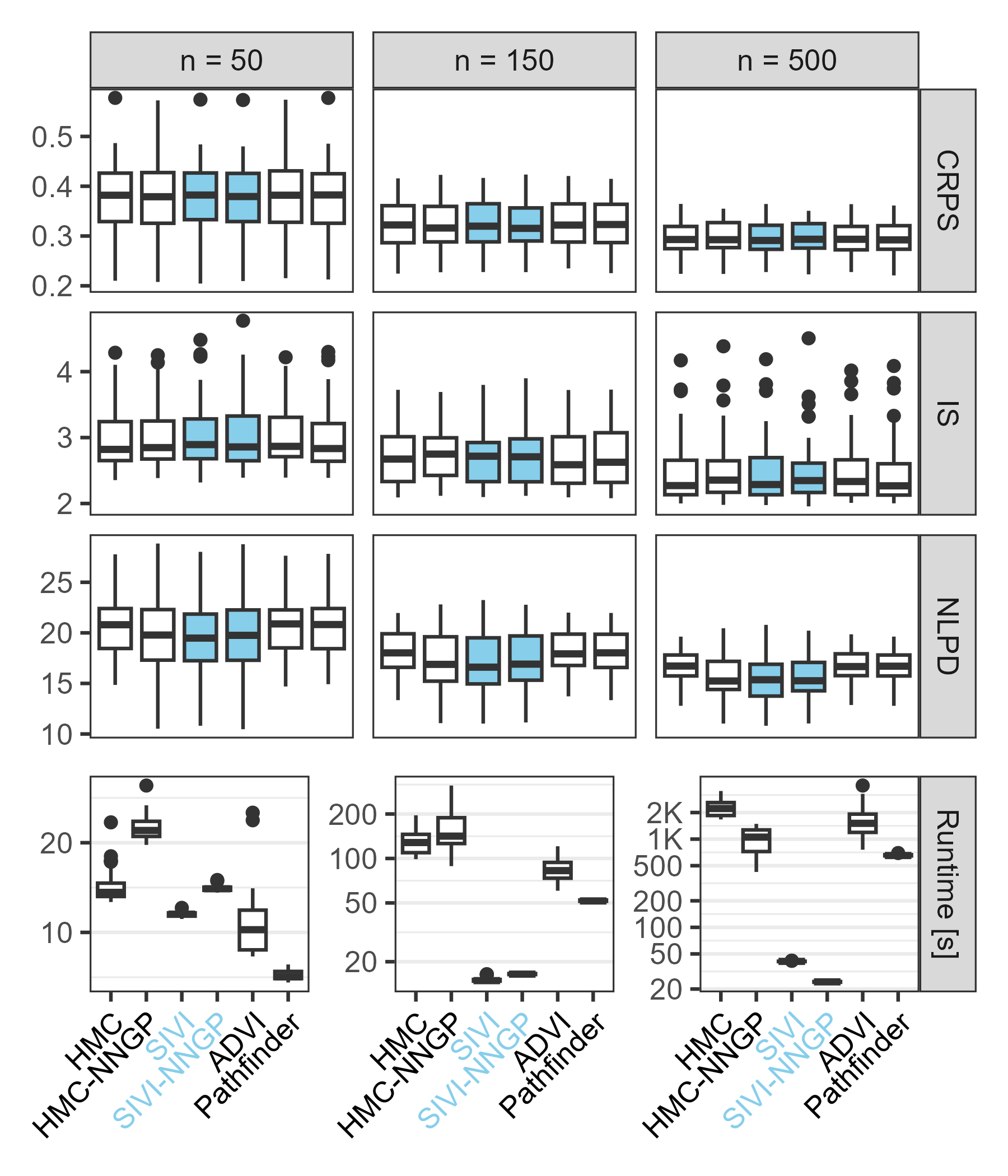}
    \end{minipage}
    \begin{minipage}[t]{0.49\textwidth}
        \centering
        \includegraphics[
            width=\linewidth,
            keepaspectratio
       ]{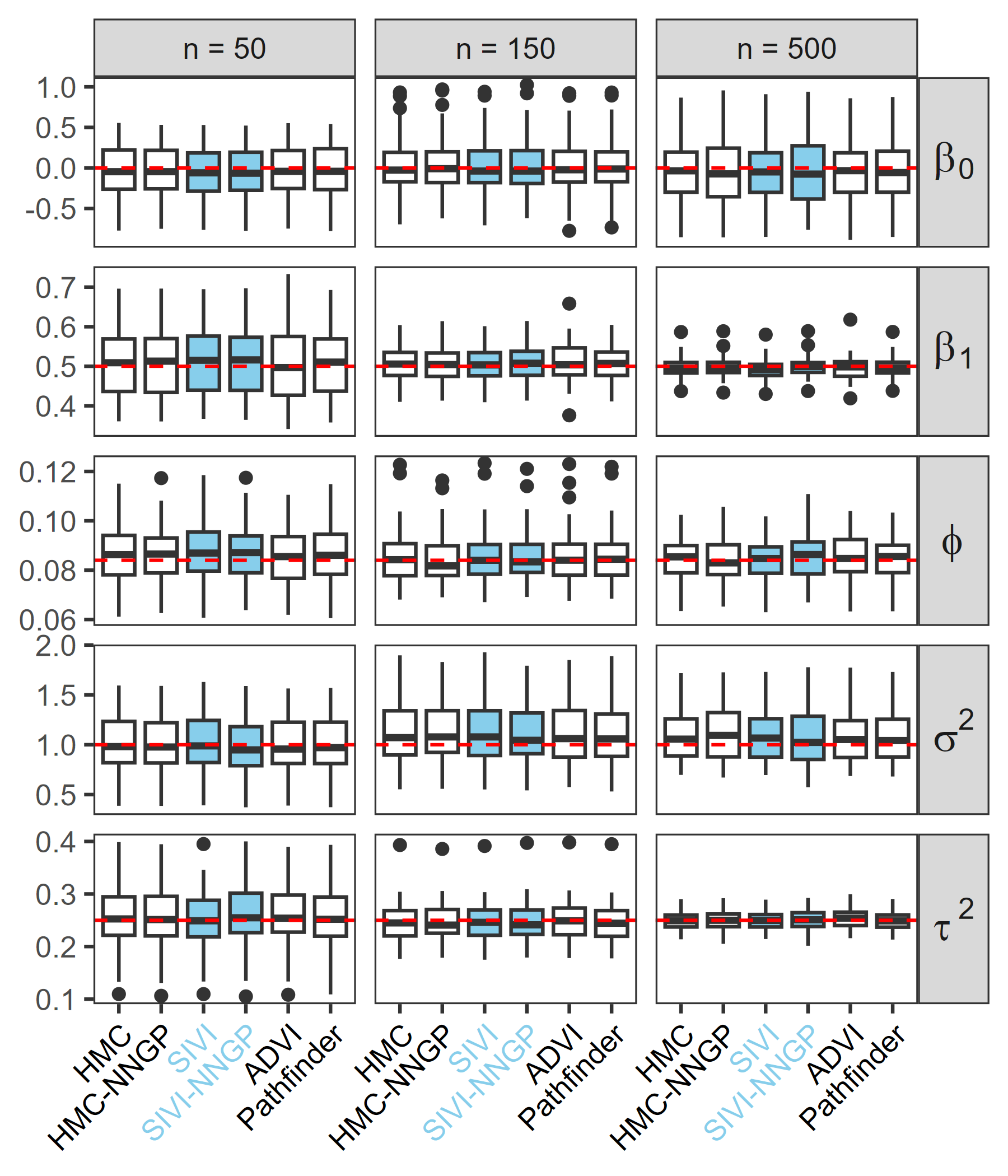}
    \end{minipage}

    \caption{Results across 50 replicates for a \textbf{marginal Gaussian} spatial model with a \textbf{Matérn 5/2 covariance} function. The left panel presents \textbf{predictive scores} and \textbf{runtime} (log scale), while the right panel presents distributions of the \textbf{posterior means} of model parameters. Predictive scores are based on 20 held-out locations. SIVI methods are highlighted in light blue, and dashed lines indicate the true parameter values. CRPS: continuous ranked probability score; IS: interval score; NLPD: negative log-predictive density. Methods with the \texttt{-NNGP} suffix use a nearest-neighbor Gaussian process prior.}
    \label{fig:Results_gaussian_mat}
\end{figure}

The left panel of Figure ~\ref{fig:Results_gaussian_mat} presents the predictive scores and runtime of the methods considered. The last row shows that SIVI becomes the fastest method once the training sample size reaches 150, with the computational advantage increasing at the largest training sample size of 500. For $n_{train} = 500$, SIVI takes, on average, 41 seconds with a dense GP prior and 24 seconds with an NNGP prior. On the other hand, HMC takes, on average, 2277 seconds ($\approx$ 38 minutes) with a dense GP prior and 1001 seconds ($\approx 16$ minutes) with an NNGP prior. All methods yield similar CRPS and IS values across sample sizes. HMC, ADVI, and Pathfinder obtain slightly higher NLPD scores than HMC-NNGP, SIVI, and SIVI-NNGP across the sample sizes. This difference may reflect the increased computational and inferential difficulty of the Matérn $5/2$ covariance function, which induces smoother spatial processes than the exponential covariance function. The distributions of the posterior means across the 50 replicates are shown in the right panel of Figure ~\ref{fig:Results_gaussian_mat}. The estimates are similar across methods for each sample size.

\begin{figure}[!ht]
\centering{
\includegraphics[width=0.8\textwidth]{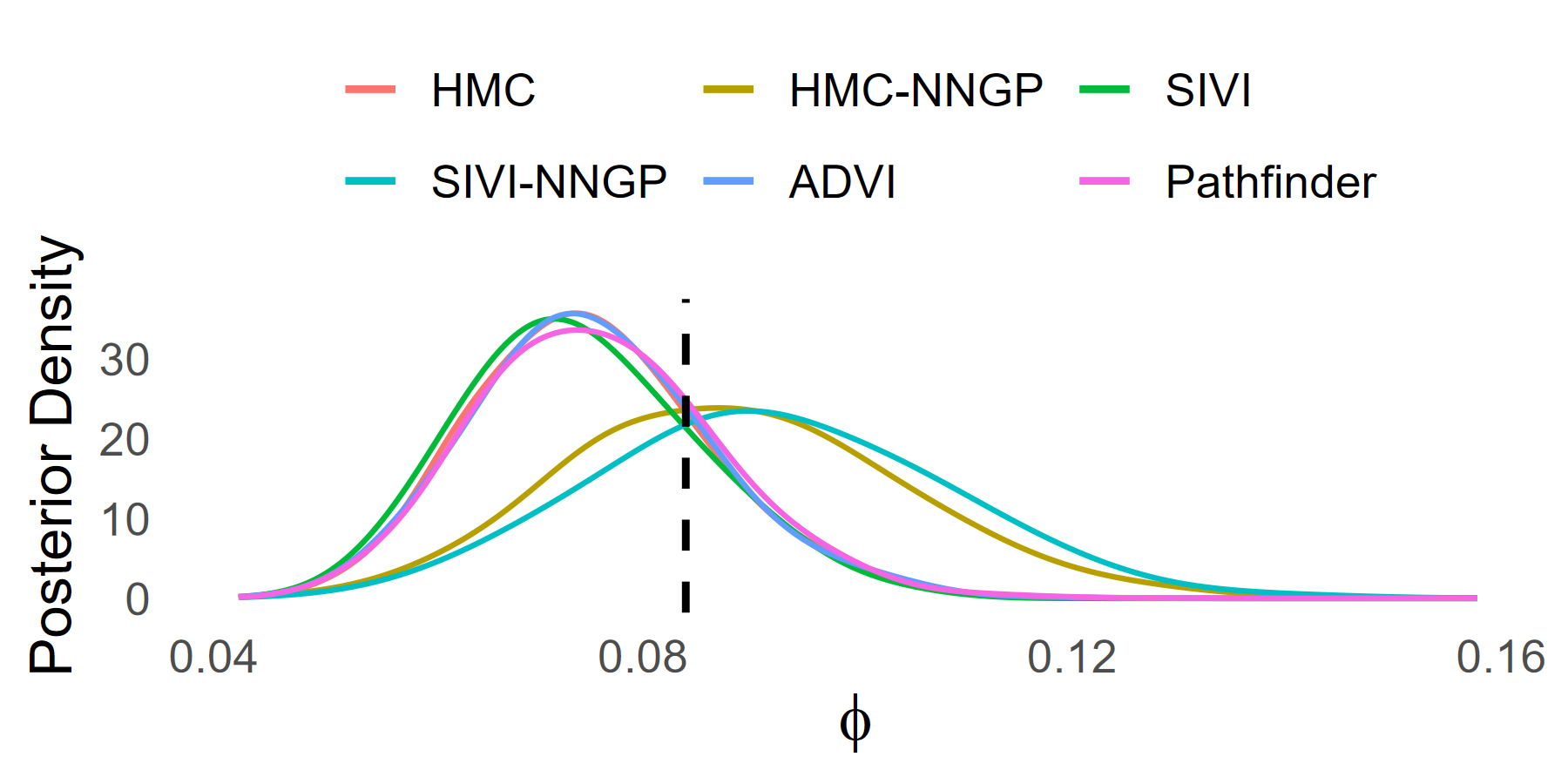}}
\caption{\textbf{Posterior density} of the range parameter $\phi$ for replicate 35 of the \textbf{marginal Gaussian} spatial model with a \textbf{Matérn 5/2 covariance} function. Methods with the \texttt{-NNGP} suffix use a nearest-neighbor Gaussian process prior.\label{fig:Dens_phi_gaussian_mat}}
\end{figure}

Figure \ref{fig:Dens_phi_gaussian_mat} shows the posterior density of the range parameter $\phi$ for replicate 35 (selected at random from replicates 1 to 50) across all considered models. Methods using a dense GP obtain similar posterior distributions, as do methods using an NNGP prior. 

\clearpage
\subsubsection{Non-Gaussian Spatial Models}

We use identical prior distributions as the ones used in the main manuscript for the exponential covariance function, with the exception of $\phi$, for which the prior distribution is $\phi \sim LN(\log(0.084),0.09)$. This gives a 95\% prior interval of $0.05-0.15$. 

\begin{figure}[!ht]
    \centering
    \begin{minipage}[t]{0.47\textwidth}
       \centering
        \includegraphics[
            width=\linewidth,
            keepaspectratio
        ]{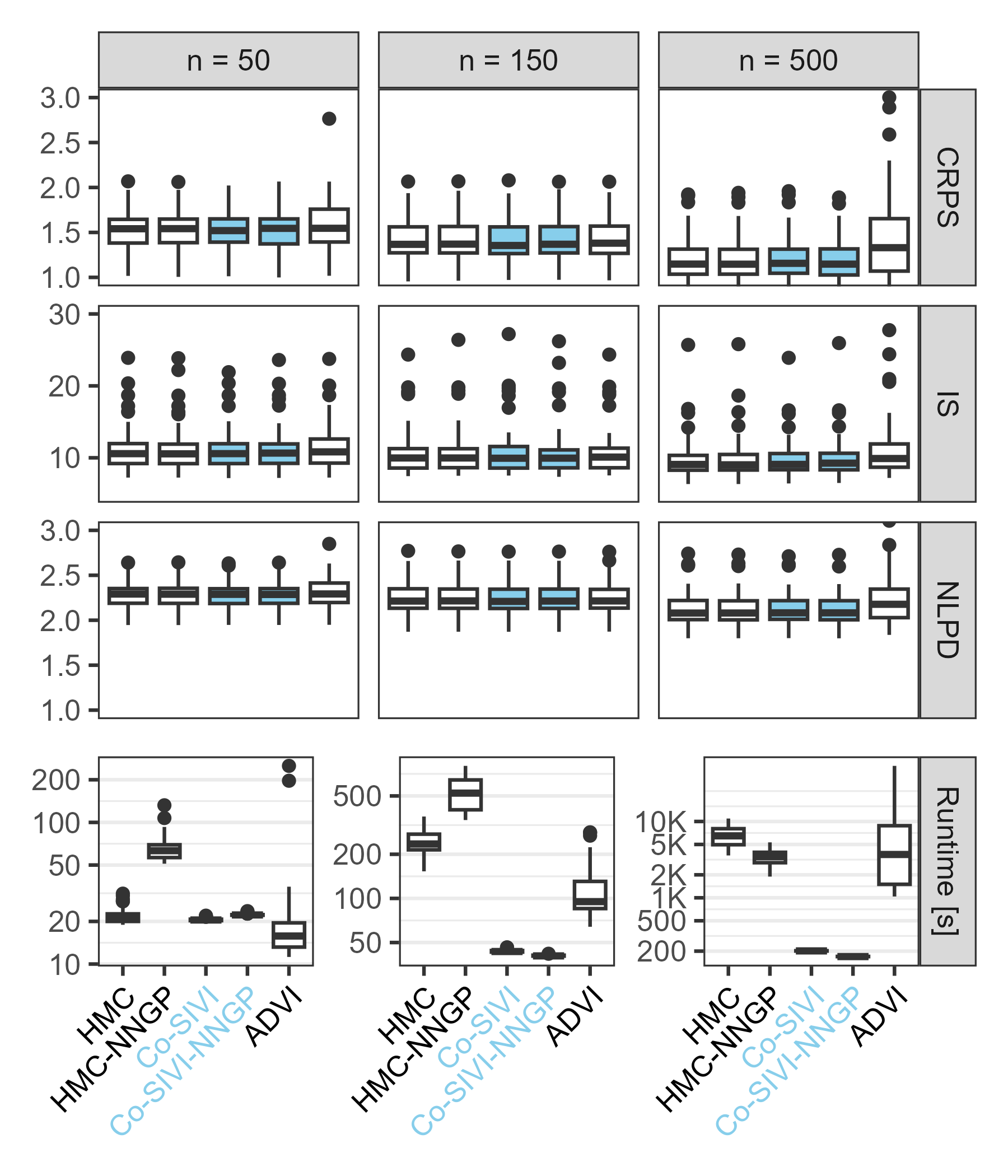}
   \end{minipage}
   \begin{minipage}[t]{0.47\textwidth}
        \centering
        \includegraphics[
            width=\linewidth,
            keepaspectratio
        ]{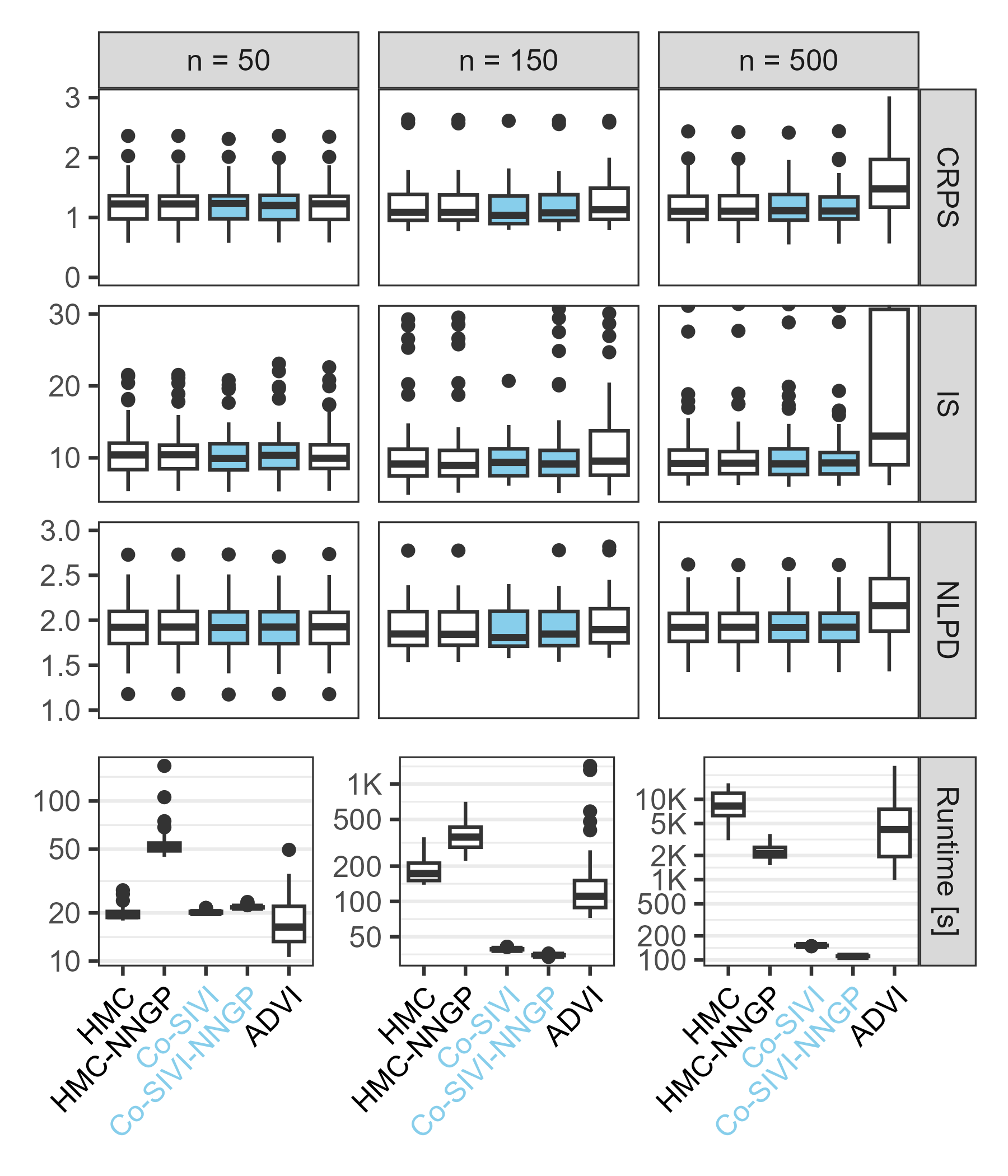}
    \end{minipage}

    \begin{minipage}[t]{0.47\textwidth}
        \centering
        \includegraphics[
            width=\linewidth,
            keepaspectratio
        ]{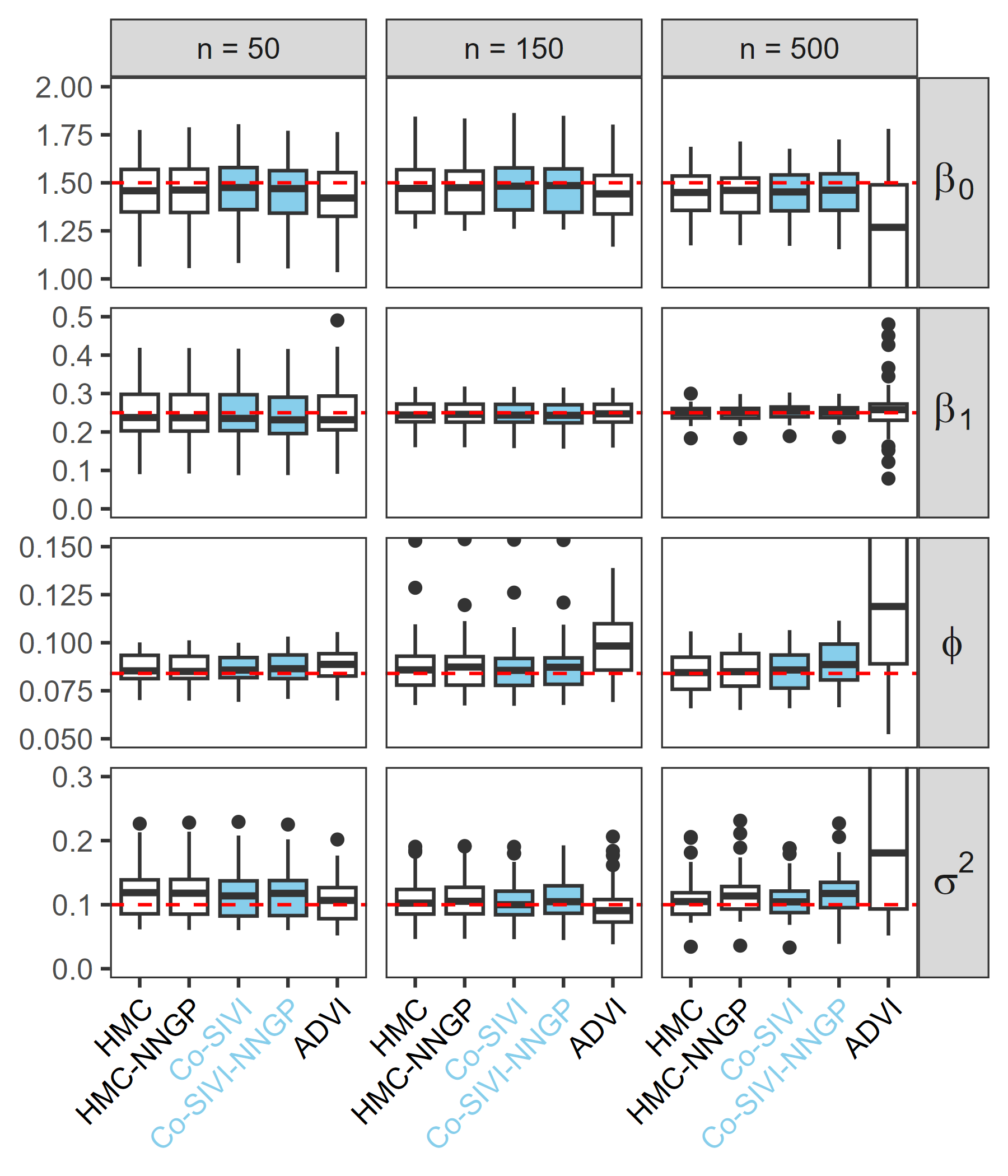}
    \end{minipage}
   \begin{minipage}[t]{0.47\textwidth}
        \centering
        \includegraphics[
            width=\linewidth,
            keepaspectratio
        ]{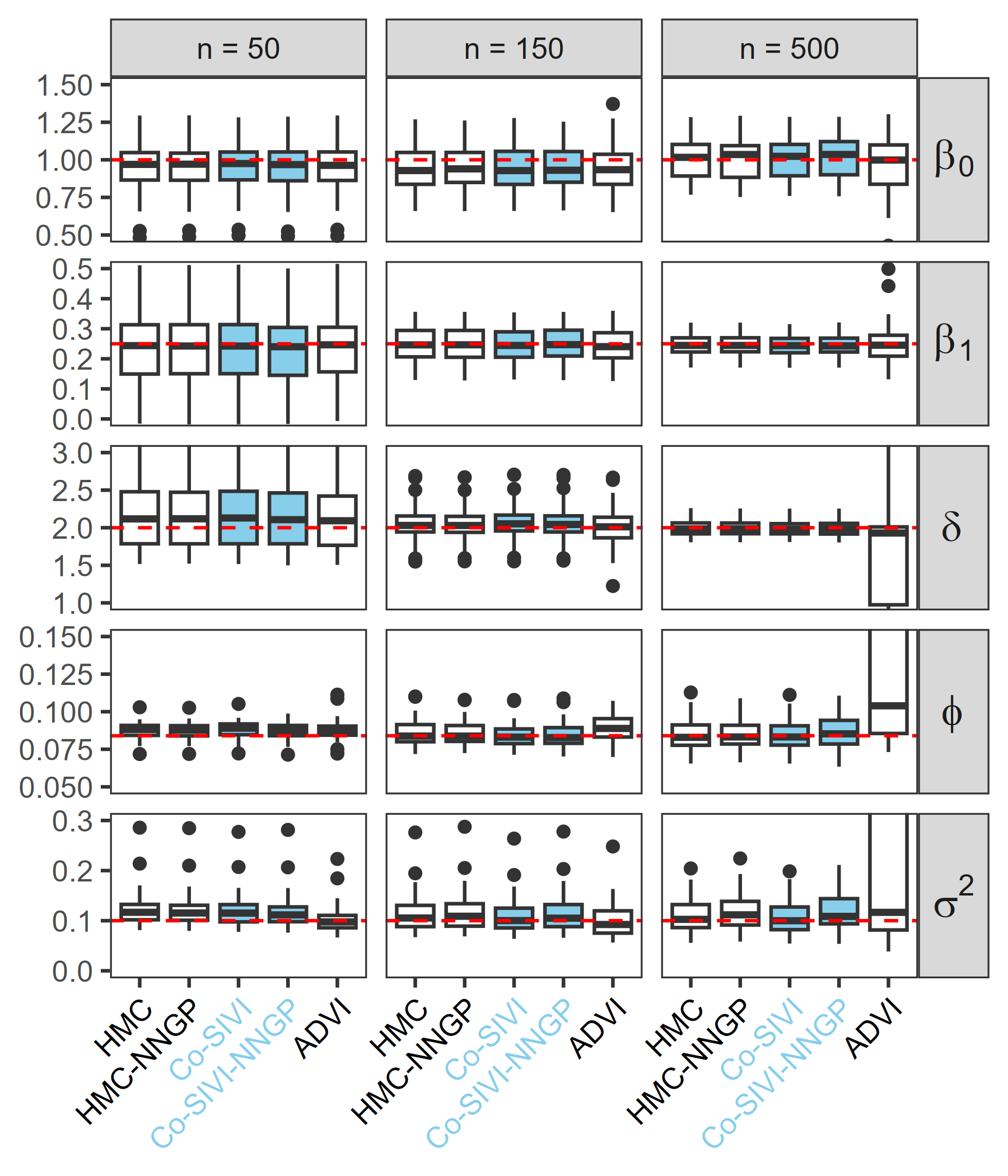}
    \end{minipage}

    \caption{Results across 50 replicates for the \textbf{Poisson} (left) and \textbf{Gamma} (right) spatial models with a \textbf{Matérn 5/2 covariance} function. The top row shows \textbf{predictive scores} and \textbf{runtime} (log scale), and the bottom row shows distributions of \textbf{posterior means}. Predictive scores are based on 20 held-out locations. Proposed methods are highlighted in light blue, and dashed lines indicate true parameter values. CRPS: continuous ranked probability score; IS: interval score; NLPD: negative log-predictive density. Methods with the \texttt{-NNGP} suffix use a nearest-neighbor Gaussian process prior.}
    \label{fig:Results_non_gaussian_mat}
\end{figure}

The predictive scores and runtime of the considered methods are presented on the top row of Figure ~\ref{fig:Results_non_gaussian_mat}, with results from the Poisson spatial model on the left and results from the Gamma spatial model on the right. For both models, Co-SIVI achieves predictive scores similar to HMC, whereas ADVI performs slightly worse on average for the largest training sample size considered, with some replicates yielding much larger prediction scores that are truncated in the figure. The runtime results also show that Co-SIVI becomes the fastest method once the training sample size reaches 150, with the computational advantage increasing at the largest training sample size of 500. For the Poisson spatial model and $n_{train}= 500$, Co-SIVI takes, on average, 202 seconds with a dense GP prior and 170 seconds with an NNGP prior. On the other hand, HMC takes, on average, 6651 seconds ($\approx$ 110 minutes) with a dense GP prior and 3449 seconds ($\approx 57$ minutes) with an NNGP prior. For the Gamma spatial model and $n_{train}=500$, Co-SIVI takes, on average,  152 seconds with a dense GP prior and 111 seconds with an NNGP prior. On the other hand, HMC takes, on average, 8896 seconds ($\approx$ 148 minutes) with a dense GP prior and 2282 seconds ($\approx $ 38 minutes) with an NNGP prior.

The distributions of the posterior means across the 50 replicates are shown on the bottom row of Figure \ref{fig:Results_non_gaussian_mat}, with the results from the Poisson spatial model on the left and results from the Gamma spatial model on the right. For both models, the estimates are similar for Co-SIVI and HMC, with a small difference between methods with a dense GP prior and methods with an NNGP prior. However, for the largest sample sizes, ADVI struggles to recover posterior means comparable to those obtained by HMC for the range parameter $\phi$, the marginal variance $\sigma^2$, and the dispersion parameter $\delta$ for the Gamma spatial model, which helps explain the larger prediction scores observed for ADVI.

\begin{figure}[!ht]
\centering{
\includegraphics[width=0.5\textwidth]{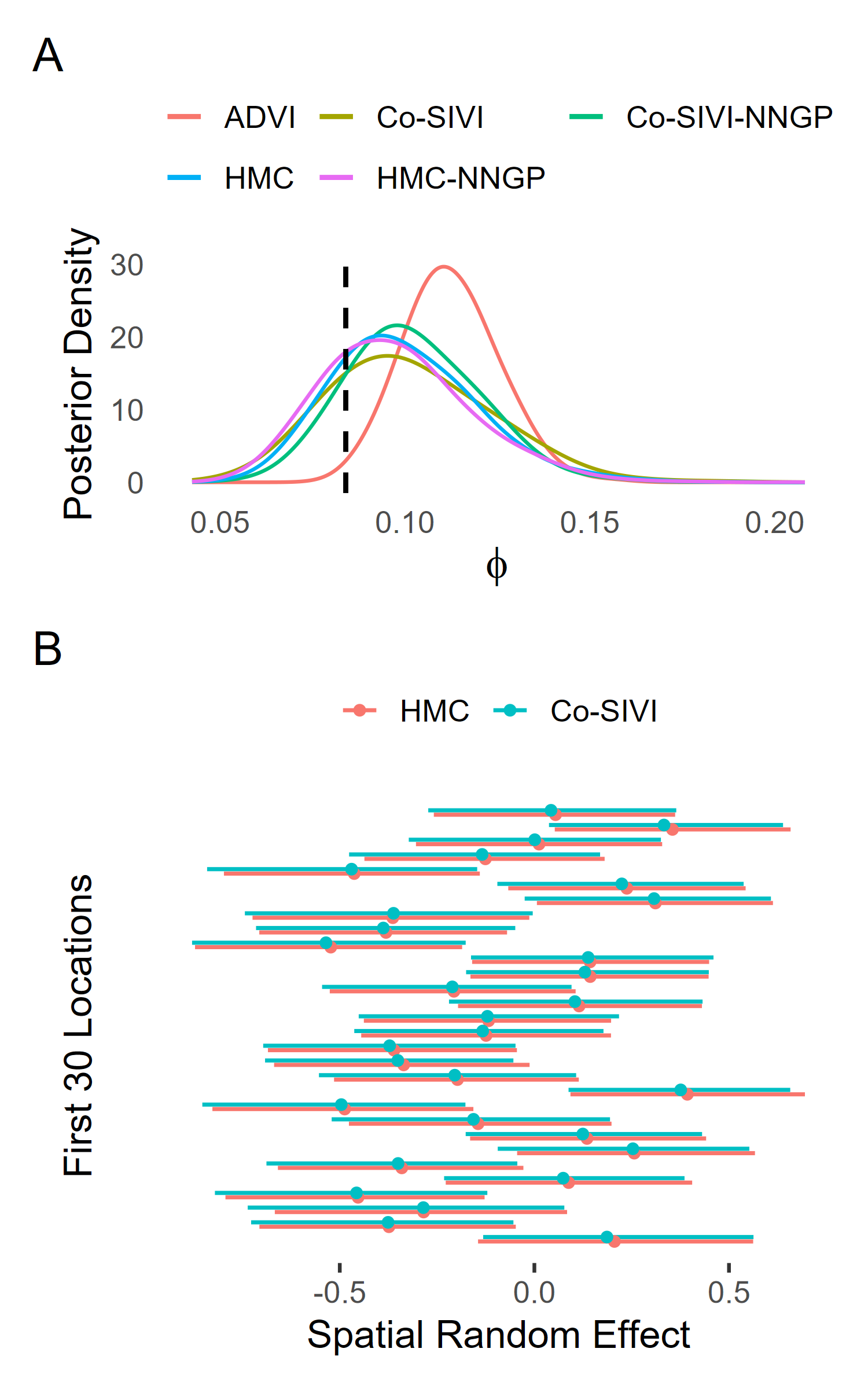}
}
\caption{\textbf{Posterior density} of the range parameter $\phi$ (A) and $95\%$ credible intervals for the first 30 spatial random effects (B) for replicate 40 of the \textbf{Poisson} spatial model with a \textbf{Matérn 5/2 covariance} function. The \texttt{-NNGP} suffix indicates use of a nearest-neighbor Gaussian process prior.\label{fig:Dens_poisson_mat}}
\end{figure}

Panel A of Figure \ref{fig:Dens_poisson_mat} shows the posterior density of the range parameter $\phi$ for replicate 40 (selected at random from replicates 1 to 50)  across all considered models. Co-SIVI and HMC methods yield similar posterior distributions, whereas ADVI does not closely approximate the posterior distribution obtained by HMC. Panel B of Figure \ref{fig:Dens_poisson_mat} shows the $95\%$ credible intervals of the first 30 random effects for Co-SIVI and HMC. Co-SIVI yields credible intervals very similar to those from HMC. 

Panel A of Figure \ref{fig:Dens_poisson_mat} shows the posterior density of the range parameter $\phi$ for replicate 19 (selected at random from replicates 1 to 50) across all considered models except ADVI, as it produces a posterior distribution far to the right. Co-SIVI and HMC methods yield similar posterior distributions. Panel B of Figure \ref{fig:Dens_poisson_mat} shows the $95\%$ credible intervals of the first 30 random effects for Co-SIVI and HMC. Co-SIVI produces credible intervals very similar to those from HMC.

\begin{figure}[!ht]
\centering{
\includegraphics[width=0.5\textwidth]{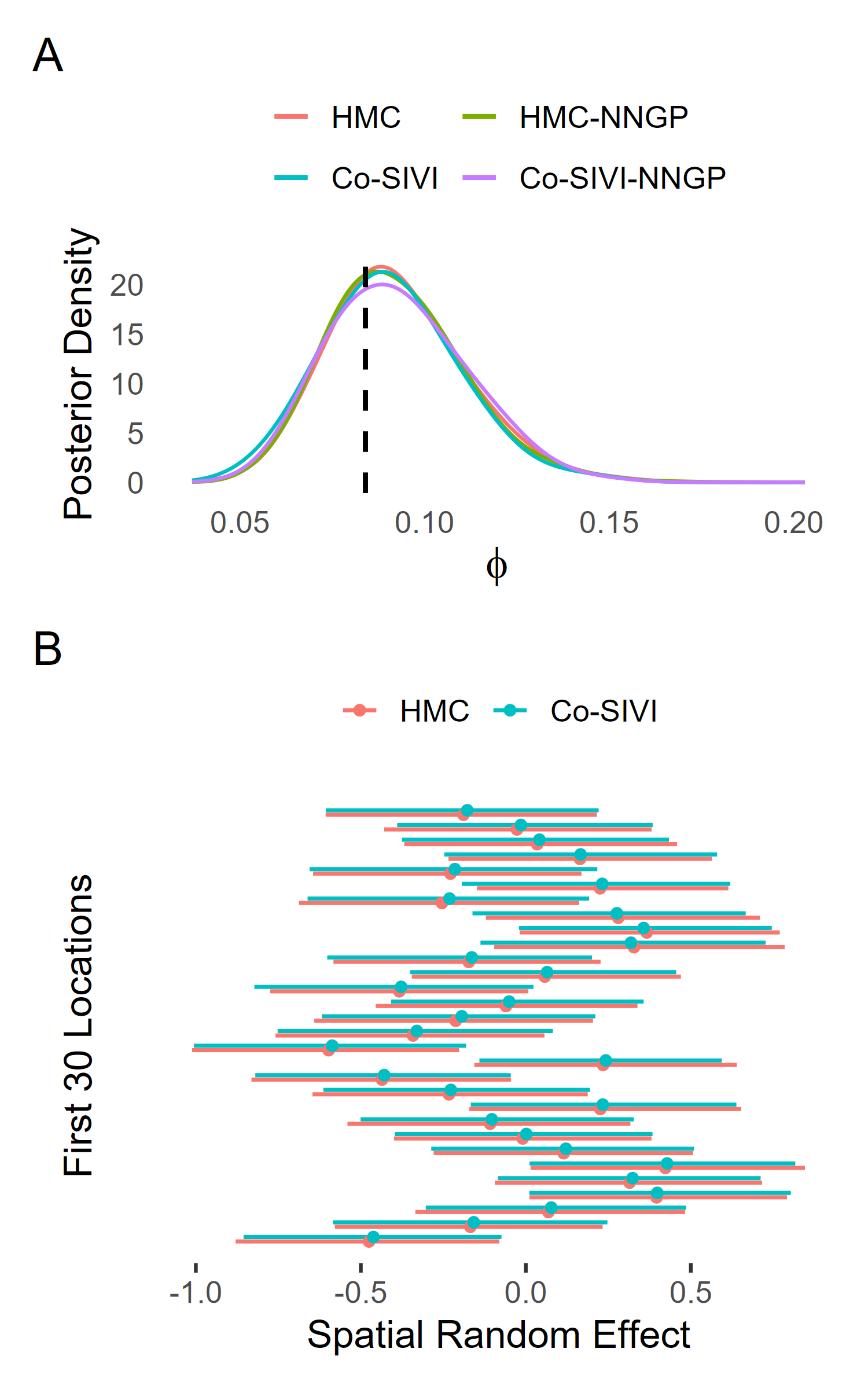}
}
\caption{\textbf{Posterior density} of the range parameter $\phi$ (A) and $95\%$ credible intervals for the first 30 spatial random effects (B) for replicate 19 of the \textbf{Gamma} spatial model with a \textbf{Matérn 5/2 covariance} function. The \texttt{-NNGP} suffix indicates use of a nearest-neighbor Gaussian process prior.\label{fig:Dens_gamma_mat}}
\end{figure}

\clearpage
\setcounter{figure}{0}
\section{Additional Results for the Housing Dataset}\label{sec:housing_extra_res}
\subsection{Posterior Densities for All Considered Methods}

\begin{figure}[!hbt]
\centering{
\includegraphics[width=\columnwidth]{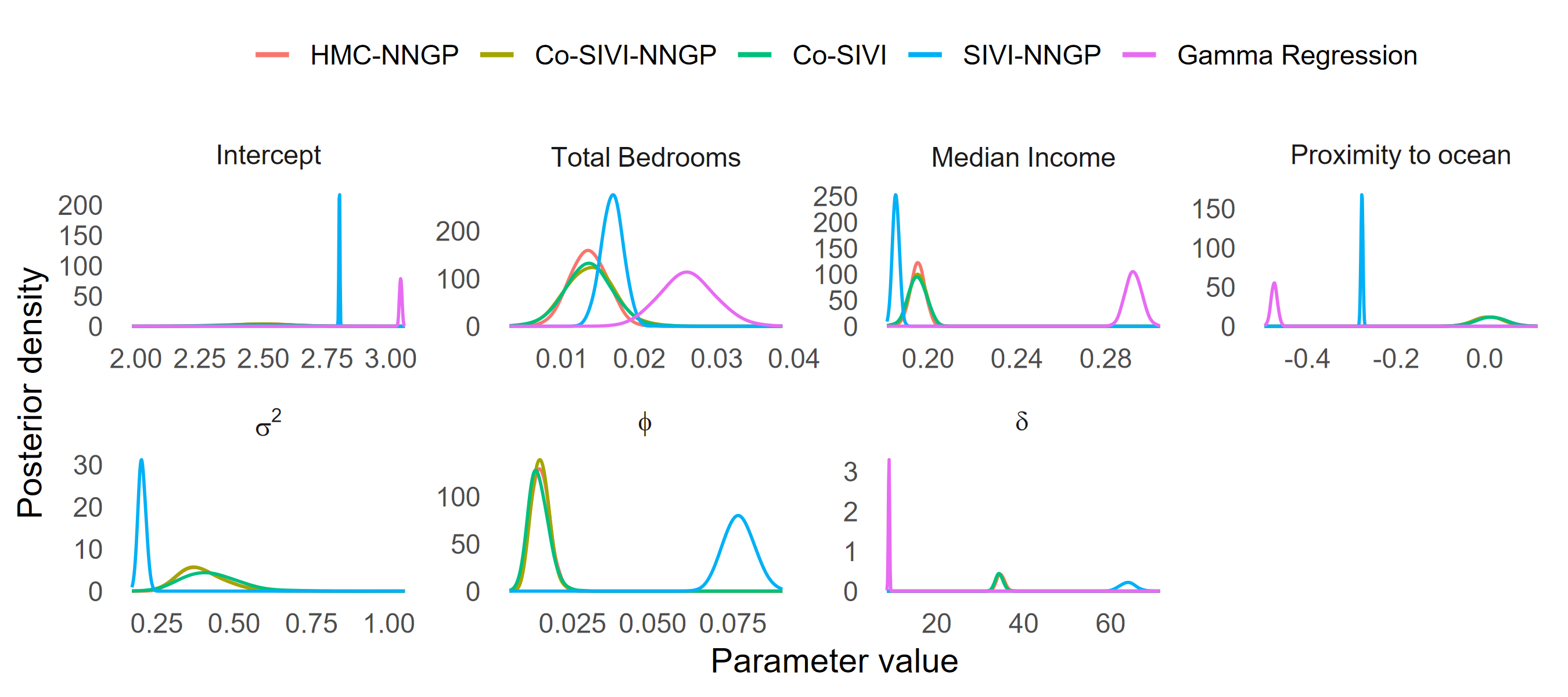}
}
\caption{\textbf{Posterior densities} for all model parameters obtained from HMC-NNGP, Co-SIVI-NNGP, Co-SIVI with a dense GP prior, SIVI-NNGP, and a Bayesian Gamma regression without random effects. The Gamma regression does not provide posterior densities for $\phi$ and $\sigma^2$.\label{fig:housing_dens_full}}
\end{figure}

Figure \ref{fig:housing_dens_full} presents the posterior densities of the model parameters obtained by HMC-NNGP, Co-SIVI-NNGP, Co-SIVI, SIVI-NNGP, and a Gamma regression. Since there are no random effects in the Gamma regression, there are no posterior densities for $\phi$ and $\sigma^2$. SIVI-NNGP and the Gamma regression produce posterior densities that differ substantially from those obtained by HMC-NNGP, Co-SIVI-NNGP, and Co-SIVI.

\subsection{Comparison of Posterior Samples for the Predicted Random Effects}
\begin{figure}[!hbt]
\centering{
\includegraphics[width=0.7\textwidth,
keepaspectratio]{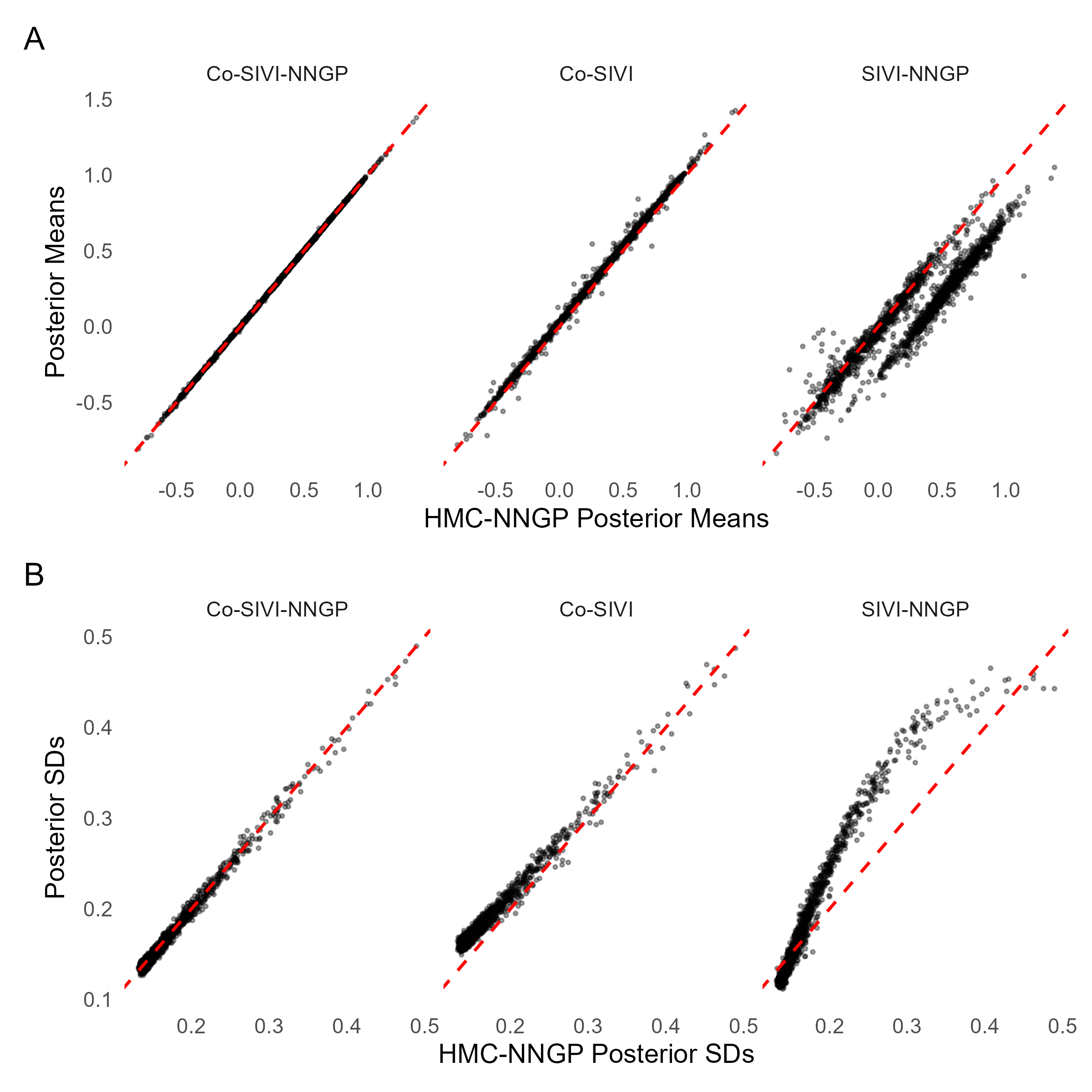}
}
\caption{Scatter plots of posterior means (A) for the predicted random effects obtained with Co-SIVI-NNGP, Co-SIVI with a dense GP prior, and SIVI-NNGP against those obtained with HMC. Corresponding scatter plots of posterior standard deviations (B) (SDs). \label{fig:Housing_re_scatter}}
\end{figure}

Figure \ref{fig:Housing_re_scatter} presents scatter plots comparing the posterior means and standard deviations (SDs) of the predicted spatial random effects obtained with each SIVI-based method against those obtained with HMC-NNGP. Co-SIVI-NNGP produces posterior means and SDs that are well aligned with those obtained by HMC-NNGP.  Co-SIVI also produces well-aligned posterior means, but slightly overestimates the posterior SDs. This difference may partly reflect the use of different priors on the spatial random effects (GP and NNGP) or differences in how uncertainty is allocated among the intercept, the spatial random effects, and their posterior dependence. On the other hand, SIVI-NNGP struggles to recover posterior means and SDs comparable to those obtained by HMC-NNGP.

\subsection{Map of the Predicted Spatial Random Effects}

\begin{figure}[!ht]
\centering{
\includegraphics[width=\textwidth]{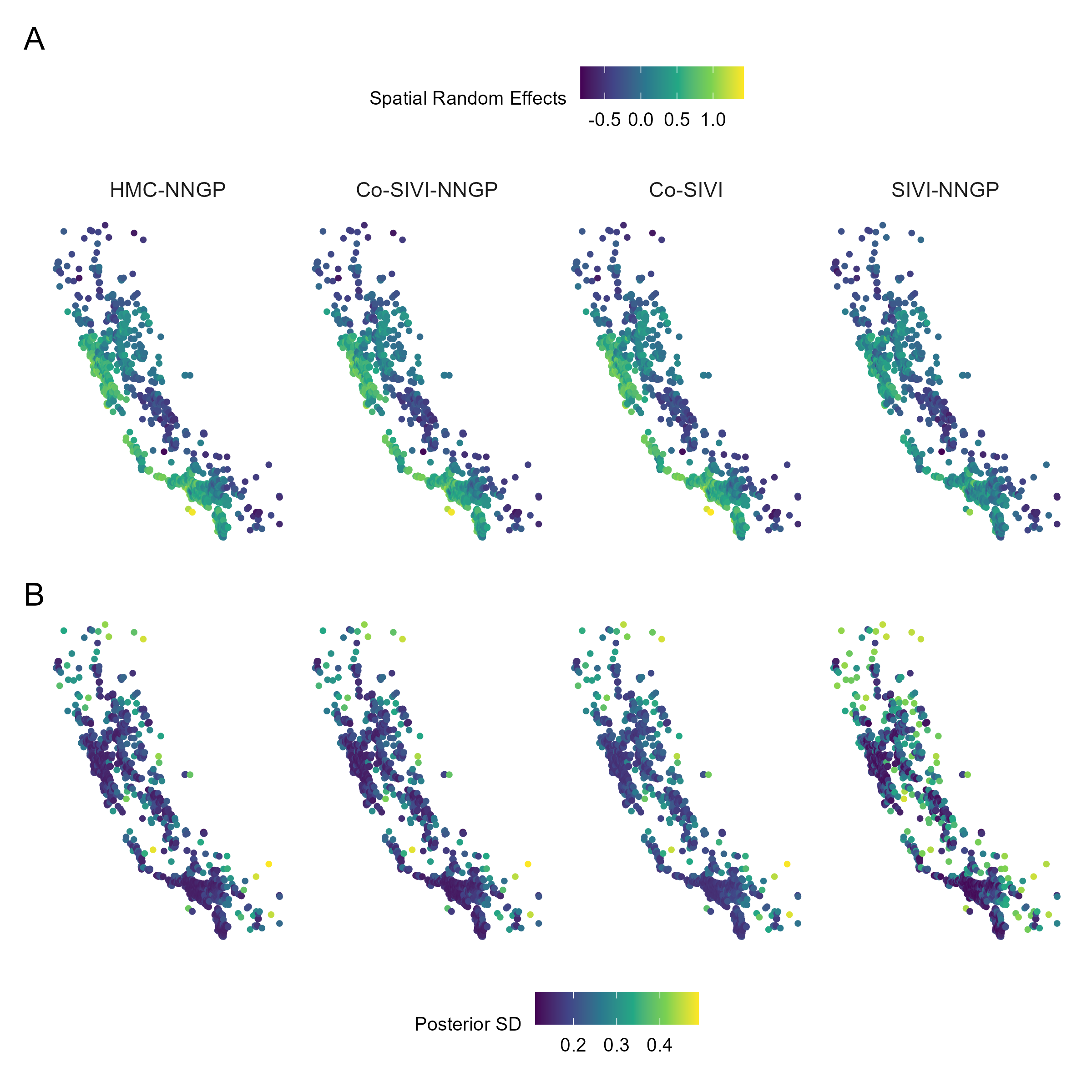}
}
\caption{Posterior means of the predicted random effects at validation locations (A) for HMC-NNGP, Co-SIVI-NNGP, Co-SIVI with a dense GP prior, and SIVI-NNGP. Corresponding posterior standard deviations(B) (SDs).\label{fig:Housing_re_map}}
\end{figure}

Figure \ref{fig:Housing_re_map} compares the posterior means and standard deviations (SDs) of the predicted spatial random effects obtained by HMC-NNGP, Co-SIVI-NNGP, Co-SIVI and SIVI-NNGP. As seen in the top row of Figure \ref{fig:Housing_re_map}, all methods recover a similar spatial pattern, with the largest and smallest effects occurring in the same regions. However, SIVI-NNGP generally yields lower point estimates than HMC, as reflected by the overall darker shading of the points.

\clearpage
\subsection{Comparison of the Predicted Values}

\begin{figure}[htbp]
\centering{
\includegraphics[width=\textwidth]{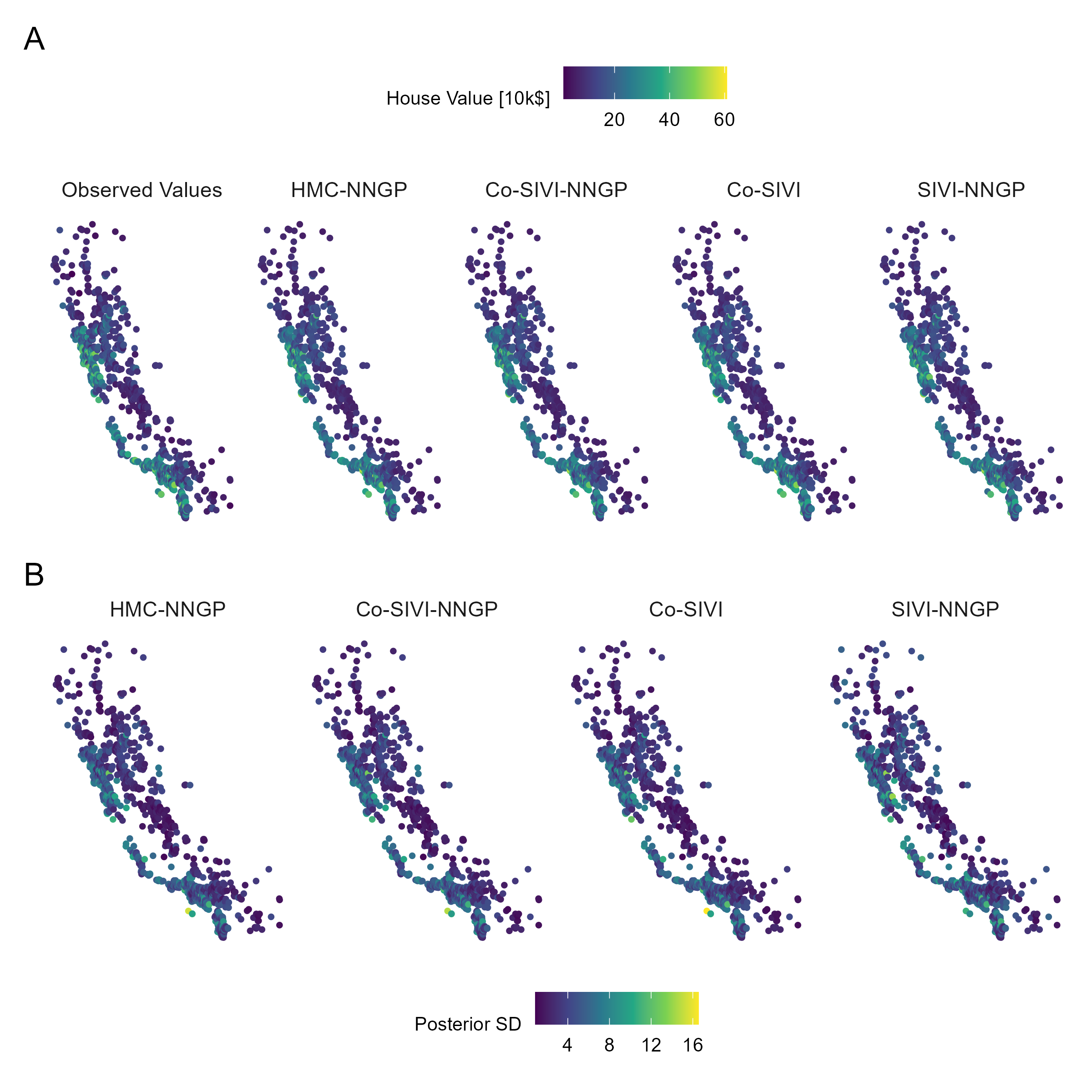}
}
\caption{Observed median house prices, in tens of thousands of US dollars, and posterior means (A) at the California validation locations for HMC-NNGP, Co-SIVI-NNGP, Co-SIVI with a dense GP prior, and SIVI-NNGP. Corresponding posterior standard deviations (B) (SDs).\label{fig:Housing_map}}
\end{figure}

\begin{figure}[htbp]
\centering{
\includegraphics[width=0.7\textwidth]{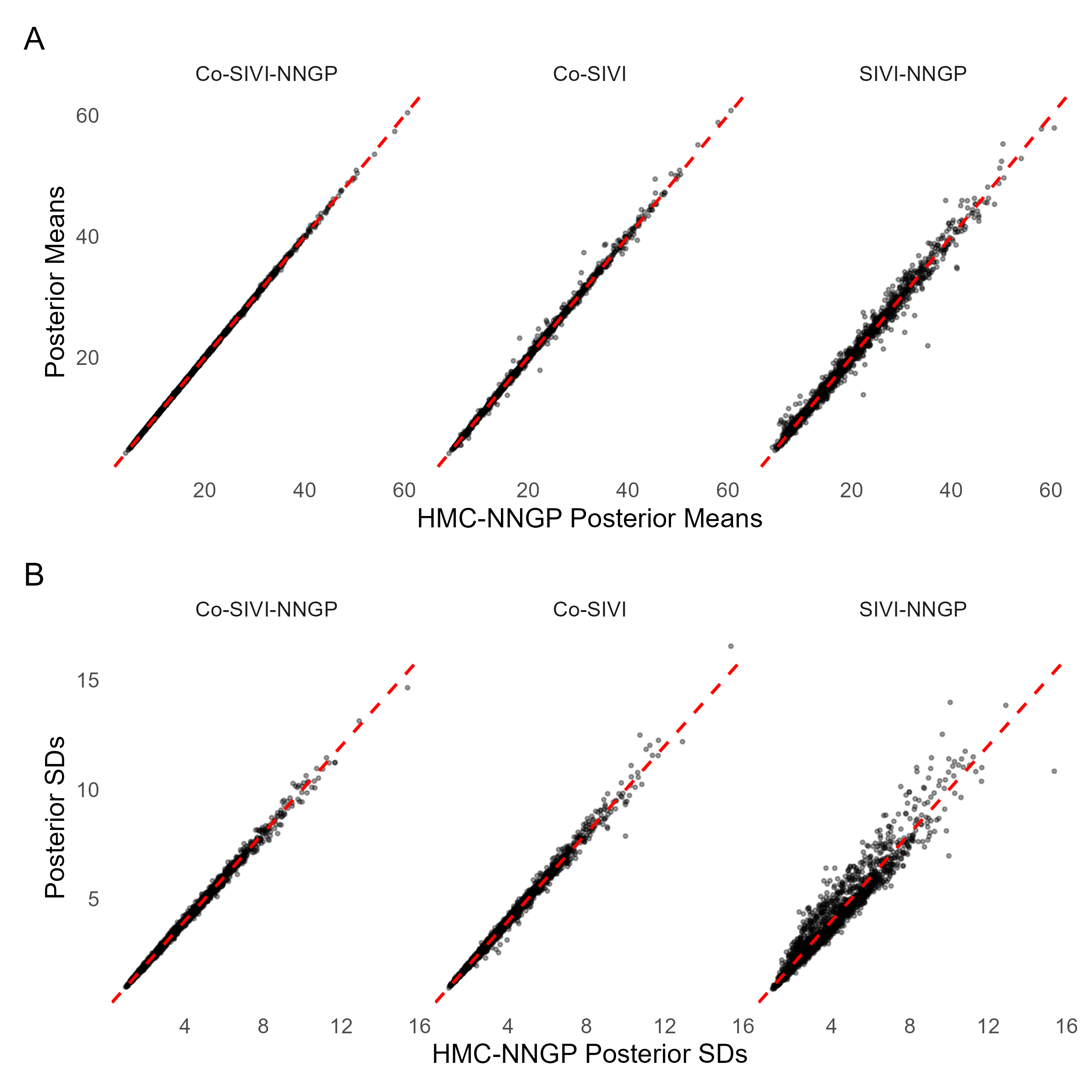}
}
\caption{Scatter plots of posterior means (A) for the predicted values obtained with Co-SIVI-NNGP, Co-SIVI with a dense GP prior, and SIVI-NNGP against those obtained with HMC. Corresponding scatter plots of posterior standard deviations (B) (SDs). \label{fig:Housing_hmc_scatter}}
\end{figure}

Figure \ref{fig:Housing_map} compares the posterior means and standard deviations (SD) of the predicted values obtained with HMC-NNGP, Co-SIVI-NNGP, Co-SIVI and SIVI-NNGP. All methods recover a similar spatial pattern, with the largest and smallest effects being in the same regions. Figure \ref{fig:Housing_hmc_scatter} presents scatter plots comparing the posterior means and SDs of the predicted values obtained with each SIVI-based method against those obtained with HMC-NNGP. The posterior means and SDs obtained with Co-SIVI-NNGP and Co-SIVI are closely aligned with those obtained with HMC-NNGP, whereas those obtained with SIVI-NNGP show a greater range around perfect agreement.

\clearpage

\bibliographystyle{plainnat}
\bibliography{bibliography}

\end{document}